\documentclass[12pt,english]{article}
\usepackage[utf8]{inputenc}
\usepackage[backend=biber,style=apa]{biblatex}
\usepackage{hyperref}
\usepackage{amssymb}
\usepackage{amsmath}
\usepackage{stmaryrd}
\usepackage{calrsfs}
\usepackage{geometry}
\usepackage{graphicx}
\usepackage{color}
\usepackage{caption}
\usepackage{subfigure}
\usepackage{float}
\usepackage{setspace}
\usepackage{empheq, bigstrut}
\usepackage{booktabs}
\usepackage{tabularx}
\usepackage{lscape}
\usepackage[active]{srcltx}
\usepackage{setspace}
\usepackage{ltxtable}
\makeatletter
\providecommand{\tabularnewline}{\\}
\newcolumntype{R}{>{\raggedleft\arraybackslash}X}
\newcommand{\PreserveBackslash}[1]{\let\temp=\\#1\let\\=\temp}

\makeatother
\usepackage{babel}
\begin{document}
\setcounter{page}{0}

\begin{singlespace}
\title{Blockchain scaling and liquidity concentration on decentralized exchanges
\thanks{We would like to thank Marius Zoican, Alfred Lehar, Gleb Kurovskiy, 
Fahad Saleh, Norman Schürhoff, Andrea Barbon, Fayçal Drissi, Ganesh Viswanath-Natraj, Roman Kozhan,  Nir Chemaya, Dingyue Liu, Peter O'Neill
and seminar participants at Warwick Business School, Gillmore Center
for Financial Technology, UCSB-ECON DeFi Seminar, World Federation of Exchanges and Oxford-Man Institute (OMI) Finance Seminar for insightful comments on this paper. This
paper has been presented at the 2023 Edinburgh Economics of Financial Technology
Conference and 6th Sydney Market Microstructure and Digital Finance Meeting. The authors declare that they have no known competing
financial interests or personal relationships that could have appeared
to influence the work reported in this paper. This research did not
receive any specific grant from funding agencies in the public, commercial,
or not-for-profit sectors.}
}

\author{Basile Caparros\thanks{ChainSolid (basile@chainsolid.com)}, Amit Chaudhary\thanks{Warwick Business School, Gillmore Centre for Financial Technology (amit.chaudhary.3@warwick.ac.uk), ChainSolid (amit@chainsolid.com) - I thank Prabal Banerjee for valuable feedback and comments.}, Olga Klein\thanks{Warwick Business School, Gillmore Center for Financial Technology (olga.klein@wbs.ac.uk)}}

\maketitle
\thispagestyle{empty}
\end{singlespace}

\begin{abstract}
\begin{singlespace}
Liquidity providers (LPs) on decentralized exchanges (DEXs) can protect themselves from adverse selection risk by updating their positions more frequently. However, repositioning is costly, because LPs have to pay gas fees for each update. We analyze the causal relation between repositioning and liquidity concentration around the market price, using the entry of blockchain scaling solutions, Arbitrum and Polygon, as our instruments. Lower gas fees on scaling solutions allow LPs to update more frequently than on Ethereum. Our results demonstrate that higher repositioning intensity and precision lead to greater liquidity concentration, which benefits small trades by reducing their slippage.
\end{singlespace}
\end{abstract}

\vspace{1.5cm}

\textbf{JEL classifications:} G14, G18, G19

\begin{singlespace}
\textbf{Keywords:} decentralized exchanges, FinTech, gas fees, liquidity concentration, market depth, slippage
\end{singlespace}

\maketitle
\thispagestyle{empty}

\pagebreak{}
\section{Introduction}

One of the major cryptoasset exchanges, FTX, filed for bankruptcy
on November 11, 2022. Crucially, FTX kept custody of its client deposits.
Thus, whereas the exact reasons for its collapse were the subject of investigation
by the US Securities and Exchange Commission (SEC), at least \$1B
of its customers' funds have vanished, according to Reuters\footnote{``Exclusive: At least \$1 billion of client funds missing at failed
crypto firm FTX'', Reuters, November 13, 2022. Available at https://www.reuters.com/markets/currencies/exclusive-least-1-billion-client-funds-missing-failed-crypto-firm-ftx-sources-2022-11-12/}. FTX was a centralized exchange (CEX), with all trades taking place
through a limit order book (LOB). In any LOB-based market,
a central institution, such as FTX, is needed to match trades and
keep records of all transactions. 

A recent alternative to CEXs are so-called decentralized exchanges
(DEXs) that operate directly on the blockchain. In contrast to CEXs,
users themselves keep custody of their assets on DEXs. They
execute their trades directly from their wallet, using smart contracts,
i.e. a set of pre-programmed rules.\footnote{\textcite{aspris2021} also argue that decentralized
exchanges have higher security, because assets are never transferred
to the custody of a third party. \textcite{lehar2021decentralized}
and \textcite{barbon2021quality} provide a comprehensive
overview of differences between CEXs and DEXs. However, allowing users
to keep custody of their own assets is the most important advantage of DEXs.} Therefore, executing directly on the blockchain provides higher level
of security for clients' funds. Indeed, there is evidence of significant
trading volume shifting to DEXs in the week of FTX collapse. According
to Reuters, ``volumes at the largest DEX, Uniswap, spiked to \$17.2
billion in the week of Nov 6-13, from just over \$6 billion the week
before''.\footnote{``Cryptoverse: Let's talk about DEX, baby'', Reuters, November 22,
2022. Available at https://www.reuters.com/markets/currencies/cryptoverse-lets-talk-about-dex-baby-2022-11-22/} 

This higher level of security on DEXs comes, however, at a cost. First,
transactions are overall slower, since they have
to be validated by ``miners'' or ``validators'' before being
recorded on the blockchain. Second, users of the blockchain have to
compensate validators with the so-called ``gas fees'' for their
efforts. Importantly, ``gas fees'' are paid not only by liquidity
demanders, but also by liquidity providers (LPs). Essentially, gas
fees represent a fixed cost and act as a ``transaction tax'' for every transaction added to
the blockchain. Liquidity demanders pay gas fees whenever they would
like to execute a trade.\footnote{Gas fees are in addition to standard exchange fees that both CEXs and
DEXs charge for trade execution. Prior findings by \textcite{barbon2021quality}
suggest that gas fees on the main blockchain for DEXs, Ethereum, indeed
represent a considerable portion of execution cost for traders on
Uniswap v2.} Liquidity providers pay gas fees whenever they deposit or withdraw liquidity
from the exchange.

In this paper, we analyze the effects of repositioning intensity,
i.e. frequency of position updates by LPs, on liquidity concentration
on the largest DEX, Uniswap v3. Much alike traditional market makers,
LPs are subject to adverse selection risk \parencite{lehar2021decentralized, capponi2021adoption, oneill2023}.
To protect themselves from being adversely selected, LPs need to update
their positions in response to changes in the market price, similar
to canceling and reposting limit orders in the LOB. With lower gas
fees, it is cheaper for LPs to update their liquidity positions. Consequently,
they can reposition more frequently, which results in better protection
from adverse selection.\footnote{\textcite{lehar2022liquidity} discuss trading mechanics
on Uniswap v3 in great detail and also argue that liquidity updating
should be more frequent with lower gas fees.} With better protection from adverse selection, LPs can earn higher
fees (for the same amount of capital deposited) by setting narrower
price ranges around the market price, i.e. their positions become
more concentrated. Thus, we expect that a higher repositioning intensity by LPs leads to higher concentration of aggregate liquidity around
the market price.

In contrast, if gas fees are high, LPs should only update their positions
if the market price sufficiently deviates from their position's price
range, as repositioning is costly. Therefore, we expect LPs to set
wider price ranges when they are not able to update frequently, in
order to reduce adverse selection risk. Lower repositioning intensity
should thus result in lower concentration of aggregate liquidity in
the pool. Overall, concentration of aggregate pool liquidity around
the market price is important, because higher liquidity concentration translates
into lower slippage and thus lower execution cost for traders.

Testing the empirical relation between repositioning intensity and
liquidity concentration might be problematic, because repositioning
intensity is potentially endogenous. A choice of an LP to update their
position can in itself depend on the current liquidity concentration
in the pool. To identify the causal effect of repositioning intensity,
we use the launch of Uniswap v3 on two Ethereum scaling solutions, Arbitrum and Polygon, as our instruments. 
Blockchain scaling solutions allow for a quicker validation
of transactions, with an average block time of 0.25-3 seconds, relative to Ethereum's 12 seconds.\footnote{There is a trade-off between quicker transaction validation and security of scaling solutions, which we discuss in detail in Section 2.2.  See also \textcite{chemaya2022} for estimation of investors' preferences for blockchain security, using a structural model.} Moreover, scaling solutions charge significantly lower gas fees of only around \$0.01 on Polygon and \$0.20-\$2 on Arbitrum per trade on Uniswap v3, as compared to an average gas fee of \$14 on Ethereum. Importantly, lower gas fees on scaling solutions allow
LPs to update their positions more frequently. Therefore, we use the launch of Uniswap v3 on scaling solutions as our instrument for an exogenous increase in repositioning
intensity of LPs.


We test our predictions above using the two most liquid pairs on Uniswap v3, ETH/USDC 0.05\% and
ETH/USDC 0.3\%, that are traded across all three chains: Ethereum, Arbitrum and Polygon. Hence, our benchmark dataset consists of six distinct liquidity pools.\footnote{The sample of pools that are traded across all three chains and are still sufficiently liquid is quite limited. We also replicate our main results for four additional pairs, BTC/ETH 0.05\%, BTC/ETH 0.3\%, UNI/ETH 0.3\% and LINK/ETH 0.3\%, that are relatively actively traded across all three chains, thus mapping into twelve additional pools. All our main results are robust and continue to hold for these four additional pairs.}  Our sample period starts in January 2022, after the launch of Uniswap v3 on both Arbitrum and Polygon, and ends in June 2023. 

We first examine trading volume, trade size and total value locked (TVL) in liquidity pools across all chains.\footnote{Total value locked (TVL) is equivalent to total market depth in
the traditional LOB.} Overall, we observe significantly larger trades, volumes and TVL on Ethereum, relative to scaling solutions. These results are largely due to higher security of Ethereum, driven by its high number of validators. 
In contrast, LPs are more reluctant to deposit large
amounts of liquidity on less secure scaling solutions, resulting in a lower
overall TVL. Similar to a separating equilibrium,
larger TVL on Ethereum attracts larger trades, for which greater liquidity is more important than high gas fees. In contrast, blockchain scaling solutions are more attractive
for smaller liquidity providers and smaller traders, who are primarily
concerned about gas fees, and less so about security.

Importantly, despite their lower TVL, we observe higher aggregate liquidity concentration around the market price on blockchain scaling solutions. Our benchmark measure of liquidity concentration is market depth within 2\% of the market price, divided by TVL (i.e. total market depth).\footnote{All our results hold if we use 1\% or 10\% as a cutoff instead of 2\%.} We hypothesize that this higher liquidity
concentration is caused by higher repositioning intensity of LPs on Arbitrum and Polygon.
As discussed above, we use the launch of Uniswap v3 on Arbitrum and Polygon as an instrument for
an exogenous increase in updating frequency of LPs. Our findings from
instrumental variable regressions provide strong supporting evidence
for our predictions, suggesting that an increase in repositioning
intensity indeed leads to greater concentration of liquidity around
the market price. A one-standard deviation increase in repositioning intensity results in a 4.38\% increase in liquidity concentration. This value is economically significant, because it represents a 43\% increase from the average liquidity concentration of 10.16\% on Ethereum. 

One potential concern could be that, whereas liquidity providers increase
intensity of their repositioning, they do not necessarily reposition
close to the new market price. Further, although a position might
be centered around the market price, it might still have a wide price
range. To address these issues, we also run instrumental variable
regressions for three measures of repositioning precision of LPs:
the average gap between the midprice of their positions and the market
price; the average range length of their positions; and the average
position precision, which combines the previous two measures.
Consistent with our expectations, we find that an increase in repositioning
precision also results in greater concentration of liquidity around
the market price.

An increase in aggregate liquidity concentration is important, because it reduces slippage, for a given TVL. We define the slippage of a trade as the difference between its average execution price and the pre-execution market price. Indeed, we find that small
trades (up to \$5K) have significantly lower slippage on scaling solutions, compared to Ethereum. In contrast, large trades have lower slippage on Ethereum due to its larger TVL.

In the last part of our analysis, we 
address a potential concern that higher liquidity concentration is not necessarily driven by higher repositioning intensity, but is rather an equilibrium outcome of liquidity provision on scaling solutions. Specifically, we use an exogenous shock to repositioning intensity within Arbitrum chain, related to the airdrop of Arbitrum's native token ARB on March 23, 2023. Uniswap was entitled to 4.3M ARB tokens, which were used to incentivize LPs to stay active and re-position in the market price range to maximize their rewards. 
Arguably, if a shock occurs within the same chain, all chain's parameters, such as chain security, expected trading size etc., should remain constant. Thus, changes in liquidity concentration can indeed be attributed to an increase in repositioning activity of LPs. Consistent with our expectations, we find that an increase in repositioning intensity on Arbitrum after the airdrop has a more pronounced effect on aggregate liquidity concentration, relative to the pre-airdrop sample.

Overall, our findings are important, because they show that blockchain
scaling solutions already represent a viable alternative to Ethereum
for small traders and liquidity providers. Specifically, their lower gas fees allow small liquidity providers
to better manage their positions, protecting themselves from
adverse selection. In turn, higher repositioning intensity and precision
of LPs leads to higher liquidity concentration, which especially benefits small traders by reducing their slippage.

Our paper contributes to the emerging literature on decentralized
exchanges. First studies on decentralized exchanges focus their analysis
on a particular subclass of automated market makers (AMMs), the constant product market makers (CPMMs), on the example
of Uniswap v2. CPMMs do not allow liquidity providers to set a price
range for their positions \parencite{aoyagi2020, aoyagi2021, park2022}.
Hence, the only decision of a liquidity provider on Uniswap v2 is
whether to provide liquidity or not in a specific pool. \textcite{lehar2021decentralized},
\textcite{capponi2021adoption} and \textcite{oneill2023}
show that liquidity providers on AMMs are subject to adverse selection
by competing arbitrageurs. In this setup, \textcite{lehar2021decentralized} and \textcite{oneill2023} show how pool size acts
as an equilibrating force in AMMs. Pools with higher adverse selection
problem will experience more liquidity withdrawals and therefore reduce
to an optimal size, in which average earned fees compensate liquidity
providers in taking adverse selection risk. \textcite{lehar2021decentralized}
do not explicitly analyze the role of gas fees, and treat them as
an important pre-commitment by liquidity providers not to withdraw
liquidity from the AMM. \textcite{barbon2021quality}
compare transaction cost and price efficiency among DEXs and CEXs,
taking gas fees into consideration. They show that transaction costs are approximately comparable on CEXs and DEXs. Whereas CEXs are superior in terms of price efficiency, DEXs eliminate custodian risk. 

To the best of our knowledge, there are currently only few papers
that analyze liquidity provision on Uniswap v3. Importantly, Uniswap
v3 allows LPs to set price ranges for their positions, similar to
limit orders in limit order books. \textcite{hasbrouck2022}
show theoretically that higher-fee pools attract more liquidity providers,
which reduces price impact of trades and increases the equilibrium
trading volume. \textcite{hasbrouck2023liqprovision} and \textcite{cartea2023predictableloss} theoretically model optimal liquidity provision on DEXs with concentrated liquidity (e.g., Uniswap v3). \textcite{chemaya2022} estimate investors' preferences for blockchain security on scaling solutions, using a structural model.\textcite{lehar2022liquidity} focus
their analysis on liquidity fragmentation across low- and high-fee
pools on Ethereum. They argue that low-fee pools require more frequent
updating in response to a higher trading volume, catering mostly to
large LPs. High-fee pools are rather attractive for passive small
(retail) LPs, due to their lower liquidity management cost. Their
model also predicts that, in presence of very low gas fees, all liquidity
consolidates in a low-fee pool. Indeed, we find strong supporting
evidence for this prediction in our paper, on an example of low- and
high-fee pools on Arbitrum and Polygon. In contrast to \textcite{lehar2022liquidity},
the main focus of our study is the effect of repositioning intensity
and precision of LPs on aggregate liquidity concentration around the
market price.

\begin{onehalfspace}
\section{The landscape of cryptoassets markets\label{sec:Institutional-Setting}}
\end{onehalfspace}

\bigskip{}

\begin{onehalfspace}
\subsection{Centralized vs decentralized exchanges}
\end{onehalfspace}

\begin{figure}[H]
\caption{\textbf{\footnotesize{}Binance BTC/USDT limit order book
snapshot.}{\footnotesize{} Bid side (in green) represents the cumulative
BTC quantity of buy limit orders, and ask side (in red) represents
the cumulative BTC quantity of sell limit orders. The displayed BTC
market price in USDT (vertical line) is the mid price, which is the
average of the best bid price and the best ask price.}}

{\small{}\vspace{1cm}
}{\small\par}

\label{Flo:binance_ob}
\begin{singlespace}
\begin{centering}
\includegraphics[scale=0.5]{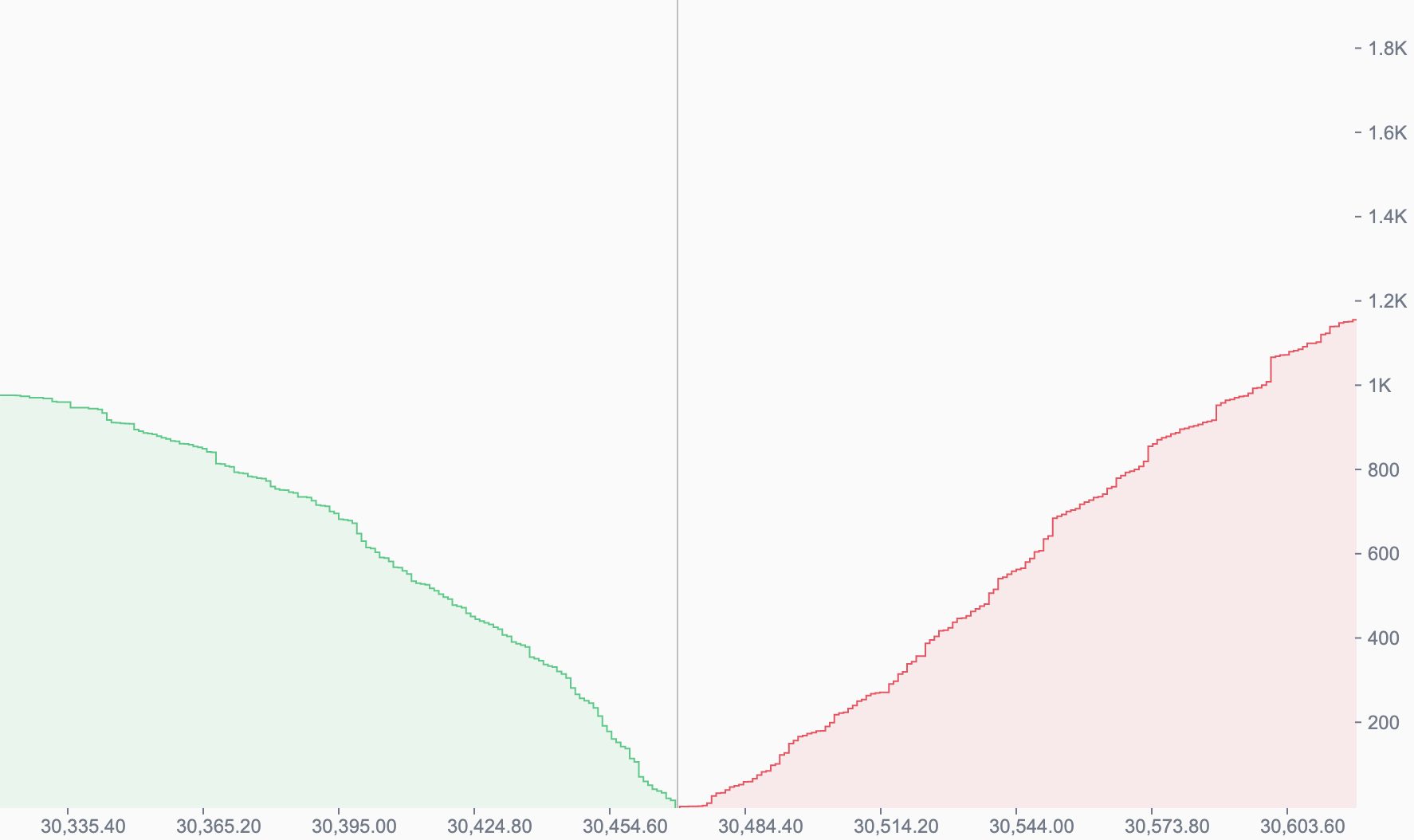}
\par\end{centering}
\end{singlespace}
{\small{}\vspace{0.5cm}
}{\small\par}

{\small{}Source: Binance.}{\small\par}
\end{figure}

Cryptoassets are currently traded on two different types of exchanges:
centralized exchanges (CEXs) and decentralized exchanges (DEXs). Centralized
exchanges, such as Binance, Kraken, Coinbase (and previously, FTX) operate using limit order books. Figure \ref{Flo:binance_ob} presents a snapshot
of Binance BTC/USDT limit order book, with the bid side (in
green) representing the cumulative BTC quantity of buy limit orders,
and the ask side (in red) representing the cumulative BTC quantity
of sell limit orders. In limit order markets, liquidity is usually
provided by professional market makers, who strategically compete
with each other by submitting limit orders. 

\begin{center}
{[}Insert Figure \ref{Flo:binance_ob} approximately here{]}
\par\end{center}

In contrast, decentralized exchanges operate directly on a blockchain,
with liquidity usually provided through an ``automated market maker''
(AMM). An AMM just follows a set of pre-programmed rules, so-called
``smart contracts'', such that there is no explicit human intervention
required when a trade (or a so-called ``swap'') is submitted. Each
asset pair, for example, ETH/USDC, comprises a separate liquidity
pool. Liquidity providers (LPs), including retail investors, can deposit
(``mint'') liquidity to the pool by adding both assets, or so-called
tokens, in a respective ratio. They can also withdraw (``burn'')
liquidity from the pool at a later point in time. Liquidity demanders (traders)
can then swap one token for another in the pool at the current
market price. Thus, in contrast to limit order markets, proper
matching of buy and sell orders is not required in an AMM. All trades
are executed against the AMM, with the market price determined by
mathematical formulas in smart contracts. Liquidity demanders pay
a pre-specified exchange fee, e.g. 0.3\%, for each trade as a compensation
to liquidity providers. Both liquidity demanders and liquidity providers
have to pay additional transaction fees, so-called ``gas fees'',
as a compensation to miners who actually record their transactions
(i.e. ``swaps'', ``mints'', ``burns'') on the blockchain. Importantly,
DEXs users keep custody of their assets, because they execute trades
from their own wallets and settlement is immediate.\footnote{\textcite{barbon2021quality} provide an excellent overview of additional
differences between CEXs and DEXs, relating to custody of assets,
fees accrual, etc.}

Uniswap is the leading decentralized exchange, with its first version,
Uniswap v1, launched on Ethereum on November 2, 2018. The main drawback
of Uniswap v1 is that it only supports trading against ETH, i.e. selling
DAI for USDC requires two transactions: first, selling DAI for ETH
and, second, ETH for USDC. To overcome this problem, Uniswap v2 was
released in May 2020. Importantly, Uniswap v2 does not allow for competition
of liquidity providers. The only decision liquidity providers make
is whether to deposit liquidity in the pool or not. The trading fee
equals 0.3\% for all pools on Uniswap v2. Fees earned from trades
are distributed pro-rata to liquidity providers, i.e. in proportion
to the amount of liquidity they provide in the pool. Therefore, all
gains (and losses) are mutualized by liquidity providers.

The latest version, Uniswap v3, was released in May 2021, implementing
two main changes to Uniswap v2. First, it allows some degree of competition
on price between liquidity providers by introducing the so-called
``concentrated liquidity''. Specifically, when depositing their
tokens, LPs can now indicate the price range, i.e. the minimum and
the maximum prices in a given pool at which their liquidity position
is active. However, there is no competition on speed available, as
all gains for LPs within the same price range are still mutualized.
Second, Uniswap v3 allows for four various fee tiers: 0.01\%, 0.05\%, 0.3\%
and 1\%. Therefore, the same pair of assets can potentially be traded
in four different pools.\footnote{In contrast to Uniswap v2, fees earned by LPs are no longer deposited in the pool as liquidity. Instead, fee earnings in Uniswap v3 are stored separately and can be withdrawn any time by LPs.}
We discuss mechanics of trading on Uniswap v3 in more
detail in Appendix B.

\begin{figure}[H]
\caption{\textbf{\footnotesize{}Monthly traded volume on CEXs and DEXs. }{\small{}Panel
A of this figure shows monthly traded volume on CEXs (in \$) from
May 2017 until April 2023, showing the market shares separately for
each exchange. Panel B shows monthly traded volume on DEXs (in \$)
from June 2020, after the launch of Uniswap v2, until April 2023.}}

{\small{}\vspace{0.5cm}
}\label{Flo:cex_dex_volume}
\begin{centering}
\textbf{\small{}Panel A: CEXs volume}{\small\par}
\par\end{centering}
{\small{}\vspace{0.5cm}
}{\small\par}
\begin{singlespace}
\begin{centering}
\includegraphics[width=\textwidth]{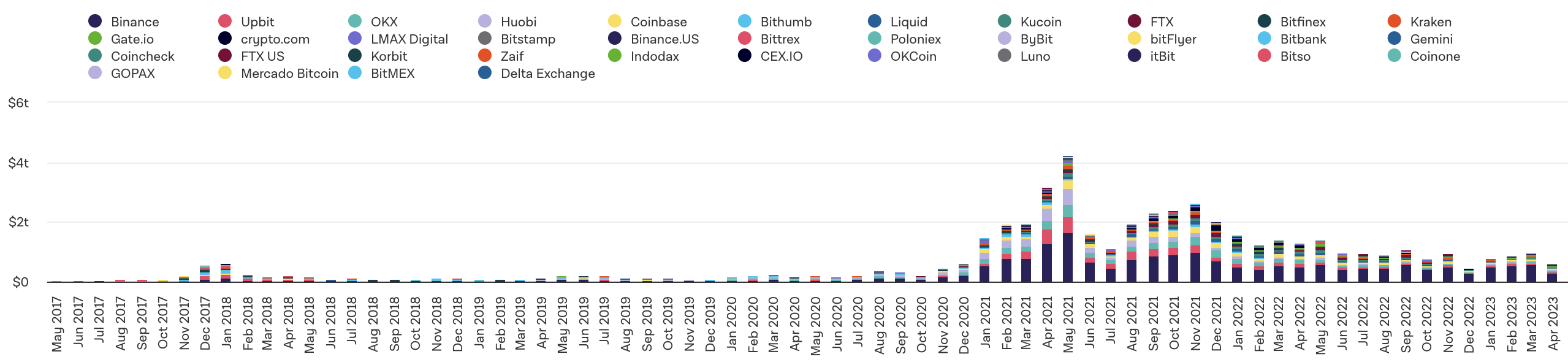}
\par\end{centering}
\end{singlespace}
{\small{}\vspace{0.5cm}
}{\small\par}
\begin{centering}
\textbf{\small{}Panel B: DEXs volume}{\small\par}
\par\end{centering}
{\small{}\vspace{0.5cm}
}{\small\par}
\begin{singlespace}
\begin{centering}
\includegraphics[width=\textwidth]{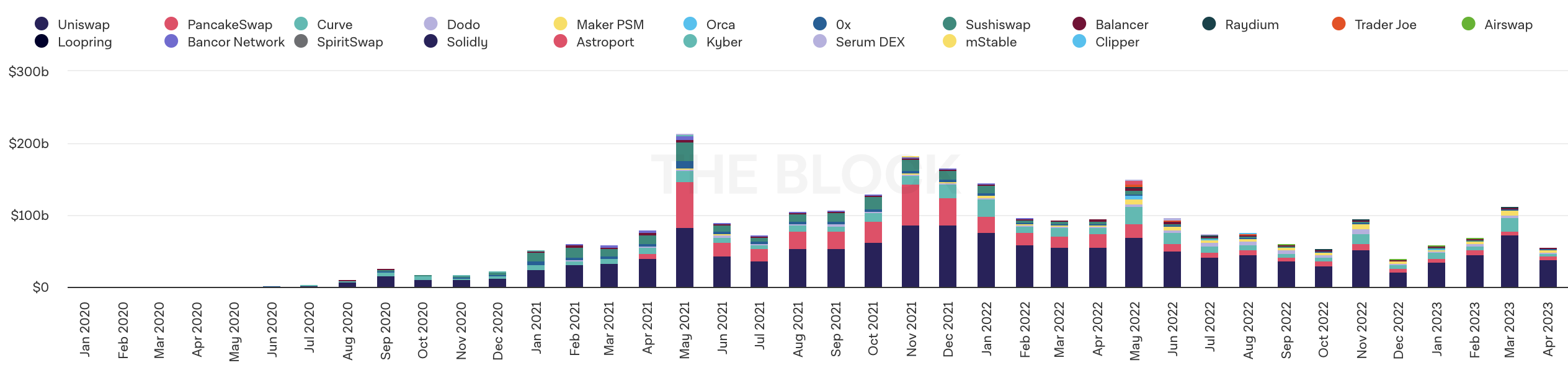}
\par\end{centering}
\end{singlespace}
{\small{}\vspace{0.5cm}
}{\small\par}

{\small{}Source: The Block,  https://www.theblock.co/data.}{\small\par}
\end{figure}

As of August 2022, \$914B were traded on CEXs, out of which \$443B were traded on Binance. Panel A of Figure \ref{Flo:cex_dex_volume} shows monthly traded volume
on CEXs (in \$) from May 2017 to April 2023\footnote{Source: The Block, https://www.theblock.co/data.}. Among all CEXs, Binance is dominating, with an average market share
of 50\%.

\begin{center}
{[}Insert Figure \ref{Flo:cex_dex_volume} approximately here{]}
\par\end{center}

Total traded volume on DEXs is 12 times lower than on CEXs as of August
2022, around \$75B. However, it fluctuates considerably, with high
values of \$182B in November 2021 and \$112B in March 2023, and low
values of \$39B in December 2022. Panel B of Figure \ref{Flo:cex_dex_volume}
shows monthly trading volume on DEXs (in \$) from January 2020 to
April 2023. Uniswap v3 is the leading DEX with a market share of 58\%
in August 2022, followed by PancakeSwap with 11\% and Curve with 8\%.
As Uniswap v3 is leading in terms of its market share among DEXs,
we focus our analysis on this DEX in our paper. 

\subsection{Blockchain scaling solutions: Arbitrum and Polygon}

Whereas Ethereum is currently the most secure blockchain with
a number of validators exceeding 500,000 as of January 2023, its
main disadvantage is low scalability. Specifically, a block is added to Ethereum every 12 seconds after the so-called Merge, i.e. when Ethereum moved to proof-of-stake consensus on September 15, 2022. Previously, with the proof-of-work consensus, block time was probabilistic and averaged between 13-15 seconds. Importantly, low scalability of Ethereum results in high
gas fees, with an estimated average of gas fees for a trade
on Uniswap v3 amounting to \$14 in 2022.\footnote{See Appendix A.}

To address the scalability issue, there exist overlays of Ethereum
that offer higher speed and, more importantly, lower gas fees. Arbitrum and Polygon
are two of the most adopted Ethereum scaling solutions. Currently, Uniswap v3 has also launched on the following Ethereum scaling solutions: Optimism, Celo, BNB Smart Chain, Base and Avalanche. We conduct our study on Arbitrum and Polygon, because they have the highest total value locked (TVL) and trading volume, in comparison to other solutions.\footnote{As of January 2023, total monthly trading volume on Arbitrum and Polygon equals \$2.1B and \$1.93B, respectively. The corresponding number for Optimism is \$1.1B. Total monthly trading volume never exceeds \$300M in 2023 for other solutions. See https://info.uniswap.org/ for current TVL and volume across all Ethereum scaling solutions.}

\subsubsection{Entry of Arbitrum}

Uniswap v3 launched on Arbitrum on August 31, 2021. Arbitrum is a “rollup”, i.e. a Layer 2 (L2) scaling solution that involves “rolling up”, or accumulating, transactions on Arbitrum in “batches”, subsequently  compressing them and posting them on the Layer 1, i.e. Ethereum.  This periodic posting allows Arbitrum to inherit Ethereum’s security, subject to the posting delay. Arbitrum uses a particular type of rollup - so-called optimistic rollup - that processes off-chain transactions optimistically, assuming validity by default and relying on occasional on-chain verification.\footnote{Another category of rollups are so-called ZK (zero-knowledge) rollups, as zkSync or Polygon zkEVM. In contrast to optimistic rollups, ZK rollups use zero-knowledge proofs for non-interactive and cryptographically secure validation of off-chain transactions.}


The main benefit  of rollups is that the Layer 1 blockchain, i.e. Ethereum, does not need to validate separate transactions, but only batches of transactions. Therefore, Arbitrum (and other rollups) can offer higher speed by processing and batching transactions off-Ethereum before settling them on the main network.\footnote{See https://docs.arbitrum.io/intro for details.} Average block time on Arbitrum was around 1-2 seconds at the beginning of 2022, subsequently dropping to around 0.25 seconds in 2023.\footnote{Source: https://arbiscan.io/chart/tx}. 

Importantly, transaction batching also allows rollups to offer much lower gas fees than Ethereum. Indeed, the gas price of a transaction on Arbitrum can be split into two parts: a part linked to the rollup network itself, which can be considered stable (it starts at 0.1 Gwei, i.e. one-billionth of ETH, and can increase with congestion), and a part linked to the posting of batches on Ethereum, which varies depending on Ethereum’s gas price. Batch data is compressed to reduce the cost of posting on Ethereum as much as possible. \footnote{For details of gas fee estimation on Arbitrum, see https://docs.arbitrum.io/devs-how-tos/how-to-estimate-gas.} Still, posting batches on Ethereum represents a significant cost for rollups, with Arbitrum offering gas fees between \$0.2 and \$2 for a trade on Uniswap v3.\footnote{Based on Uniswap v3 frontend, https://app.uniswap.org/swap?chain=arbitrum}
 
\subsubsection{Entry of Polygon}

Uniswap v3 launched on Polygon PoS (proof-of-stake) four months later than on Arbitrum, in December 2021. Polygon PoS is different to rollups, because it is a sidechain that requires its own security and decentralization efforts. In contrast to Arbitrum and other rollups, it does not automatically derive its security from the main blockchain, i.e. Ethereum. Specifically, Polygon is secured by a proof-of-stake consensus mechanism. However, Polygon still depends on Ethereum, because all the staking management is defined on
Ethereum\footnote{While validators are required to stake MATIC tokens to secure the
network, users stake them in an Ethereum smart contract while bridging
assets to Polygon.} The overall number of validators on Polygon is lower, compared to Ethereum, but it still
has a decent level of decentralization with over 100 unique validators.
Whereas security of Polygon is not as robust as Ethereum's, fewer validators can achieve consensus more quickly, hence leading
to a shorter time between blocks and higher overall throughput.\footnote{\textcite{chemaya2022} discuss in detail the trade-off between security and
scalability of Ethereum scaling solutions, including Polygon.} Average block time on Polygon is around 2-3 seconds, compared to Ethereum's 12 seconds.\footnote{See https://polygonscan.com/chart/blocks.}
Thus, Polygon can handle
around 700 transactions per second (TPS)\footnote{Source: Polygon's blog, https://polygon.technology/blog/what-do-you-prefer-maximum-security-or-cheaper-transactions.},
up to 70 times Ethereum's TPS\footnote{Source: Binance Academy, https://academy.binance.com/en/glossary/transactions-per-second-tps.}.
Further, Polygon's gas fees are the lowest, around \$0.01 per transaction.\footnote{Polygon's base gas fee is around \$0.01 per transaction, but users
can also add so-called gas tips to miners to prioritize the order
of their transaction within the block. Still, most of transactions
on Polygon only cost a few cents in gas fees.} 

Polygon's and Arbitrum's liquidity pools are separate to those on Ethereum, i.e. liquidity
is fragmented across pools. Thus, the same liquidity pool, ETH/USDC 0.05\% for example, can exist on multiple chains, Ethereum, Polygon, Arbitrum and other scaling solutions. In our following analysis,
we use the launch of Uniswap v3 on Arbitrum and Polygon as an instrument to identify
the effect of lower gas fees on liquidity and its distribution around the market price.

\section{Data and summary statistics }
We analyze the two most liquid pairs on Uniswap v3, ETH/USDC 0.05\% and
ETH/USDC 0.3\%, that are traded across all three chains: Ethereum, Arbitrum and Polygon. Hence, our benchmark sample consists of six distinct pools.\footnote{All pools on Uniswap v3 use wrapped
ETH (WETH) instead of Ethereum's native ETH as they only support ERC-20 tokens.} For comparability across chains, we always use ETH as the base token and USDC as the quote token, i.e all data are presented for the ETH/USDC pair. ETH/USDC 0.05\% is a low-fee pool that charges 5bp for each trade, or swap (in addition to the gas fee on the respective chain). ETH/USDC 0.3\% is a high-fee pool that charges 30bp for each swap (in addition to the gas fee).\footnote{\textcite{lehar2022liquidity} show that liquidity on Ethereum is fragmented across low- and high-fee pools due to different economies of scale across LPs.} 

We also replicate our main results for four additional pairs, BTC/ETH 0.05\%, BTC/ETH 0.3\%, UNI/ETH 0.3\% and LINK/ETH 0.3\%, in Tables \ref{Flo:intensity_other} and \ref{Flo:precision_ia_other} in the Internet Appendix.  We choose these pairs because they are among few that are relatively actively traded across all three chains, mapping into twelve additional pools. For our analysis, we require a pair to be traded across all three chains. Further, we need it to be relatively liquid to be able to identify the effect of repositioning intensity. Hence, the sample of pools that satisfy the two criteria above is quite limited.

Our sample period starts after the launch of Uniswap v3
on Polygon, on January 1, 2022, and ends on June 30, 2023.\footnote{Prior to 2022, there was not enough trading volume on any of blockchain scaling solutions. For this reason, we start our sample at the beginning of 2022.} We download historical logs of Ethereum, Arbitrum and Polygon transactions (swaps, mints and burns on Uniswap v3), using The Graph queries. 

Panel A of Table \ref{Flo:average_vol_tvl} compares volume
and liquidity for both low- and  high-fee pools across Ethereum, Arbitrum and Polygon.
Specifically, we report the average trading volume over the previous 24 hours and the average liquidity size, measured as total value locked (TVL), both in \$M.\footnote{Total value locked (TVL) on DEXs corresponds to aggregate market depth on CEXs, i.e. total liquidity available on both bid and ask sides of the limit order book.} 

\begin{center}
{[}Insert Table \ref{Flo:average_vol_tvl} approximately here{]}
\par\end{center}

Unsurprisingly, both volume and TVL on Ethereum
significantly exceed those on scaling solutions. For example, the average 24-hour volume for Ethereum's most liquid pool, ETH/USDC 0.05\%, is \$513.73M, around 10 times higher than Arbitrum's \$51.93M and 18 times higher than Polygon's \$29.03M. Ethereum's
average TVL of \$201.19M is around 11 times higher than Arbitrum's \$18.04M and around 20 times higher than Polygon's
TVL of \$10.26M. Notably, the turnover, measured as the ratio of
volume to TVL, is approximately the same across all three chains (as reported in the last column of
Table \ref{Flo:average_vol_tvl}). We observe much lower trading volume and TVL for the high-fee pool,
ETH/USDC 0.3\%, especially on Arbitrum and Polygon. These findings are consistent with theoretical predictions
of \textcite{lehar2022liquidity} that lower gas fees
lead to lower fragmentation of liquidity across high- and low-fee
pools. As gas fees on Arbitrum and Polygon are relatively cheap, it is
optimal for LPs to actively manage their positions in the low-fee
pool. Thus, most of liquidity provision takes place on the low-fee
pool on Arbitrum and Polygon, and not on the high-fee pool.

We further provide summary statistics of trade sizes and trade frequencies
in Panels B and C of Table \ref{Flo:average_vol_tvl}. Panel B shows that the average trade size on Ethereum is significantly
larger for both pools. For example, the average trade size for
ETH/USDC 0.05\% on Ethereum is around \$72K. In contrast, Arbitrum's and Polygon's
average trade size for this pool are much lower, around \$4K and \$2.5K, respectively. We observe
even lower trade sizes on Arbitrum and Polygon for ETH/USDC
0.3\%, most likely because of their overall lower
TVL on these scaling solutions. In contrast, we observe that the high-fee pool has a larger average trade size of \$116K on Ethereum, compared to the low-fee pool. This finding
is consistent with the theory of \textcite{hasbrouck2022} that pools with higher fees attract more liquidity providers,
increasing their TVL and attracting larger trades.

Panel C of Table \ref{Flo:average_vol_tvl} reports the average daily number
of trades as well as the average daily number of purchases and sales
across the three chains. Strikingly, both Arbitrum and Polygon have almost twice as many trades,
around 12.5K per day, compared to Ethereum's 7K for ETH/USDC 0.05\%.
For ETH/USDC 0.3\%, the number of trades is approximately the
same across all chains. For all pools, the order flow is balanced,
i.e. the average daily number of buys is approximately equal to the
average daily number of sells. We also report the average time between
the trades (in seconds) in the last column. Overall, we conclude that
the higher volume on Ethereum is mostly driven by larger trades
being executed on this chain. In contrast, Arbitrum and Polygon are rather used
by smaller traders who trade more frequently.

As discussed in Section 2.2, Ethereum's larger liquidity can
be explained by its higher security. Liquidity providers would
like to minimize the risk of losing their funds, especially so for
large liquidity deposits. Therefore, Ethereum is mostly used for large
liquidity deposits, resulting in higher TVL. Large traders
are also attracted to Ethereum, both due to its higher security and
larger liquidity, which helps them reduce the slippage of their trades. Since
gas fees are fixed, they are not of primary concern for large traders.
In contrast, blockchain scaling solutions are rather attractive for smaller traders, who
are primarily concerned about gas fees, and less so about security.
Thus, we observe a separating equilibrium, in which large traders
and LPs choose Ethereum due to its higher security, whereas
small traders and LPs choose Arbitrum and Polygon due to their lower gas fees. 

Table \ref{Flo:sumstat} reports summary statistics of control variables, which we use later in our regressions: trading volume over the previous 24 hours, $Volume$, (in \$M); the 1-minute return of ETH/USD, $Return$ (in bp); and the realized volatility of ETH/USD, $Volatility$, computed as the square root of the sum of squared 1-minute returns over the previous 24 hours (in \%). Both $Volatility$ and $Return$ are based on a single time series for ETH/USD from Binance. 1-minute returns are strongly balanced, with positive and negative returns being of approximately the same magnitude, with both the average and the median $Return$ close to zero. The average daily realized volatility of 3.53\% of ETH/USD corresponds to an annualized volatility of 67.44\%. Appendix E provides a detailed description of all variable definitions.

\section{Repositioning and liquidity concentration\label{sec:Liquidity-concentration}}

In this section, we test the effect of lower gas fees, provided by blockchain scaling solutions, on liquidity concentration around the current pool price. We start with formulating our hypotheses in Section 4.1. We then test the causal relation between repositioning intensity and liquidity concentration, using instrumental variable regressions, in Section 4.2. Section 4.3 examines the causal relation between repositioning precision and liquidity concentration.

\subsection{Hypotheses development}\label{subsec:hypothesis-dev}

Higher speed of transaction processing and lower gas fees on blockchain scaling solutions allow liquidity providers to
update their positions more frequently. Should a permanent price change
occur, LPs can withdraw and re-deposit their positions, re-setting the price range around the
new market price more quickly and at a cheaper cost. Thus, higher
speed and lower gas fees help LPs better protect their positions from
adverse selection by arbitrageurs. Being able to better protect themselves,
LPs can earn higher fees (for the same deposited amount) by concentrating
their positions around the market price. 

In contrast, when updating is costly, we expect LPs to post their
liquidity on wider price ranges. In absence of updating, posting on a wider range provides
better protection from adverse selection relative to a narrow price
range.\footnote{\textcite{cartea2023predictableloss} also show theoretically that wider
position ranges protect the value of LP's assets. However, they do
not model explicitly the ability of LPs to update their positions
and the role of gas fees, i.e. cost of updating.} Therefore, we expect less frequent updating when gas fees are higher,
with LPs updating their positions only if the market price significantly
deviates from the position range. Wider liquidity positions should
then result in less concentrated aggregate liquidity around the market price.

\begin{figure}[H]
\caption{\textbf{\footnotesize{}Protecting from adverse selection by LPs. }{\footnotesize{}Panel
A shows a scenario when an LP's position is adversely selected by
an arbitrageur. We assume that ETH price (in USDC) experiences a permanent
increase from 1,000 to 2,000. The arbitrageur's buy trade moves the
market price to the new level, depleting all ETH reserves in the LP's
position and leaving him only with USDC. The LP experiences an adverse
selection loss of \$500K, compared to the scenario, in which he would
just hold his initial portfolio of ETH and USDC. Panel B shows that
the LP could partially protect himself from adverse selection by redistributing
the same amount of capital on a wider price range. Thus, his position
would be less concentrated around the market price. Posting on a wider
price range would reduce LP's losses to \$300K, because average execution
price for his ETH reserves would be higher, compared to Panel A. In
contrast, Panel C assumes that the LP continuously monitors the market
and is able to update his position even before the arbirtrageur's
trade arrives. In this scenario, the LP is able to fully avoid adverse
selection loss and makes a profit of \$500K.}}

{\small{}\vspace{0.5cm}
}\label{Flo:adv_selection}
\begin{centering}
\textbf{\small{}Panel A: LP is adversely selected}{\small\par}
\par\end{centering}
{\small{}\vspace{0.1cm}
}{\small\par}
\begin{singlespace}
\begin{centering}
\includegraphics[scale=0.5]{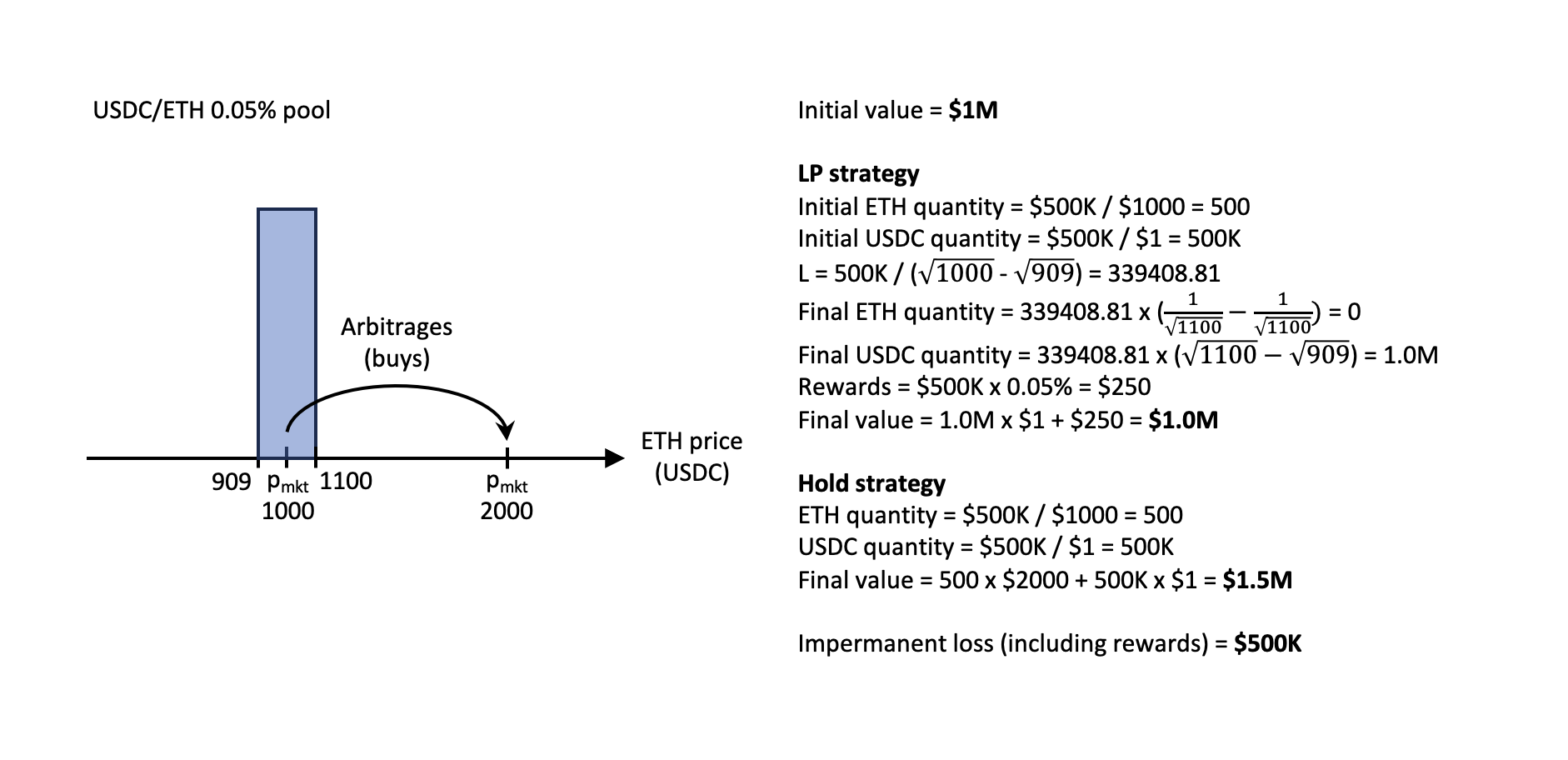}
\par\end{centering}
\end{singlespace}
{\small{}\vspace{0.1cm}
}{\small\par}
\begin{centering}
\textbf{\small{}Panel B: Protecting by setting a wider initial price
range}{\small\par}
\par\end{centering}
{\small{}\vspace{0.1cm}
}{\small\par}
\begin{singlespace}
\centering{}\includegraphics[scale=0.5]{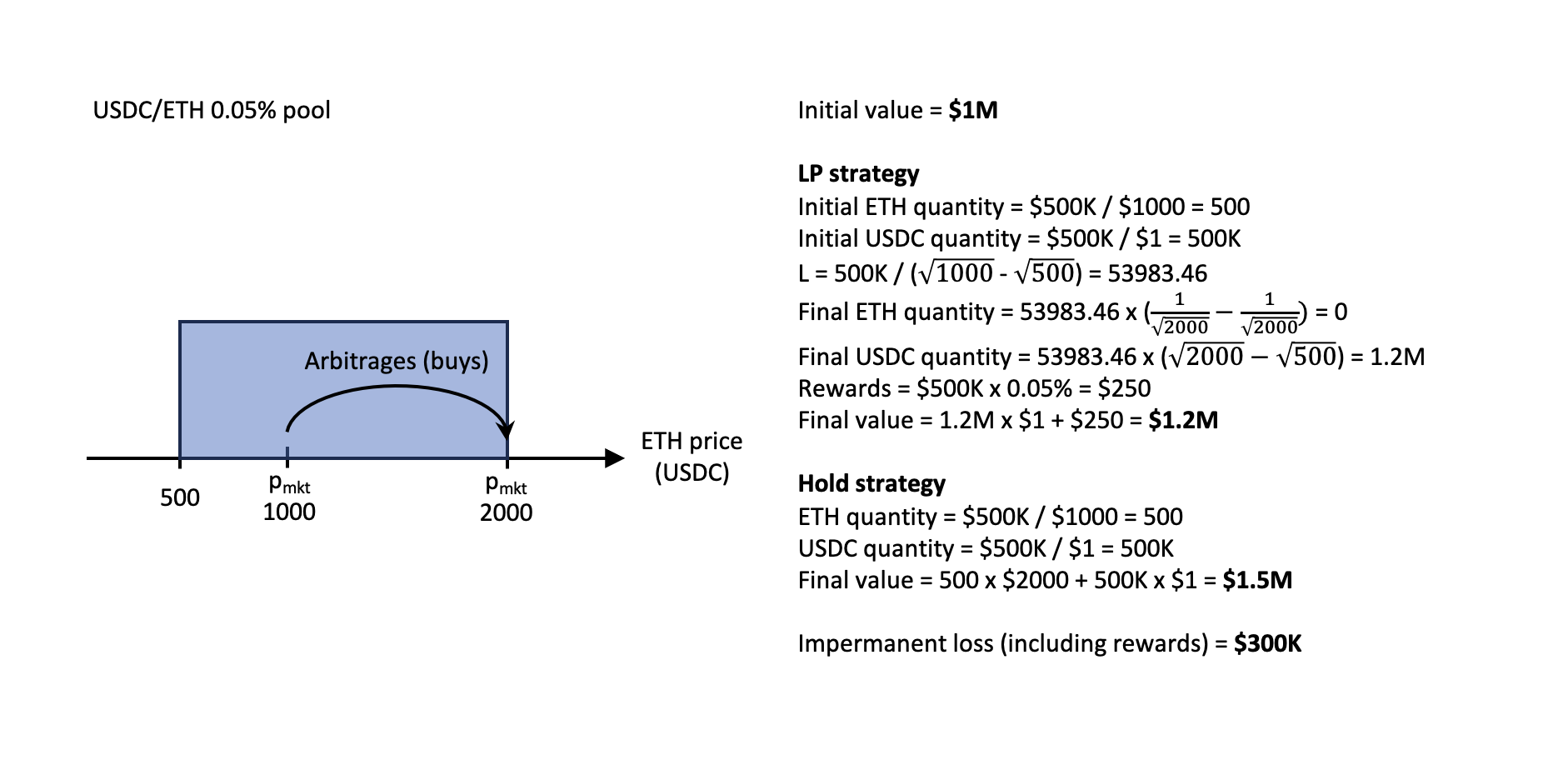}
\end{singlespace}
\end{figure}

\begin{center}
{[}Insert Figure \ref{Flo:adv_selection} approximately here{]}
\par\end{center}

Figure \ref{Flo:adv_selection} further illustrates our predictions
with a stylized numerical example. In Panel A, we assume that ETH
price (in USDC) experiences a permanent increase from 1,000 to 2,000.
Without updating, an LP's position is adversely selected by an arbitrageur,
who submits a buy trade. The arbitrageur's trade moves the market
price to the new level, depleting all ETH reserves in the LP's position
and leaving him only with USDC. The LP experiences an adverse selection
loss of \$500K, compared to the scenario, in which he would just buy and hold
his initial portfolio of ETH and USDC, without providing liquidity on a DEX. 

Panel B shows that the LP could
partially protect himself from adverse selection by redistributing
the same amount of capital on a wider price range. Thus, his position
would be less concentrated around the market price. Posting on a wider
price range would reduce LP's losses to \$300K, because average execution
price for his ETH reserves would be higher, compared to Panel A. In
contrast, Panel C assumes that the LP continuously monitors the market
and is able to update his position even before the arbitrageur's
trade arrives. In this scenario, the LP is able to fully avoid adverse
selection loss and makes a profit of \$500K. It is optimal for him
to set a narrower price range, i.e. make his position more concentrated,
in order to earn higher rewards. 

Being able to immediately update is the first-best solution for any
LP. However, updating is costly and requires continuous monitoring.
Therefore, we expect LPs to reposition more frequently when updating
is relatively cheaper, i.e. on blockchain scaling solutions. In the
following, we refer to the position updating frequency of LPs as repositioning
intensity. We formalize our first hypothesis as follows:

\textbf{Hypothesis 1.} \textit{Compared to Ethereum, higher repositioning intensity on blockchain scaling solutions leads to higher aggregate liquidity
concentration around the current market price.}

Whereas we use repositioning intensity as our benchmark measure, it could be that liquidity providers do
not necessarily update their positions close to the new market price. Specifically, repositioning intensity does not take into account whether liquidity positions
are centered around the market price or not. Further, a position
might be centered around the market price, but it could also have
a wide price range, which should not necessarily result in higher
liquidity concentration. To address these issues, we also focus our
analysis on repositioning precision, which we define as the distribution
of individual liquidity positions around the market price. 

Specifically, we expect repositioning precision to be higher on blockchain scaling solutions. With lower gas fees and better protection from adverse selection, we expect LPs to make their positions more concentrated around the market price, i.e. we expect them to re-post on a narrow price range, with the mid-price of the range close to the current market price. More concentrated individual positions allow LPs to maximize their rewards from liquidity provision. Importantly, more concentrated individual positions should result in higher aggregate liquidity concentration around the current market price. Based on these predictions, we formalize our second hypothesis as follows: 

\textbf{Hypothesis 2.} \textit{Compared to Ethereum, higher repositioning precision  on blockchain scaling solutions leads to higher aggregate liquidity
concentration around the current market price.}

Higher liquidity concentration around the market price is important, because it affects slippage. In this paper, we define slippage of a trade as the difference between the average execution price and the observed pre-execution market price.\footnote{See Section 5 as well as Appendix C and D for details of slippage computation on Uniswap v3.} For a given TVL, higher liquidity
concentration around the market price should reduce the slippage of any trade.\footnote{For pools with large TVL in the current price range, higher liquidity concentration might not have any effect on slippage of a small trade. However, higher liquidity concentration is still beneficial for larger trades that would otherwise exhaust liquidity in the current price range.} In contrast, less concentrated aggregate liquidity should increase slippage for a given TVL. 

Importantly, Arbitrum's and Polygon's TVL is
significantly lower than Ethereum's (see Panel A of Table \ref{Flo:average_vol_tvl}). Ethereum's large TVL, which corresponds to overall
market depth, is still of first-order importance for slippage of large trades, relative to liquidity concentration. Hence, we expect large trades to have overall lower slippage on Ethereum due to its higher TVL. Based on these differences in observed TVL, slippage is only comparable for smaller trades (up to \$5K) between Ethereum and blockchain scaling solutions.\footnote{Panel B of Table \ref{Flo:average_vol_tvl} shows that the 75th percentile of trades on both Arbitrum and Polygon never exceeds \$5K, with average trade sizes of \$2K-\$4K.} Therefore, we formalize our third hypothesis as follows:

\textbf{Hypothesis 3.} \textit{Higher liquidity concentration on blockchain scaling solutions should result in lower slippage for small trades (up to \$5K), compared to Ethereum. For large trades (above \$5K), slippage should be lower on Ethereum due to its higher TVL.}

\subsection{Repositioning intensity and liquidity concentration}
 
\textbf{Liquidity concentration.} We start by comparing aggregate liquidity
concentration around the market price across Ethereum and blockchain scaling solutions. 
Specifically, we define
liquidity concentration within x\% of the market price, as the market depth within x\% of $p_{mkt}$, divided by $TVL$.
Market depth within x\% of $p_{mkt}$ is computed as the dollar value of the liquidity between $\frac{p_{mkt}}{1+x\%}$ and $p_{mkt} \cdot (1+x\%)$. Table \ref{Flo:unistats_liqconcentration} reports average levels of liquidity concentration across Ethereum, Arbitrum and Polygon, separately for low-fee and high-fee pools. 

\begin{center}
{[}Insert Table \ref{Flo:unistats_liqconcentration} approximately
here{]}
\par\end{center}

Column (1) reports average levels of liquidity concentration within
1\% of the market price. Columns
(2) and (3) report corresponding statistics for liquidity concentration
within 2\% and 10\% of the market price, respectively. Liquidity concentration
mechanically increases with the percentage band, i.e. market depth as percentage of TVL within 10\% of the market price is by construction
higher than within 1\% of the market price. More surprisingly, only
around 24\%-42\% of all liquidity is concentrated within 10\% for
all pools. Hence, around two thirds of TVL are further away than 10\%
from the market price across all pools. Consistent with our expectations,
we find strong evidence for higher liquidity concentration on blockchain scaling solutions across all percentage bands, compared to Ethereum.
We observe the differences
between average liquidity concentration on Arbitrum and Ethereum, $ \Delta Arb - Eth$, of 0.93\%-5.36\%. The corresponding numbers for differences between Polygon and Ethereum,  $ \Delta Pol-Eth$, are 1.18\%-14.54\%. T-statistics of the two-tailed t-test with the null-hypothesis
of difference equaling zero all exceed 100, i.e. all differences are statistically significant at the 1\% level. 

\begin{figure}[H]
\caption{\textbf{\footnotesize{}Average range liquidity, scaled by TVL: ETH/USDC 0.05\%.}{\footnotesize{} This figure shows average liquidity within each price range, scaled by TVL, for ETH/USDC 0.05\% pool, separately for Ethereum, Arbitrum and Polygon. Averages for each price range are constructed from hourly liquidity snapshots over our sample period January 1, 2022 - June 30, 2023. The dotted vertical line shows the average market price over our sample period.}}

{\small{}\vspace{1cm}
}{\small\par}

\label{Flo:liq_conc}
\begin{singlespace}
\begin{centering}
\includegraphics[scale=0.35]{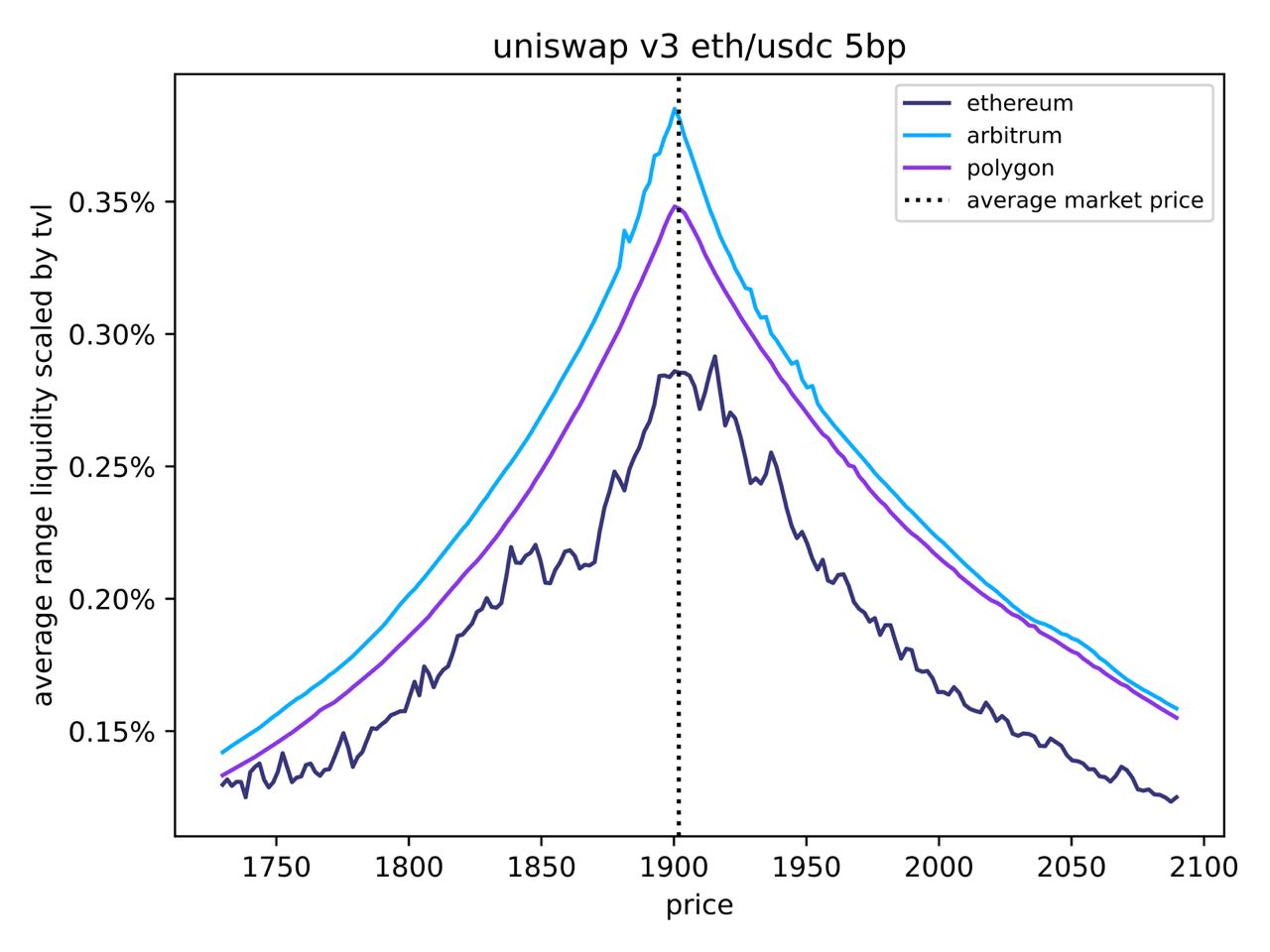}
\par\end{centering}
\end{singlespace}
{\small{}\vspace{0.5cm}
}{\small\par}

\end{figure}

Figure \ref{Flo:liq_conc} further illustrates the distribution of aggregate liquidity around the market price for ETH/USDC 0.05\% pool, separately for Ethereum, Arbitrum and Polygon. Specifically, we use hourly snapshots to calculate average liquidity within each price range, scaled by TVL (in \%), over our sample period January 1, 2022 - June 30, 2023. Consistent with our findings in Table \ref{Flo:unistats_liqconcentration}, we observe higher liquidity concentration around the market price for Arbitrum and Polygon, relative to Ethereum, i.e. liquidity distribution is more "peaked" for blockchain scaling solutions. 

\begin{center}
{[}Insert Figure \ref{Flo:liq_conc} approximately
here{]}
\par\end{center}

\textbf{Liquidity provision and repositioning intensity.} We further
examine differences in liquidity provision and repositioning intensity
across Ethereum and blockchain scaling solutions in Table \ref{Flo:unistats_mintsburns}.
For all our analyses of liquidity provision, we filter out liquidity
mints and burns, for which the mid price of their range, computed
as $\sqrt{p_{lower}\cdot p_{upper}}$, lies further than 20\% away
from the market price. We treat these mints and burns as outliers
and exclude them from our analysis.\footnote{These positions represent around 13\% of our overall sample. Our main
findings continue to hold irrespective of the cutoff that we use to
filter out outliers (for example, 10\% or 30\% away from the current
price). Minting and burning with the mid price far away from the current
market price could potentially be used by ``wash LPs'', who aim
to artificially increase TVL in the pool for improving its overall
statistics. } We further exclude so-called ``just-in-time'' (JIT) liquidity positions,
which are used to provide ``flash'' liquidity within the same block
to large trades on Ethereum towards the end of our sample.\footnote{\textcite{lehar2022liquidity} also exclude JIT positions
from their analysis. JIT positions involve minting liquidity just
before the large trade and subsequently burning the position, all
within the same block.} JIT positions represent around 27\% of our sample on Ethereum and
less than 2.5\% on Arbitrum and Polygon. We exclude them, since this
``flash'' liquidity does not affect TVL and thus, liquidity concentration,
which is our main variable of interest. Further, JIT liquidity provision
only benefits a few large traders on Ethereum, and is generally not
used to provide liquidity for small trades. 

\begin{center}
{[}Insert Table \ref{Flo:unistats_mintsburns} approximately here{]}
\par\end{center}

We report the average daily number of mints for each chain in column
(1) and the average daily number of burns in column (4). Compared to Ethereum, we observe
significantly more frequent minting and burning of liquidity on Arbitrum and Polygon
for ETH/USDC 0.05\% pool. A mint on Arbitrum (Polygon) takes place on average
every 4.57 (2.63) minutes and a burn every 6.05 (3.88) minutes, as compared to Ethereum's
17.73 and 23.40 minutes, respectively (see columns 2 and 5). However, the average daily minted value on Arbitrum (Polygon) of around \$11.75M (\$2.95M) is significantly lower than \$41.19M
minted daily on Ethereum (column 3). Average daily burned values on Arbitrum and Polygon are of approximately the same magnitude as daily minted values, and also significantly lower than Ethereum's \$39.9M (column 6).

For ETH/USDC 0.3\%, the daily number of mints  is
approximately the same across Ethereum and blockchain scaling solutions, with a mint occurring every
20 minutes on each of the chains.  The daily minted value is much lower on Arbitrum and Polygon, which is
consistent with our prior findings of overall lower TVL for high-fee
pools on blockchain scaling solutions. Consequently, the daily burned value is also lower
for high-fee pools on Arbitrum and Polygon. These findings further support predictions
of \textcite{lehar2022liquidity} that all liquidity
mostly consolidates in the low-fee pool in presence of low gas fees. 

More frequent minting and burning liquidity on blockchain scaling solutions, especially for the low-fee pool, suggests more intense
liquidity repositioning. 
We next measure repositioning intensity, $Intensity$, more explicitly. Specifically, we define repositioning as a burn, followed by a mint, in the same pool by the same liquidity provider within next five minutes. In the following, we refer to mints, associated with repositioning, as "repositioning mints". We then compute repositioning intensity as the dollar value of repositioning mints over a 5-minute interval, divided by the total dollar value minted over the same interval:

\begin{center}
\begin{equation}
Intensity=\frac{RepositioningMinted_{\$}}{TotalMinted_{\$}}, 
\end{equation}
\par\end{center}

$Intensity$ can take values between 0 and 1, with higher values
associated with more intense repositioning. We use five minutes as our benchmark time interval for computing  intensity, because it represents the average time between all mints and burns in our sample. 
However, we also check that all our results are robust if we use  the median time of one minute between all mints and burns in our sample instead. 

Column (7) of Table \ref{Flo:unistats_mintsburns} presents the average
5-minute repositioning intensity (in \%) across Ethereum, Arbitrum and Polygon.
For ETH/USDC 0.05\%, the repositioning intensity on Arbitrum (Polygon) is 40.64\% (35.90\%), which is more than double as high, compared to Ethereum's 15.89\%. For ETH/USDC 0.3\% we observe overall lower levels of intensity across all three chains,
as compared to the low-fee ETH/USDC 0.05\% pool. Indeed, high-fee pools are mostly used by more passive liquidity providers (LPs), consistent with prior findings of \textcite{lehar2022liquidity}. Importantly, $Intensity$
on both Arbitrum and Polygon of 17.10\%-24.78\% is still two to three times higher for the high-fee pool, relative to Ethereum's 7.29\%. Overall, as expected, we observe significantly more repositioning
taking place on blockchain scaling solutions, driven by higher speed of transaction processing and lower gas
fees. 

\textbf{Instrumental variable (IV) regressions. }
So far, we find that both liquidity concentration and repositioning intensity is significantly higher on blockchain scaling solutions, relative to Ethereum.  In this section, we would like to explicitly test our Hypothesis 1, showing the causal
relation between repositioning intensity and liquidity concentration
on DEXs. The straightforward approach to test this association is by
regressing liquidity concentration on repositioning intensity and
other variables, controlling for market conditions. However, the choice
of an LP to update their position, i.e. burn and subsequently mint at
a new price range, is potentially endogenous, and can in itself depend
on the current liquidity concentration in the pool. Hence, the slope
coefficient on repositioning intensity from standard OLS estimation
would represent a biased estimate of its causal effect on liquidity
concentration.

To identify the causal effect of repositioning intensity, we use
the launch of Uniswap v3 on Arbitrum and Polygon as our instrument for an exogenous increase in
updating frequency of LPs.  For any instrument to be valid, it has
to satisfy two criteria. First, the instrument must be correlated with
the endogenous explanatory variable, i.e. it should induce change
in the explanatory variable. Second, it must satisfy the exclusion
restriction, i.e. it should not be correlated with the error term
in the explanatory equation. The first condition holds,
because we indeed observe more frequent repositioning by LPs on blockchain scaling solutions as a result of much lower gas fees (see Table \ref{Flo:unistats_mintsburns}).  For exclusion restriction to hold, Uniswap's launch on blockchain scaling solutions should not affect liquidity concentration other than through
its effect on repositioning intensity. We argue that such correlation
with the error term is rather unlikely, because it would mean that
Uniswap v3 chose its entry date on Arbitrum (Polygon) strategically and was able
to accurately predict a market-wide increase in liquidity concentration.\footnote{We also discuss potential alternative explanations of our findings in Section \ref{sec:Alternative_explanations}.} Our setup is similar to \textcite{hendershott2011}, who use the introduction of NYSE Autoquote
as an instrument for an exogenous increase in algorithmic trading,
and investigate its causal effect on liquidity.

To test the causal relation between repositioning intensity and liquidity
concentration, we next run instrumental variable regressions. We first estimate the following first-stage regression,
separately for each pool:

\begin{eqnarray*}
Intensity_{t} & = & \alpha+\beta_{1}Arbitrum/Polygon{}_{t}+\\
 & + & \beta_{2}Volume{}_{t}+\beta_{3}Volatility{}_{t}+\beta_{4}|Return|{}_{t}\\
 & + & HourFE+DayFE+\varepsilon_{i,t},
\end{eqnarray*}

where $Arbitrum$ ($Polygon$) takes value of 1 for Arbitrum (Polygon) pool, and equals zero for the corresponding Ethereum pool. $Intensity$ and $|Return|$ are measured over 5-minute intervals.
$Volume$ is the volume over previous 24 hours and $Volatility$ is the
realized volatility over previous 24 hours, both measured at the end of
each 5-minute interval. All regressions include hour- and day-fixed
effects and allow standard errors to cluster at the day level.

\begin{center}
{[}Insert Table \ref{Flo:intensity} approximately here{]}
\par\end{center}

Panel A of Table \ref{Flo:intensity} reports the results
for ETH/USDC 0.05\%. The coefficient of 0.24 on $Arbitrum$ in Model (1) implies
that repositioning intensity is 24\% higher on Arbitrum for this pool, relative to Ethereum. This coefficient is 
consistent with our previous statistics on $ \Delta Arb - Eth$ from Table \ref{Flo:unistats_mintsburns}.
We find that repositioning intensity on Polygon is also significantly higher by 20\%, relative to Ethereum (Model 3). We next include both Arbitrum and Polygon pools in one regression. In Model (5), we define $BlockScaling$ as an indicator variable that takes  value of 1 for Arbitrum and Polygon pools, and zero for the Ethereum pool, used as a benchmark. On average, we observe 21\% higher repositioning intensity on blockchain scaling solutions, relative to Ethereum.   

Models (2), (4) and (6) report corresponding results for the second-stage regression: 

\begin{eqnarray*}
Conc_{t} & = & \alpha+\beta_{1}Intensity{}_{t}+\\
 & + & \beta_{2}Volume{}_{t}+\beta_{3}Volatility{}_{t}+\beta_{4}|Return|{}_{t}\\
 & + & HourFE+DayFE+\varepsilon_{i,t},
\end{eqnarray*}

Liquidity concentration, $Conc$, is measured within 2\% of the current market price at the end of each 5-minute interval.
The set of instruments consists of all explanatory variables, except
that we use $Arbitrum$ in place of $Intensity$ in Model (2). In Models (4) and (6), we use $Polygon$ and $BlockScaling$ in place of $Intensity$, respectively.   Overall, Models (2), (4) and (6) show that an increase in repositioning intensity significantly
increases liquidity concentration within 2\% of the market price. The IV estimates of 0.09-0.11 on $Intensity$
variable mean that a 1\% increase
in $Intensity$ increases liquidity concentration by around 10\% for ETH/USDC
0.05\% pool. The average standard deviation for $Intensity$ is 0.4384
for this pool, such that a one-standard deviation change in $Intensity$
is associated with a $0.4384\cdot0.10=0.0438$ or 4.38\% change in
liquidity concentration. This value is economically significant, because it represents a 43\% increase
from the mean liquidity concentration (within 2\% of the market price)
of 10.16\% on Ethereum.


We also observe
a significant effect of repositioning intensity on liquidity concentration for the high-fee pool in Panel B,
with IV estimates  on $Intensity$ of 0.08 for Arbitrum and 0.22 for Polygon.
Table \ref{Flo:unistats_mintsburns} shows that an increase in the repositioning intensity is higher on Polygon for this pool, hence we observe a stronger effect, relative to Arbitrum. 

To analyze which subset of LPs are actively engaging in repositioning, we next split LPs by their aggregate minted and burned dollar value in each pool. In other words, for each LP, we sum up their total minted and burned liquidity (in \$) over our entire sample period, separately for each chain and each pool.  We then define large LPs as those in the top quartile of our sample distribution, i.e. those with the highest dollar value minted and burned. We replicate our previous analysis from Table \ref{Flo:intensity} for a subset of large LPs. Unsurprisingly, they are practically identical to our benchmark findings in Table \ref{Flo:intensity} (see Table \ref{Flo:intensity_act} in the Internet Appendix). Overall, these findings suggest that repositioning is indeed largely done by the largest (professional) liquidity providers on each chain. 

We also replicate our analysis from Table \ref{Flo:intensity} for four additional pairs: BTC/ETH 0.05\%, BTC/ETH 0.3\%, UNI/ETH 0.3\% and LINK/ETH\%. We choose these pairs because they are among few that are relatively actively traded across all three chains, mapping into twelve additional pools. All of these pairs have ETH as their quote token, and are therefore more volatile relative to our benchmark ETH/USDC pair. Table \ref{Flo:intensity_other} in the Internet Appendix presents the results of pooled regressions across all four pairs. Overall, all our previous findings for our benchmark pair, ETH/USDC, continue to hold. On average, we observe 15\% higher repositioning intensity on blockchain scaling solutions, relative to Ethereum. Further, an increase in repositioning intensity significantly
increases liquidity concentration by around 29\%.

\textbf{Robustness checks.} Table \ref{Flo:intensity_rob} presents robustness checks of our main findings. Panel A presents results for the low-fee pool, and Panel B for the high-fee pool. Models (1) and (2) report results of the second-stage IV regressions with $BlockScaling$ as an instrument for the low-fee pool (similar to our benchmark Model 6 in Table \ref{Flo:intensity}), using liquidity concentration within 1\% and within 10\% of the current market price as the dependent variable, respectively. 
We observe that the effect of repositioning intensity on liquidity concentration
increases with the percentage band, i.e. it affects liquidity concentration
within 10\% (2\%) to a larger extent than within 2\% (1\%). This effect is 
likely due to more repositioning taking place within 10\% (2\%) of the
market price, as opposed to repositioning within 2\% (1\%). 

\begin{center}
{[}Insert Table \ref{Flo:intensity_rob} approximately here{]}
\par\end{center}

Model (3) replicates our analysis from benchmark Model (6) in Table \ref{Flo:intensity}, using an alternative measure for repositioning intensity, $IntensFreq$, defined as the ratio of the number of repositioning mints to the total number of mints within a 5-minute interval. The IV estimate on $IntensFreq$ of 0.24 for the low-fee pool is even higher, compared to the benchmark estimate of 0.11 from Panel A of Table \ref{Flo:intensity}. Model (4) uses another alternative measure of $Intensity$, based on the 1-minute repositioning mints, which is the median time between all mints and burns in our sample. In this analysis, we  classify repositioning mints as those that are preceded by burns of the same LP within the previous minute. The IV estimate of 0.16 is of comparable economic magnitude to our benchmark estimate. In Model (5), we use aggregation, based on 10-minute intervals instead of
5-minute intervals. The results are practically identical to our benchmark estimate.  We also observe similar patterns for all robustness tests for the high-fee pool in Panel B. 

To sum up, we find strong empirical evidence for our Hypothesis 1. Our instrumental variable regressions show that higher repositioning intensity on blockchain scaling solutions indeed leads to higher aggregate liquidity
concentration around the current market price, both for the low-fee and the high-fee pools.

\subsection{Repositioning precision and liquidity concentration}

\textbf{Repositioning precision.} We next test our Hypothesis 2 about the causal link of repositioning precision and aggregate liquidity concentration around the current market price. We use three measures for repositioning precision:
position gap, position length and position precision. Our first
measure is position gap, $Gap$, computed as $\left|\frac{p_{mid}}{p_{mkt}}-1 \right|$,
where $p_{mid}$ is the mid price of the position. For each position,
posted on the range $[p_{lower},p_{upper}]$, we compute $p_{mid}$
as $\sqrt{p_{lower}\cdot p_{upper}}$. The lower the position gap, the closer the mid price of minted positions to the market price. Therefore, $Gap$ is inversely related to repositioning precision. We 
\pagebreak{}

report the average position gap (in \%) across all repositioning mints in Column (1) of Table \ref{Flo:unistats_gap_length}.

\begin{center}
{[}Insert Table \ref{Flo:unistats_gap_length} approximately here{]}
\par\end{center}

For ETH/USDC 0.05\%, we observe the average position gap of 1.86\%
on Ethereum, i.e. liquidity providers re-position their mints on average
1.86\% away from the market price. Consistent with our prior expectations,
we find that the average position gap on blockchain scaling solutions is significantly
lower by 0.8\%-1\%, which represents an improvement in repositioning
precision of around 50\%. For the high-fee pool, the average position
gap on Ethereum is higher, 3.41\%, because liquidity providers do not reposition as often for this pool, compared to the low-fee pool. However, we also observe significant
precision improvement on blockchain scaling solutions for this pool by around one third (i.e. by 1\%-1.3\%). 

Our second measure of repositioning precision is position length, $Length$, computed as $\frac{\lvert p_{upper}-p_{lower}\rvert}{p_{mid}}$. The shorter the position length, the narrower the position is. 
Similar to $Gap$, it is inversely related to repositioning precision.
We report the average position length (in \%) across all repositioning mints in Column (2) of Table \ref{Flo:unistats_gap_length}. For the low-fee pool, we observe that
positions are significantly narrower by 7.38\% on Arbitrum and 5.16\% on Polygon, relative to the average value of around 18\% on Ethereum.
For the high-fee pool, positions are overall substantially wider on Ethereum,
in the range of 30\%. This result is consistent with more passive
liquidity providers (with lower repositioning intensity) posting
wider ranges on high-fee pools to protect themselves from adverse
selection. Importantly, we observe substantially narrower positions
in this pool on blockchain scaling solutions by around 14\%. This substantial reduction
in position length for the high-fee pool is due to lower gas fees on
Arbitrum and Polygon. With low gas fees, it is cheaper, even for relatively more
passive (i.e. retail) liquidity providers, to update their positions,
protecting themselves from adverse selection. 

The previous two measures look separately at the mid price of the position
range and on its length. Our third measure, position precision, $Precision$, combines the previous two to address a potential criticism that positions
with lower gaps could potentially be wider, or, alternatively, narrower
positions could be posted further away from the market price. We compute
$Precision$ as $\frac{1}{Gap\cdot Length}$
and further scale it to lie in the range {[}0,1{]} as $1-1.0001^{\frac{-1}{Gap\cdot Length}}.$ $Precision$ has the lowest possible value of 0 and  the highest
possible value of 1, or 100\%. As for other measures, we compute the average precision
across all repositioning mints.

Column (3) of Table \ref{Flo:unistats_gap_length}
shows that position precision is indeed significantly higher on
blockchain scaling solutions, relative to Ethereum. For ETH/USDC 0.05\%, $Precision$ is higher by
18\%-20\%, relative to the average value of
51\% on Ethereum. As expected, position precision is generally
lower for the high-fee pool with more passive liquidity providers, around 20\% on Ethereum.  Importantly, it almost doubles on blockchain scaling solutions, increasing by 17\% on Arbitrum and by 22\% on Polygon.

Overall, our univariate results confirm that liquidity providers re-position on Arbitrum and Polygon not only closer to the market price, but also at significantly narrower ranges.
Thus, lower gas fees result in both higher repositioning
intensity and higher repositioning precision around the market price. 

\textbf{Instrumental variable (IV) regressions. } We next test the causal relation between repositioning
precision and liquidity concentration. Similar to repositioning intensity, repositioning precision is endogenous. To address
this issue, we use the launch of Uniswap v3 on blockchain sclaing solutions as our instrument for an exogenous
increase in repositioning precision. Since LPs are able to reposition
more frequently on Arbitrum and Polygon, they are better able to protect themselves
from adverse selection. Hence, for the same amount of capital deposited,
LPs can maximize their rewards from liquidity provision by improving the
precision of their positions, i.e. minting positions that are more
concentrated around the market price. Indeed, our findings
from Table \ref{Flo:unistats_gap_length} show that repositioning
precision is significantly higher on both Arbitrum and Polygon, relative to Ethereum.

We re-estimate our previous instrumental variables regressions, with
$Blockscaling$ now used as an instrument for repositioning precision. Table \ref{Flo:precision}
reports results for instrumental variable regressions, separately for three measures of repositioning precision.\footnote{Table \ref{Flo:precision_ia} in the Internet Appendix reports results separately for Arbitrum and Polygon. To conserve space, we only present combined results for $Blockscaling$ in Table \ref{Flo:precision}.}

\begin{center}
{[}Insert Table \ref{Flo:precision} approximately here{]}
\par\end{center}



Models (1) and (2) show results with position gap as a measure of repositioning precision. Model (1) reports results of the first-stage IV regression, with $Gap$ as the dependent variable and $Blockscaling$ as the main explanatory variable. Consistent with previous findings in Table \ref{Flo:unistats_gap_length}, we observe significantly lower position gaps on blockchain scaling solutions, relative to Ethereum. Model (2) shows results of the second-stage IV regressions, with liquidity concentration within 2\% of the market price as the dependent variable.\footnote{All our results also hold if we use liquidity concentration within 1\% or 10\% as the dependent variable.}  Consistent with Hypothesis 2, we find that higher position gaps significantly reduce liquidity
concentration around the current market price. The average standard deviation for $Gap$ is
0.023 for ETH/USDC 0.05\%. Therefore, the IV estimate of -4.15
means that a one-standard deviation increase in $Gap$ is
associated with a $0.023\cdot-4.15=-0.0954$ or 9.54\% decrease
in liquidity concentration. As expected, we also observe a significant negative effect of $Length$ (Models 3 and 4) and a significant positive effect of $Precision$ (Models 5 and 6) on liquidity
concentration for both pools. 


We also replicate our analysis from Table \ref{Flo:precision} for four additional pairs: BTC/ETH 0.05\%, BTC/ETH 0.3\%, UNI/ETH 0.3\% and LINK/ETH\%. Table \ref{Flo:precision_ia_other} in the Internet Appendix presents the results of pooled regressions across all four pairs. Consistent with our benchmark results, we find a significant negative effect of $Gap$ and $Length$ and a significant positive effect of $Precision$ on liquidity concentration within 2\% of the market price. 

Consistent with our Hypotheses 1 and 2, our results in this section show that both repositioning
intensity and precision increase liquidity concentration around the
market price. As both these variables are potentially
endogenous, we use the launch of Uniswap v3 on Arbitrum and Polygon as our instruments to identify
the causal effect of repositioning intensity and precision on liquidity concentration. An increase in liquidity concentration is important because it reduces slippage, for a given TVL. In the next section, we explicitly compare slippage for trades of different sizes on Ethereum and blockchain scaling solutions.


\section{Slippage: Ethereum vs blockchain scaling solutions\label{sec:Slippage}}

In this paper, we define slippage of a trade as the difference between
the average execution price, $p_{avg}=\frac{\Delta y}{\Delta x}$ and the pre-execution pool price, $p_{mkt}$:

\begin{center}
$Slippage=\left|\frac{p_{avg}}{p_{mkt}}-1\right|$ 
\par\end{center}

In other words, slippage shows by how much the
execution price is worse than the previously displayed market price.\footnote{The term \textquotedblleft slippage\textquotedblright{} on Uniswap
is defined more broadly as the percentage difference between the quoted
price at the time of submitting the transaction and the actual execution
price. For instance, a sandwich attack could also cause slippage.
However, our definition abstracts from ``sandwich attacks'', assuming
that nothing happens between submission and execution, and refers
to the difference of the average execution price relative to the market price.} Slippage is greater for larger trades and, as in \textcite{kyle1985continuous}, is
inversely related to market depth. Appendix C discusses mechanics of trade execution and derivations of average execution prices on Uniswap v3. Appendix D provides a numerical example
of slippage computation on DEXs. 

We compute slippage, $Slippage$, for
hypothetical trades of sizes [\$100, \$500, \$1K, \$5K, \$10K, \$50K, \$100K] at the end of each 5-minute interval over our sample period (January 1, 2022 - June 30, 2023). Panel A of Table \ref{Flo:slippage} presents summary
statistics for $Slippage$ (in bp), separately
for each chain. 

\begin{center}
{[}Insert Table \ref{Flo:slippage} approximately here{]}
\par\end{center}

Given Ethereum's overall larger TVL, it is not surprising that
its average slippage of 0.3 bp for the low-fee pool is significantly lower than Arbitrum's 8.8 bp and Polygon's
4.25 bp. The median slippage of 0.60-0.73 bp on blockchain sclaing solutions is also higher,
compared to Ethereum's 0.05 bp. As expected, average and median slippage values are higher for the high-fee pool  across all chains, due to its lower TVL. However, these statistics represent
the average across both small hypothetical trades of \$100, \$500, etc.
and large hypothetical trades of up to \$100K. As Arbitrum and Polygon are mostly
used by smaller traders, the main focus of our analysis will be the
comparison of slippage for small trades across Ethereum and scaling solutions.

Before conditioning on trade size, we first confirm our findings from Panel A in a multivariate setup. Specifically, we estimate following OLS regressions:

\begin{eqnarray*}
Slippage_{i,t} & = & \alpha+\beta_{1}BlockScaling{}_{i,t}+\beta_{2}Size{}_{i,t}+\\
 & + & \beta_{3}Buy{}_{i,t}+\beta_{4}Volume{}_{i,t}+\beta_{5}Volatility{}_{i,t}+\beta_{6}|Return|{}_{i,t}\\
 & + & HourFE+DayFE+\varepsilon_{i,t}.
\end{eqnarray*}

The dependent variable, $Slippage$, shows the slippage for a trade of
a given size for pool $i$ at the end of each 5-minute interval $t$ (from the end-of-minute
snapshot of liquidity distribution). As before, $BlockScaling$, equals 1 for transactions on Arbitrum and Polygon, and zero for Ethereum.
Therefore, trades on Ethereum serve as a benchmark sample. The vector of
control variables includes trade size (in \$K), $Size$; the direction
of the trade, $Buy$, that equals 1 for purchases of token X and 0
for its sales; trading volume over previous 24 hours (in \$M), $Volume$;
realized volatility of ETH/USD over previous 24 hours,  $Volatility$; and absolute return
of ETH/USD over the previous minute, $|Return|$. All regressions
include hour- and day-fixed effects and allow standard errors to cluster
at the day level.

Panel B of Table \ref{Flo:slippage} reports results for ETH/USDC
0.05\% pool. The coefficient on $BlockScaling$ in Model (1) shows that
slippage on blockchain scaling solutions is on average 1.41 bp higher, relative to the
average slippage on Ethereum, after including control variables. Unsurprisingly, slippage
is increasing in the size of the trade. It does not differ significantly between buys
of token X and its sells. Further, slippage
is decreasing in the trading volume, implying that LPs deposit more
liquidity if they observe higher volume over previous 24 hours. This
finding is consistent with our prior expectations, because LPs can
earn higher fees in periods of higher volume. Finally, slippage is
increasing in both the absolute return of ETH/USD over the previous
five minutes and the realized volatility over previous 24 hours. These findings
are also in line with our expectations, suggesting that LPs are more
likely to withdraw liquidity from the pool during times of higher
uncertainty. We observe similar findings for the high-fee pool in Panel C, with slippage on average higher by 21.61 bp on blockchain scaling solutions, 

We next test our Hypothesis 3, which predicts that higher liquidity concentration on blockchain scaling solutions should result in lower slippage for small trades. Hence, we condition our analysis on trade size, using the \$1K and the \$5K cutoffs to split trades into small and large categories. We use these cutoffs, because the 75th percentile of trades on both Arbitrum and Polygon never exceeds \$5K, with average trade sizes of \$2K-\$4K (see Panel B of Table \ref{Flo:average_vol_tvl}). 

In Model (2), we add a dummy variable $Large$ that equals 1 for large trades that exceed the \$1K cutoff, and
zero otherwise. We also add its interaction term with $BlockScaling$,
$BlockSc\cdot Large$, that captures the relative difference in slippage
of large trades on blockchain scaling solutions, relative to Ethereum. The coefficient
on $BlockScaling$ now captures the relative difference in slippage of
small trades between Ethereum and blockchain scaling solutions, and is of main interest
in our analysis. As $Large$ is mechanically related to $Size$, we
omit $Size$ from the vector of our control variables in Model (2).

We observe a positive and significant coefficient on $Large$ in Model
(2), suggesting that large trades on Ethereum (i.e. those in excess
of \$1K) have on average 0.51 bp higher slippage in the low-fee pool, compared to small
trades. Importantly, we observe coefficients of different signs on
$BlockScaling$ and $BlockSc\cdot Large$. The negative coefficient on
$BlockScaling$ shows that slippage of small trades (less than \$1K) on scaling solutions is by 4.6 bp lower, compared to small trades on Ethereum.
The positive coefficient on the interaction term, $BlockSc\cdot Large$,
shows that large trades on scaling solutions have a significantly higher slippage
of 10.53 bp, relative to large trades on Ethereum. Hence, the overall
positive coefficient on $BlockScaling$ in Model (1) is driven by significantly
higher slippage for larger trades. Once we condition on trade size,
we observe significantly lower slippage for small trades on Arbitrum and Polygon.

Model (3) reports similar results, using \$5K cutoff
to split trades into small and large categories. Models (4) and (5) report results separately for Arbitrum and Polygon, using \$1K cutoff. Overall, we observe economically stronger effect for Arbitrum, most likely because Arbitrum pools are more liquid and have higher TVL, compared to Polygon (see Table \ref{Flo:average_vol_tvl}). All results also hold for the high-fee pool (Panel C), except that the coefficient on $Polygon$ is no longer significant. Indeed, ETH/USDC 0.3\% pool on Polygon has the lowest trading volume and TVL, such that we do not see any improvement in slippage for the high-fee pool on Polygon. However, this is in line with theoretical predictions of \textcite{lehar2022liquidity} that all liquidity should consolidate on the low-fee pool in presence of low gas fees. 

Overall, our findings in this section strongly support our Hypothesis 3 that
small trades on scaling solutions have significantly lower slippage, relative
to small trades on Ethereum. Importantly, the total execution cost of small trades
is further diminished by low gas fees on Arbitrum and Polygon. In contrast, large
trades have significantly lower slippage on Ethereum due to its larger liquidity (higher TVL). Whereas gas fees are higher on Ethereum, they are not
of first-order importance for large trades, because they represent
a fixed cost. Thus, the effect of lower slippage outweighs the effect of
higher gas fees for large trades on Ethereum. 

\section{Alternative explanations\label{sec:Alternative_explanations}}

In this section, we address potential alternative explanations of our results. Indeed, it could be that higher liquidity concentration on scaling solutions is driven not necessarily by higher repositioning activity, but is an equilibrium outcome of liquidity provision. \textcite{hasbrouck2023liqprovision} model equilibrium liquidity provision for any price range on Uniswap v3. Specifically, their Proposition 3.1 shows that equilibrium liquidity provision in a given interval increases in the expected fee revenues and in the ex-fee return to liquidity providers from holding a dynamic portfolio of two assets (e.g. ETH/USDC), $R_{_{P\&L}}^{i}$. The ex-fee return to liquidity providers is driven not only by fluctuations in ETH/USDC price, but also by changes in the quantity of ETH and USDC in a given price range, i.e. $R_{_{P\&L}}^{i}$ is decreasing in adverse selection.

Importantly, this equilibrium outcome is most likely  different across Ethereum and scaling solutions due to differences in the distribution of expected trade sizes and trading volume. Fee levels  (0.05\% or 0.3\%) are fixed on Uniswap v3 and remain the same across all three chains. In absence of repositioning and assuming that arbitrageurs are not constrained in their capital, the ex-fee return to LPs should be the same across all chains (for any given amount of trading volume in a price interval).\footnote{Proposition 4.8 of \textcite{hasbrouck2023liqprovision} shows that the ex-fee return to LPs is comparable to holding a covered call position, i.e. holding ETH and shorting an ETH/USDC call option against that ETH position. If the ETH/USDC price is following geometric Brownian motion, Black-Scholes formula for option valuation can be applied. Assuming perfect arbitrage,  volatility of ETH/USDC should be the same across all chains. The remaining parameters, such as strike price, interest rate and option maturity are the same by construction. Then, the ex-fee return to LPs should be equal across chains, holding trading volume constant.} However, the expected trading volume and trade sizes differ across Ethereum and scaling solutions due to differences in underlying security of chains, potential network effects etc. Given overall smaller expected trade sizes on Arbitrum and Polygon, it might be indeed optimal for LPs to concentrate liquidity around the current tick range on scaling solutions. Liquidity positions that lie further away from the current tick range are less likely to be active (similar to limit orders that are less likely to be executed if they are posted further away from the best bid/ask in the limit order book). In contrast, liquidity positions that are further away from the current tick range are more likely to become active on Ethereum due to its larger expected trade sizes.  

Whereas equilibrium liquidity distributions most likely differ across Ethereum and scaling solutions, our aim is to show that repositioning activity indeed plays an important role for aggregate liquidity concentration on DEXs. To identify the effect of repositioning activity even further, we use an exogenous shock within Arbitrum chain, related to the airdrop of Arbitrum token on March 23, 2023. Arguably, if a shock to repositioning activity occurs within the same chain, all other underlying parameters that potentially affect the distribution of the trade size (e.g., chain security), remain constant. In this case, changes in liquidity concentration can indeed be mostly attributed to an increase in repositioning activity of liquidity providers. 

During the airdrop, around 1.162B of Arbitrum's native token, ARB, was distributed to users of the platform and another 113M to Decentralized Autonomous Organizations (DAOs). Out of 113M, Uniswap was entitled to 4.3M, which makes it the third largest DAO recipient after Treasure and GMX.\footnote{See https://docs.arbitrum.foundation/airdrop-eligibility-distribution for details of ARB distribution.} Uniswap, alongside other DEXs, incentivized their liquidity pools on Arbitrum by distributing ARB tokens to LPs as “rewards”. Importantly, Uniswap's rewards to LPs were linked not only to the aggregate liquidity provided, but also to the concentration of LP's positions around the market price. The suggested formula for computation of a reward score is increasing in the fees earned by each LP position, relative to total fees earned by all LPs in the pool.\footnote{See https://gov.uniswap.org/t/rfc-gamma-strategies-distribute-at-least-1-3-of-arb-airdrop-as-liquidity-incentives/21345/2}. Hence, it incentivizes LPs to stay active and re-position in the current price range to maximize their rewards. 


Overall, we expect that an increase in repositioning intensity on Arbitrum after the airdrop should have a more pronounced effect on aggregate liquidity concentration, relative to the pre-airdrop sample. To test this prediction, we re-estimate our IV regressions, separately before and after the airdrop, in Table \ref{Flo:intensity_airdrop}. 

\begin{center}
{[}Insert Table \ref{Flo:intensity_airdrop} approximately here{]}
\par\end{center}

Models (1)-(3) report results for the pre-airdrop sample and Models (4)-(6) for the post-airdrop sample, using $Arbitrum$ as our instrument variable.  As expected, we observe a significant increase in repositioning intensity on Arbitrum after the airdrop. Before the airdrop, repositioning intensity for the low-fee pool on Arbitrum is 22\% higher, relative to Ethereum (Model 1 of Panel A).  After the Airdrop, this difference increases to 35\% (Model 4). Importantly, the effect of repositioning intensity on aggregate liquidity concentration increases from 4\% before the airdrop to around 29\% afterwards (Models 2 and 5). We observe an even stronger increase in the effect of repositioning precision on liquidity concentration, from 5\% before the airdrop to 68\% afterwards.\footnote{To conserve space, we do not report first-stage IV regressions for repositioning precision, but we confirm that it increases significantly after the airdrop (available upon request).} All our findings also hold for the high-fee pool (Panel B). 

To sum up, whereas we cannot possibly rule out differences in equilibrium liquidity distributions across Ethereum and blockchain scaling solutions, an exogenous shock to repositioning activity within Arbitrum helps us further identify the causal effect of repositioning activity on aggregate liquidity concentration on DEXs.
 
\section{Conclusions}

Liquidity providers (LPs) on decentralized exchanges (DEXs) can protect
themselves from adverse selection by either setting a wide price
range for their position or by updating it more frequently, in response
to changes in the market price. However, updating is costly, because
every repositioning from an LP requires the payment of a fixed cost (a
gas fee). Blockchain scaling solutions,
such as Arbitrum and Polygon, allow for more frequent updating by LPs due to their lower gas fees. With more frequent updating,
LPs are better able to track the market price and protect themselves
from adverse selection. Thus, for the same amount of capital deposited
in the pool, they can maximize their rewards by making their positions
more concentrated around the market price. In this paper, we use the
launch of Uniswap v3 on Arbitrum and Polygon as our instrument to show that higher repositioning
intensity and precision of LPs leads to higher aggregate liquidity concentration around the market price. 

Higher liquidity concentration around the market price is important,
because it reduces slippage, especially for small trades. Indeed, we
find that slippage of small trades (up to \$5K) is significantly
smaller on blockchain scaling solutions, relative to the incumbent Ethereum. Thus, liquidity
pools on scaling solutions provide better execution terms for small (retail) traders.
However, these benefits come at a cost of lower security on scaling solutions,
relative to Ethereum. Due to its higher security, Ethereum
attracts larger liquidity providers, resulting in higher TVL. Consequently,
slippage of large trades is lower on Ethereum as its liquidity pools are deeper. High gas fees on Ethereum are of lower importance for large
traders due to their fixed-cost nature. Thus, similar to a separating equilibrium, we observe that
large traders and liquidity providers are attracted to Ethereum,
whereas small (e.g. retail) traders and liquidity providers are better off
using blockchain scaling solutions.

\pagebreak
\printbibliography

\pagebreak

\section*{\raggedright{Figures}}

\pagebreak{}

\pagebreak{}

\begin{center}
\textbf{\small{}Panel C: Protecting by updating the position}{\small\par}
\par\end{center}

\begin{singlespace}
\begin{center}
\includegraphics[scale=0.5]{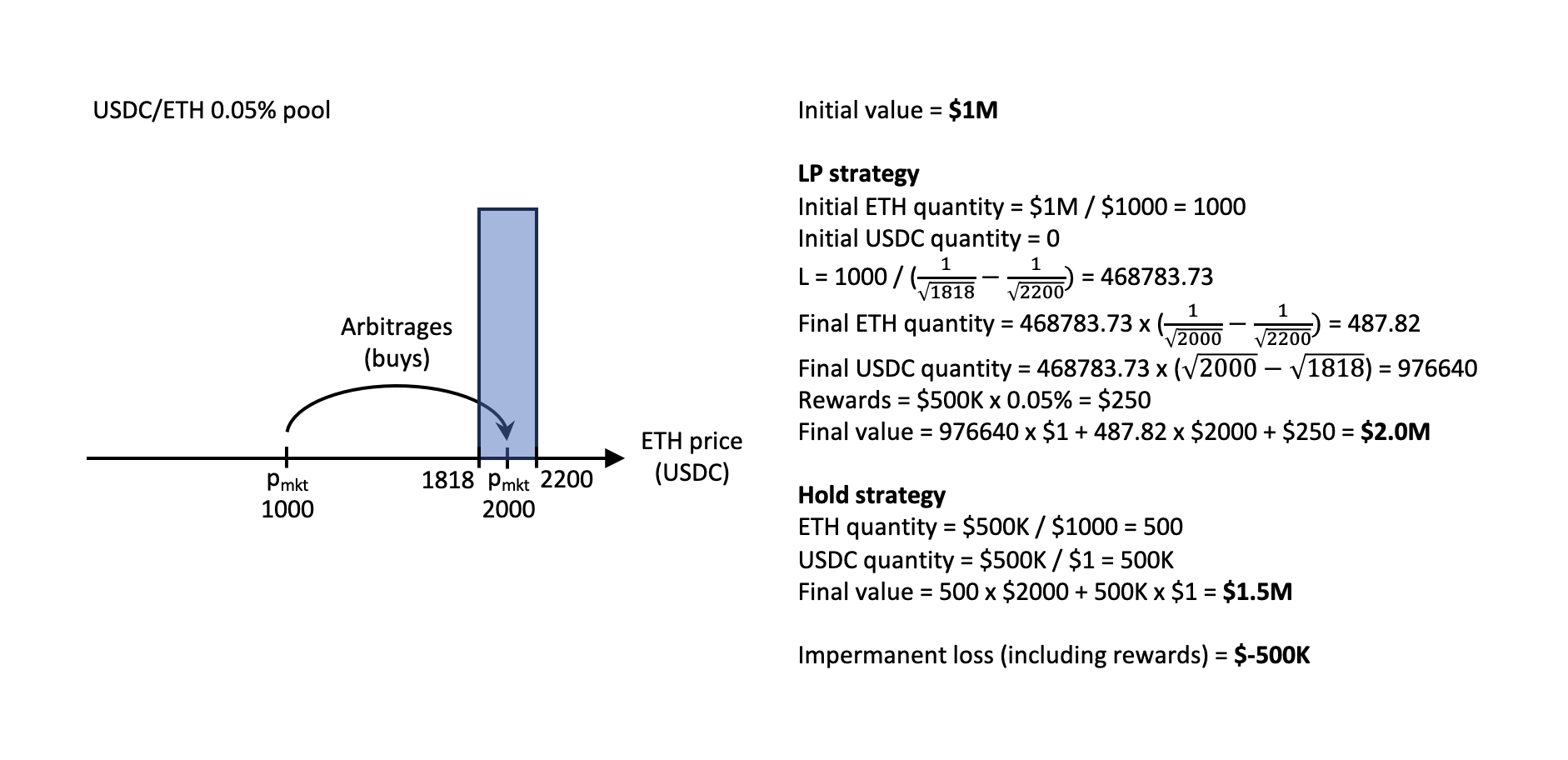}
\par\end{center}
\end{singlespace}

\pagebreak{}

\pagebreak{}

\section*{\raggedright{Tables}}

\begin{table}[H]
\caption{\textbf{\footnotesize{}Summary statistics: volume, TVL and trade size.}{\footnotesize{} Panel A of this table presents average 24-hour volume (in \$M) and average total value locked (TVL) (in \$M) on Uniswap v3 across Ethereum, Arbitrum and Polygon for two liquidity pools: ETH/USDC 0.05\% and ETH/USDC 0.3\%. Panel B reports summary statistics of trade sizes. Panel C reports the average daily numbers of trades,
purchases and sales as well as the average time (in seconds) between
two trades (delta). Our sample period ranges from January 1, 2022 to June 30, 2023. Appendix E provides a detailed description
of all variable definitions.}{\small{}\bigskip{}}}

\label{Flo:average_vol_tvl}
\begin{centering}
\textbf{\small{}Panel A: Volume and TVL}{\small\par}
\par\end{centering}
{\small{}\bigskip{}
}{\small\par}
\centering{}{\small{}\begin{tabularx}{13cm}{cccc}
\hline
\textbf{Pool}   & \textbf{Volume 24h (\$M)} & \textbf{TVL (\$M)} & \textbf{Volume/TVL} \\
\hline
ETH/USDC 0.05\% &                           &                    &                     \\
Ethereum        & 513.73                    & 201.19             & 2.55                \\
Arbitrum        & 51.93                     & 18.04              & 2.88                \\
Polygon         & 29.03                     & 10.26              & 2.83                \\
\hline
ETH/USDC 0.3\%  &                           &                    &                     \\
Ethereum        & 57.83                     & 176.59             & 0.33                \\
Arbitrum        & 1.65                      & 4.48               & 0.37                \\
Polygon         & 0.95                      & 1.89               & 0.50 
\\\hline
\end{tabularx}}{\small\par}

{\small{}\bigskip{}
}{\small\par}
\begin{centering}
\textbf{\small{}Panel B: Trade size}{\small\par}
\par\end{centering}
{\small{}\bigskip{}
}{\small\par}
\centering{}{\small{}\begin{tabularx}{12cm}{cccccc}
\hline
\textbf{Pool}   & \textbf{Mean (\$K)} & \textbf{SD} & \textbf{25\%} & \textbf{50\%} & \textbf{75\%} \\
\hline
ETH/USDC 0.05\% &                     &             &               &               &               \\
Ethereum        & 72.19               & 255.61      & 0.82          & 4.55          & 42.67         \\
Arbitrum        & 4.12                & 10.10       & 0.06          & 0.96          & 4.64          \\
Polygon         & 2.42                & 5.09        & 0.07          & 0.77          & 2.86          \\
\hline
ETH/USDC 0.3\%  &                     &             &               &               &               \\
Ethereum        & 116.19              & 241.02      & 1.31          & 30.00         & 144.65        \\
Arbitrum        & 2.95                & 6.76        & 0.08          & 1.34          & 3.33          \\
Polygon         & 1.74                & 3.37        & 0.38          & 0.88          & 1.83         
\\\hline
\end{tabularx}}{\small\par}
\end{table}

\begin{center}
\pagebreak\textbf{\small{}Panel C: Trade frequency}{\small\par}
\par\end{center}
{\small{}\vspace{-1cm}
}{\small\par}
\begin{singlespace}
\begin{center}
{\footnotesize{}\begin{tabular}{ccccc}
\hline
\textbf{Pool}   & \textbf{Daily trades} & \textbf{Daily buys} & \textbf{Daily sells} & \textbf{Delta (s)} \\
\hline
ETH/USDC 0.05\% &                       &                     &                      &                    \\
Ethereum        & 7130.90               & 3495.97             & 3634.93              & 12.14              \\
Arbitrum        & 12668.05              & 6146.88             & 6521.17              & 6.83               \\
Polygon         & 12048.61              & 5651.39             & 6397.23              & 7.18               \\
\hline
ETH/USDC 0.3\%  &                       &                     &                      &                    \\
Ethereum        & 498.17                & 245.12              & 253.05               & 173.75             \\
Arbitrum        & 560.95                & 285.41              & 275.53               & 154.31             \\
Polygon         & 545.50                & 272.64              & 272.86               & 158.67            
\\\hline
\end{tabular}}{\footnotesize\par}
\par\end{center}
\end{singlespace}

\begin{table}[H]
\caption{\textbf{\footnotesize{}Summary statistics: control variables. }{\footnotesize{}This table
presents summary statistics for the following control variables: the traded volume over the previous 24 hours, $Volume$ (in \$M); the realized volatility of
ETH/USD over the previous 24 hours, $Volatility$, (in \%); and the 1-minute
return of ETH/USD, $Return$ (in bp). $Volatility$ and $Return$ are based on
a single time series for ETH/USD from Binance, which is independent
from Uniswap v3. Appendix E provides a detailed description
of all variable definitions. Our sample period ranges from January 1, 2022 to June 30, 2023.}{\small{}\vspace{1cm}
}}

\label{Flo:sumstat}
\centering{}{\small{}\begin{tabular}{cccccc}
\hline
                    & \textbf{Mean} & \textbf{SD} & \textbf{25\%} & \textbf{50\%} & \textbf{75\%} \\ \hline
Volume 24h (\$M)    &               &             &               &               &               \\
ETH/USDC 0.05\%     &               &             &               &               &               \\
Ethereum            & 513.73        & 338.53      & 277.40        & 466.48        & 644.16        \\
Arbitrum            & 51.93         & 49.65       & 17.54         & 33.15         & 70.16         \\
Polygon             & 29.03         & 18.28       & 16.50         & 25.52         & 37.34         \\
ETH/USDC 0.3\%      &               &             &               &               &               \\
Ethereum            & 57.83         & 62.88       & 17.31         & 41.96         & 76.64         \\
Arbitrum            & 1.65          & 1.46        & 0.63          & 1.08          & 2.21          \\
Polygon             & 0.95          & 0.94        & 0.30          & 0.71          & 1.28          \\ \hline
Volatility 24h (\%) & 3.53          & 1.87        & 2.31          & 3.16          & 4.26          \\
Return 1min (bp)    & 0.00          & 10.53       & -3.87         & 0.00          & 3.80  \\ \hline       
\end{tabular}}{\small\par}
\end{table}

\begin{table}[H]
\caption{\textbf{\footnotesize{}Liquidity concentration.
}{\footnotesize{}This table reports average levels of liquidity concentration across Ethereum, Arbitrum and Polygon for two liquidity pools: ETH/USDC 0.05\% and ETH/USDC 0.3\%.
We define liquidity concentration within x\%
of the market price, as the ratio of market depth within x\% of
$p_{mkt}$, divided by $TVL$. Market depth within x\% of $p_{mkt}$
is computed as the \$ value of the liquidity between $\frac{p_{mkt}}{1+x\%}$
and $p_{mkt}\cdot(1+x\%).$ Appendix E provides a detailed description
of all variable definitions. Column (1) reports average levels of
liquidity concentration within 1\% of the market price. Columns (2) and (3) report
corresponding statistics for liquidity concentration within 2\% and
10\% of the market price, respectively. We also report the difference between averages on Arbitrum and Ethereum ($\Delta Arb-Eth$) as well as Polygon and Ethereum ($\Delta Pol-Eth$). T-statistics
of the two-tailed t-test with the null-hypothesis of differences
equaling zero exceed 100. {*}{*}{*}, {*}{*}, and {*}
indicate statistical significance at the 1\%, 5\%, and 10\% level,
respectively.}{\small{}\bigskip{}
}}

\label{Flo:unistats_liqconcentration}
\centering{}{\small{}\begin{tabular}{ccccccc}
\hline
\textbf{Pool}   & \multicolumn{6}{c}{\textbf{Liquidity   concentration (\%)}}                                             \\
\textbf{}       & \multicolumn{2}{c}{\textbf{1\%}} & \multicolumn{2}{c}{\textbf{2\%}} & \multicolumn{2}{c}{\textbf{10\%}} \\
\hline
ETH/USDC 0.05\% &                 &                &                  &               &                  &                \\
Ethereum        & 5.43            &                & 10.16            &               & 35.28            &                \\
Arbitrum        & 6.35            &                & 11.85            &               & 39.64            &                \\
$\Delta Arb-Eth$    & 0.93            & ***            & 1.70             & ***           & 4.36             & ***            \\
Polygon         & 6.61            &                & 12.45            &               & 42.44            &                \\
$\Delta Pol-Eth$      & 1.18            & ***            & 2.29             & ***           & 7.16             & ***            \\
\hline
ETH/USDC 0.3\%  &                 &                &                  &               &                  &                \\
Ethereum        & 2.80            &                & 5.53             &               & 24.20            &                \\
Arbitrum        & 3.81            &                & 7.39             &               & 29.56            &                \\
$\Delta Arb-Eth$      & 1.01            & ***            & 1.85             & ***           & 5.36             & ***            \\
Polygon         & 5.24            &                & 10.20            &               & 38.74            &                \\
$\Delta Pol-Eth$     & 2.44            & ***            & 4.67             & ***           & 14.54            & ***  
\\\hline
\end{tabular}}{\small\par}
\end{table}

\begin{table}[H]
\caption{\textbf{\footnotesize{}Liquidity provision and repositioning intensity.} {\footnotesize{}This table reports average levels
of liquidity provision across Ethereum, Arbitrum and Polygon for two liquidity pools: ETH/USDC 0.05\% and ETH/USDC 0.3\%. We report the average daily number
of mints (column 1); the average duration between two mints (in minutes)
(column 2); the average daily minted value (in \$M) (column 3); the
average daily number of burns (column 4); the average duration between
two burns (in minutes) (column 5); the average daily burned value (in \$M) (column 6); and the average repositioning intensity over 5-minute
intervals (in \%) (column 7). We define repositioning as a burn followed by a mint in the same pool by the same liquidity provider within next 5 minutes. We measure repositioning intensity as the value of repositioning mints (in \$) over a 5-minute interval divided by the total value minted (in \$) over the same interval. Appendix E provides a detailed description
of all variable definitions. We also report the difference between averages on Arbitrum and Ethereum ($\Delta Arb-Eth$) as well as Polygon and Ethereum ($\Delta Pol-Eth$). T-statistics of the two-tailed
t-test with the null-hypothesis of difference equaling zero are
reported in parentheses.}{\small{}\bigskip{}
}}

\label{Flo:unistats_mintsburns}
\centering{}{\footnotesize{}\begin{tabular}{cccccccc}
\hline
\textbf{Pool}   & \textbf{Mints}    & \textbf{Delta} & \textbf{Minted}    & \textbf{Burns}    & \textbf{Delta} & \textbf{Burned}    & \textbf{Intensity} \\
\textbf{}       & \textbf{(\#/day)} & \textbf{(min)} & \textbf{(\$M/day)} & \textbf{(\#/day)} & \textbf{(min)} & \textbf{(\$M/day)} & \textbf{(\%)}      \\
\hline
ETH/USDC 0.05\% &                   &                &                    &                   &                &                    &                    \\
Ethereum        & 81.21             & 17.73          & 41.19              & 61.54             & 23.40          & 39.90              & 15.89              \\
Arbitrum        & 315.27            & 4.57           & 11.75              & 237.83            & 6.05           & 11.60              & 40.64              \\
$\Delta Arb-Eth$      & 234.06            & -13.17         & -29.44             & 176.29            & -17.34         & -28.31             & 24.75              \\
         & (137.45)            &                & (-29.00)            & (106.40)            &                & (-28.19)             & (96.64)              \\
Polygon         & 546.64            & 2.63           & 2.95               & 371.46            & 3.88           & 2.82               & 35.90              \\
$\Delta Pol-Eth$      & 465.43            & -15.10         & -38.24             & 309.92            & -19.52         & -37.08             & 20.01              \\
        & (246.81)            &                & (-40.04)             & (205.67)            &                & (-39.32)             & (84.01)              \\
\hline
ETH/USDC 0.3\%  &                   &                &                    &                   &                &                    &                    \\
Ethereum        & 74.18             & 19.41          & 6.53               & 32.65             & 44.10          & 5.42               & 7.29            \\
Arbitrum        & 69.92             & 20.59          & 0.26               & 31.66             & 45.48          & 0.24               & 17.10            \\
$\Delta Arb-Eth$     & -4.25             & 1.18           & -6.27              & -0.99             & 1.38           & -5.18              & 9.81             \\
         & (-4.69)             &                & (-26.33)             & (-2.45)             &                & (-22.76)             & (36.14)              \\
Polygon         & 72.94             & 19.74          & 0.30               & 46.88             & 30.72          & 0.28               & 24.78              \\
$\Delta Pol-Eth$      & -1.24             & 0.33           & -6.23              & 14.22             & -13.38         & -5.14              & 17.49              \\
       & (-1.67)             &                & (-26.16)             & (30.70)             &                & (-22.58)             & (60.39)         
\\\hline
\end{tabular}}{\footnotesize\par}
\end{table}

\begin{table}[H]
\caption{\textbf{\footnotesize{}Repositioning intensity and liquidity concentration: IV regressions.}{\footnotesize{} This table presents
results of instrumental variable regressions that test causal effect
of repositioning intensity of LPs, $Intensity$, on aggregate liquidity concentration around the market price, $Conc$. Panel A reports results for the low-fee pool (ETH/USDC 0.05\%), and Panel B for the high-fee pool (ETH/USDC 0.3\%). $Arbitrum$ ($Polygon$) takes value of 1 for Arbitrum (Polygon) pools, and zero for Ethereum pools. Model (1) reports the results for the first-stage regression,
with repositioning intensity of LPs, $Intensity$, as the dependent
variable and $Arbitrum$ as the main explanatory variable, which is used as an instrument for the endogenous
$Intensity$. Model (2)
reports results for the second-stage regression, with liquidity concentration within 2\% of the market price, $Conc$, as the dependent variable. We estimate liquidity concentration at the end
of every 5-minute interval and the repositioning intensity over every 5-minute interval. The set of instruments consists
of all explanatory variables, except that $Intensity$ is replaced
with $Arbitrum$. The table shows corresponding results for Polygon pools in Models (3) and (4), with $Polygon$ used as an instrument for the endogenous
$Intensity$. Models (5) and (6) use $BlockScaling$ that takes value of 1 for both Arbitrum and Polygon pools, and zero for the corresponding Ethereum pool, as an instrument. 
The vector of control variables consists of $Volume$, 
$Volatility$ and $|Return|$. See Appendix E for a detailed description of variable definitions.
All regressions include hour- and day-fixed effects, with standard
errors clustered at the day level. T-statistics of the two-tailed
t-test with the null-hypothesis of a coefficient equaling zero are
reported in parentheses. {*}{*}{*}, {*}{*}, and {*} indicate statistical
significance at the 1\%, 5\%, and 10\% level, respectively.}{\small{}\bigskip{}
}}

\label{Flo:intensity}
\begin{centering}
{\footnotesize{}\begin{tabularx}{\textwidth}{l *{6}{R@{ }l}} \\
\toprule
 \multicolumn{11}{l}{ \textbf{Panel A: ETH/USDC 0.05\%}} \\  \hline 
 & \multicolumn{4}{c}{Arbitrum} & \multicolumn{4}{c}{Polygon} & \multicolumn{4}{c}{BlockScaling}  \\ 
 & Int & & Conc & & Int & & Conc & & Int & & Conc &    \\ 
 &  & & 2\% & &  & & 2\% & &  & & 2\% &    \\ 
& (1) & & (2) & & (3) & & (4) & & (5) & & (6) &  \\ 
\midrule
$Arbitrum$          &        0.24&         ***&            &            &            &            &            &            &            &            &            &            \\
                    &     (35.55)&            &            &            &            &            &            &            &            &            &            &            \\
$Polygon$           &            &            &            &            &        0.20&         ***&            &            &            &            &            &            \\
                    &            &            &            &            &     (33.60)&            &            &            &            &            &            &            \\
$BlockScaling$      &            &            &            &            &            &            &            &            &        0.21&         ***&            &            \\
                    &            &            &            &            &            &            &            &            &     (43.78)&            &            &            \\
$Intensity$         &            &            &        0.09&         ***&            &            &        0.11&         ***&            &            &        0.11&         ***\\
                    &            &            &     (59.67)&            &            &            &     (66.04)&            &            &            &     (69.65)&            \\
$Volume$            &        0.00&            &        0.01&         ***&       -0.01&         ***&        0.01&         ***&        0.01&         ***&        0.01&         ***\\
                    &      (1.27)&            &     (48.33)&            &     (-3.42)&            &     (59.75)&            &      (3.85)&            &     (57.25)&            \\
$Volatility$        &       -0.65&            &       -1.52&         ***&        0.45&            &       -1.57&         ***&       -1.35&         ***&       -1.41&         ***\\
                    &     (-1.19)&            &    (-41.02)&            &      (1.02)&            &    (-48.61)&            &     (-2.88)&            &    (-51.38)&            \\
$|Return|$          &       23.82&         ***&       -2.98&         ***&       18.11&         ***&       -2.45&         ***&       22.45&         ***&       -3.09&         ***\\
                    &     (16.86)&            &    (-27.79)&            &     (17.13)&            &    (-29.57)&            &     (19.14)&            &    (-38.43)&            \\
\midrule
$Observations$     &     114,817&            &     114,817&            &     148,416&            &     148,416&            &     221,268&            &     221,268&            \\
$HourFE$ & Yes & & Yes & & Yes & & Yes & & Yes & & Yes &  \\ 
$DateFE$ & Yes & & Yes & & Yes & & Yes & & Yes & & Yes &  \\ 
\bottomrule
\end{tabularx}}{\footnotesize\par}
\par\end{centering}
{\small{}\bigskip{}
}{\small\par}
\end{table}


\pagebreak{}

\begin{singlespace}
\centering{}{\footnotesize{}\begin{tabularx}{\textwidth}{l *{6}{R@{ }l}} \\
\toprule
 \multicolumn{11}{l}{ \textbf{Panel B: ETH/USDC 0.3\%}} \\  \hline 
 & \multicolumn{4}{c}{Arbitrum} & \multicolumn{4}{c}{Polygon} & \multicolumn{4}{c}{BlockScaling}  \\ 
& Int & & Conc & & Int & & Conc & & Int & & Conc &    \\ 
 &  & & 2\% & &  & & 2\% & &  & & 2\% &    \\ 
& (1) & & (2) & & (3) & & (4) & & (5) & & (6) &  \\ 
\midrule
$Arbitrum$          &        0.12&         ***&            &            &            &            &            &            &            &            &            &            \\
                    &     (12.57)&            &            &            &            &            &            &            &            &            &            &            \\
$Polygon$           &            &            &            &            &        0.16&         ***&            &            &            &            &            &            \\
                    &            &            &            &            &     (28.67)&            &            &            &            &            &            &            \\
$BlockScaling$      &            &            &            &            &            &            &            &            &        0.13&         ***&            &            \\
                    &            &            &            &            &            &            &            &            &     (22.40)&            &            &            \\
$Intensity$         &            &            &        0.08&         ***&            &            &        0.22&         ***&            &            &        0.15&         ***\\
                    &            &            &     (29.03)&            &            &            &     (40.97)&            &            &            &     (43.08)&            \\
$Volume$            &       -0.14&         ***&        0.06&         ***&       -0.00&            &        0.04&         ***&        0.01&            &        0.07&         ***\\
                    &     (-3.98)&            &     (42.50)&            &     (-0.19)&            &     (17.74)&            &      (0.53)&            &     (46.28)&            \\
$Volatility$        &        2.97&         ***&       -1.22&         ***&        0.29&            &       -1.08&         ***&        0.32&            &       -1.53&         ***\\
                    &      (4.86)&            &    (-39.04)&            &      (0.70)&            &    (-15.60)&            &      (0.70)&            &    (-35.22)&            \\
$|Return|$          &        9.15&         ***&       -0.96&         ***&       17.87&         ***&       -4.15&         ***&       17.49&         ***&       -2.78&         ***\\
                    &      (8.47)&            &    (-11.58)&            &     (12.54)&            &    (-18.77)&            &     (13.93)&            &    (-20.79)&            \\
\midrule
$ Observations$     &      53,145&            &      53,145&            &      57,472&            &      57,472&            &      81,810&            &      81,810&            \\
$ HourFE$ & Yes & & Yes & & Yes & & Yes & & Yes & & Yes &  \\ 
$ DateFE$ & Yes & & Yes & & Yes & & Yes & & Yes & & Yes &  \\ 
\bottomrule
\end{tabularx}
}{\footnotesize\par}



\end{singlespace}

\pagebreak{}

\begin{table}[H]
\caption{\textbf{\footnotesize{}Repositioning intensity and liquidity concentration: robustness checks.}{\footnotesize{} This table presents robustness checks for our instrumental variable regressions. Panel A reports results for the low-fee pool (ETH/USDC 0.05\%), and Panel B for the high-fee pool (ETH/USDC 0.3\%).  All models report results of the second-stage IV regressions with $BlockScaling$ as an instrument for $Intensity$, similar to Model (6) in Table \ref{Flo:intensity}. $BlockScaling$ takes value of 1 for both Arbitrum and Polygon pools, and zero for the corresponding Ethereum pool. Models (1) and (2) use liquidity concentration within 1\% and within 10\% of the current market price as the dependent variable, respectively. Model (3) uses an alternative measure for repositioning intensity, $IntensFreq$, defined as the ratio of the number of repositioning mints to the total number of mints within a 5-minute interval. Model (4) uses an alternative measure of $Intensity$, based on the 1-minute repositioning mints. In this analysis, we classify repositioning mints as those that are preceded by burns of the same LP within the previous minute. Model (5) uses 10-minute intervals for estimation of repositioning intensity and liquidity concentration, instead of 5-minute intervals. The vector of control variables consists of $Volume$, 
$Volatility$ and $|Return|$. See Appendix E for a detailed description of variable definitions.
All regressions include hour- and day-fixed effects, with standard
errors clustered at the day level. T-statistics of the two-tailed
t-test with the null-hypothesis of a coefficient equaling zero are
reported in parentheses. {*}{*}{*}, {*}{*}, and {*} indicate statistical
significance at the 1\%, 5\%, and 10\% level, respectively.
}{\small{}\bigskip{}
}}

\label{Flo:intensity_rob}
\begin{centering}
{\footnotesize{}\begin{tabularx}{\textwidth}{l *{5}{R@{ }l}} \\
\toprule
 \multicolumn{11}{l}{ \textbf{Panel A: ETH/USDC 0.05\%}} \\  \hline 
 & Conc & & Conc & & IntFreq & & Repostn & &  Aggr&  \\ 
 & $ 1\%$ & & $ 10\%$ & &  & &  1-min & & 10-min &    \\ 
& (1) & & (2) & & (3) & & (4) & & (5) &  \\ 
\midrule
$Intensity$         &        0.06&         ***&        0.34&         ***&            &            &        0.16&         ***&        0.10&         ***\\
                    &     (69.57)&            &     (80.57)&            &            &            &     (71.87)&            &     (63.62)&            \\
$IntensFreq$        &            &            &            &            &        0.24&         ***&            &            &            &            \\
                    &            &            &            &            &     (31.90)&            &            &            &            &            \\
$Volume$            &        0.00&         ***&        0.02&         ***&        0.01&         ***&        0.01&         ***&        0.01&         ***\\
                    &     (57.69)&            &     (47.80)&            &     (33.37)&            &     (69.84)&            &     (53.59)&            \\
$Volatility$        &       -0.84&         ***&       -2.82&         ***&       -1.35&         ***&       -1.44&         ***&       -1.39&         ***\\
                    &    (-52.21)&            &    (-36.63)&            &    (-26.90)&            &    (-51.67)&            &    (-45.54)&            \\
$|Return|$          &       -1.85&         ***&       -8.68&         ***&       -6.39&         ***&       -3.71&         ***&       -2.01&         ***\\
                    &    (-39.32)&            &    (-37.14)&            &    (-27.97)&            &    (-38.80)&            &    (-32.58)&            \\
\midrule
$ Observations$     &     221,268&            &     221,268&            &     214,875&            &     221,267&            &     149,871&            \\
$ HourFE$ & Yes & & Yes & & Yes & & Yes & & Yes &  \\ 
$ DateFE$ & Yes & & Yes & & Yes & & Yes & & Yes &  \\ 
\bottomrule
\end{tabularx}
}{\footnotesize\par}
\par\end{centering}
{\small{}\bigskip{}
}{\small\par}
\end{table}

\pagebreak{}

\begin{singlespace}
\centering{}{\footnotesize{}\begin{tabularx}{\textwidth}{l *{5}{R@{ }l}} \\
\toprule
 \multicolumn{11}{l}{ \textbf{Panel B: ETH/USDC 0.3\%}} \\  \hline 
 & Conc & & Conc & & IntFreq & & Repostn & &  Aggr&  \\ 
 & $ 1\%$ & & $ 10\%$ & &  & &  1-min & & 10-min &    \\ 
& (1) & & (2) & & (3) & & (4) & & (5) &  \\ 
\midrule
$Intensity$         &        0.08&         ***&        0.42&         ***&            &            &        0.21&         ***&        0.15&         ***\\
                    &     (42.66)&            &     (42.56)&            &            &            &     (44.66)&            &     (40.52)&            \\
$IntensFreq$        &            &            &            &            &        0.17&         ***&            &            &            &            \\
                    &            &            &            &            &     (37.48)&            &            &            &            &            \\
$Volume$            &        0.04&         ***&        0.21&         ***&        0.07&         ***&        0.07&         ***&        0.07&         ***\\
                    &     (43.93)&            &     (49.60)&            &     (40.42)&            &     (54.16)&            &     (44.07)&            \\
$Volatility$        &       -0.80&         ***&       -4.46&         ***&       -1.52&         ***&       -1.49&         ***&       -1.56&         ***\\
                    &    (-33.59)&            &    (-36.66)&            &    (-31.26)&            &    (-35.58)&            &    (-32.65)&            \\
$|Return|$          &       -1.51&         ***&       -7.55&         ***&       -3.33&         ***&       -3.59&         ***&       -2.11&         ***\\
                    &    (-20.74)&            &    (-20.20)&            &    (-20.62)&            &    (-21.11)&            &    (-20.42)&            \\
\midrule
$ Observations$     &      81,810&            &      81,810&            &      81,133&            &      81,810&            &      67,874&            \\
$ HourFE$ & Yes & & Yes & & Yes & & Yes & & Yes &  \\ 
$ DateFE$ & Yes & & Yes & & Yes & & Yes & & Yes &  \\ 
\bottomrule
\end{tabularx}
}{\footnotesize\par}


\end{singlespace}

\pagebreak{}

\begin{table}[H]
\caption{\textbf{\footnotesize{}Repositioning precision}{\footnotesize{}
This table reports average levels of position gap (\%), position length
(\%) and position precision (\%) across Ethereum, Arbitrum and Polygon for two liquidity pools: ETH/USDC 0.05\% and ETH/USDC 0.3\%. We define
repositioning gap, $Gap$, as the average gap between the mid price
of repositioning mints, $p_{mid}$, and $p_{mkt}$,
computed as $\left|\frac{p_{mid}}{p_{mkt}}-1 \right|$. $p_{mid}$ is computed as $\sqrt{p_{lower}\cdot p_{upper}}$. We define a repositioning mint as a mint, preceded by a burn in the same pool by the same liquidity provider within previous 5 minutes.
Repositioning length, $Length$, is the average range length of repositioning mints, computed as $\frac{\lvert p_{upper}-p_{lower}\rvert}{p_{mid}}$. Repositioning precision, $Precision$, is the 
average precision of repositioning mints, computed and scaled as
$1-1.0001^{\frac{-1}{Gap\cdot Length}}$. Appendix E provides a detailed description of all variable
definitions. Column (1) reports average levels of repositioning gap. Columns (2) and (3) report
corresponding statistics for repositioning length and repositioning precision,
respectively. We also report the difference between averages on Arbitrum and Ethereum ($\Delta Arb-Eth$) as well as Polygon and Ethereum ($\Delta Pol-Eth$). T-statistics of the two-tailed
t-test with the null-hypothesis of difference equaling zero are
reported in parentheses. {*}{*}{*}, {*}{*}, and {*} indicate statistical significance
at the 1\%, 5\%, and 10\% level, respectively. }{\small{}\bigskip{}
}}

\label{Flo:unistats_gap_length}
\centering{}{\small{}\begin{tabular}{ccccccc}
\hline
\textbf{Pool}   & \multicolumn{2}{c}{\textbf{Gap (\%)}} & \multicolumn{2}{c}{\textbf{Length (\%)}} & \multicolumn{2}{c}{\textbf{Precision (\%)}} \\
\hline
ETH/USDC 0.05\% &                     &                 &                      &                   &                       &                     \\
Ethereum        & 1.86                &                 & 18.07                 &                   & 50.77                 &                     \\
Arbitrum        & 0.85                &                 & 10.69                &                   & 70.89                 &                     \\
$\Delta Arb-Eth$      & -1.01               & ***             & -7.38                 & ***               & 20.12                 & ***                 \\
         & (-49.01)              &                 & (-63.20)               &                   & (53.98)                 &                     \\
Polygon         & 1.09                &                 & 12.91                &                   & 68.60                 &                     \\
$\Delta Pol-Eth$     & -0.77               & ***             & -5.16                 & ***               & 17.83                 & ***                 \\
         & (-32.70)              &                 & (-46.43)                &                   & (45.23)                 &                     \\
\hline
ETH/USDC 0.3\%  &                     &                 &                      &                   &                       &                     \\
Ethereum        & 3.41               &                 & 29.28                &                   & 20.02                 &                     \\
Arbitrum        & 2.09                &                 & 15.04                &                   & 36.63                 &                     \\
$\Delta Arb-Eth$     & -1.32               & ***             & -14.24               & ***               & 16.61                 & ***                 \\
      & (-17.14)              &                 & (-21.76)                &                   & (20.50)                 &                     \\
Polygon         & 2.44                &                 & 15.11                &                   & 42.12                 &                     \\
$\Delta Pol-Eth$     & -0.97               & ***             & -14.17               & ***               & 22.10                 & ***                 \\
        & (-12.94)               &                 & (-27.87)                &                   & (28.70)                 & 
\\\hline
\end{tabular}}{\small\par}
\end{table}

\pagebreak{}

\begin{table}[H]
\caption{\textbf{\footnotesize{}Repositioning
precision and liquidity concentration: IV regressions.}{\footnotesize{}This table presents
results of instrumental variable regressions that test causal effect
of repositioning precision of LPs on liquidity concentration around the market price. We use three measures of repositioning precision: $Gap$, $Length$ and $Precision$. See Appendix E for
a detailed description of variable definitions. Panel A reports results for the low-fee pool (ETH/USDC 0.05\%), and Panel B for the high-fee pool (ETH/USDC 0.3\%).  Model (1) reports the results of the first-stage IV regression with $Gap$ as the dependent variable. Model (2) reports the results of the second-stage regression with liquidity concentration within 2\% as the
dependent variable. $Blockscaling$ is used as an instrument for the endogenous repositioning
precision (i.e. $Gap$). Models (3) and (4) present corresponding results for $Length$ as a measure of repositioning precision. Models (5) and (6) present results for $Precision$. 
The vector of control variables
consists of $Volume$, $Volatility$ and $|Return|$.  All regressions include
hour- and day-fixed effects, with standard errors clustered at the
day level. T-statistics of the two-tailed t-test with the null-hypothesis
of a coefficient equaling zero are reported in parentheses. {*}{*}{*},
{*}{*}, and {*} indicate statistical significance at the 1\%, 5\%,
and 10\% level, respectively. }{\small{}\bigskip{}
}}

\label{Flo:precision}
\begin{centering}
{\footnotesize{}\begin{tabularx}{\textwidth}{l *{6}{R@{ }l}} \\
\toprule
 \multicolumn{13}{l}{ \textbf{Panel A: ETH/USDC 0.05\%}} \\  \hline 
 & Gap & & Conc & & Length & & Conc & & Precision & & Conc &    \\ 
 &  & & $ 2\%$ & & & & $ 2\%$ & & & & $ 2\%$ &    \\ 
& (1) & & (2) & & (3) & & (4) & & (5) & & (6) &  \\ 
\midrule
$BlockScaling$      &       -0.01&         ***&            &            &       -0.05&         ***&            &            &        0.12&         ***&            &            \\
                    &     (-9.26)&            &            &            &    (-13.83)&            &            &            &     (11.85)&            &            &            \\
$Gap$               &            &            &       -4.15&         ***&            &            &            &            &            &            &            &            \\
                    &            &            &    (-15.82)&            &            &            &            &            &            &            &            &            \\
$Length$            &            &            &            &            &            &            &       -0.47&         ***&            &            &            &            \\
                    &            &            &            &            &            &            &    (-27.70)&            &            &            &            &            \\
$Precision$         &            &            &            &            &            &            &            &            &            &            &        0.18&         ***\\
                    &            &            &            &            &            &            &            &            &            &            &     (23.28)&            \\
$Volume$            &       -0.00&         ***&        0.00&         ***&       -0.00&          **&        0.01&         ***&        0.02&         ***&        0.00&         ***\\
                    &     (-5.45)&            &      (7.53)&            &     (-2.56)&            &     (30.86)&            &      (3.76)&            &     (16.52)&            \\
$Volatility$        &        0.26&         ***&       -0.46&         ***&        0.87&         ***&       -1.13&         ***&       -4.65&         ***&       -0.69&         ***\\
                    &      (6.02)&            &     (-4.15)&            &      (4.16)&            &    (-21.77)&            &     (-6.21)&            &    (-10.96)&            \\
$|Return|$          &       -0.41&         ***&       -2.28&         ***&       -4.60&         ***&       -2.84&         ***&        6.71&         ***&       -1.77&         ***\\
                    &     (-8.99)&            &    (-13.45)&            &    (-13.27)&            &    (-21.15)&            &      (7.66)&            &    (-16.04)&            \\
\midrule
$ Observations$     &     102,625&            &     102,625&            &     102,625&            &     102,625&            &     102,625&            &     102,625&            \\
$ HourFE$ & Yes & & Yes & & Yes & & Yes & & Yes & & Yes &  \\ 
$ DateFE$ & Yes & & Yes & & Yes & & Yes & & Yes & & Yes &  \\ 
\bottomrule
\end{tabularx}
}{\footnotesize\par}
\par\end{centering}
{\small{}\bigskip{}
}{\small\par}
\end{table}

\pagebreak{}

\begin{singlespace}
\centering{}
{\footnotesize{}\begin{tabularx}{\textwidth}{l *{6}{R@{ }l}} \\
\toprule
 \multicolumn{13}{l}{ \textbf{Panel B: ETH/USDC 0.3\%}} \\  \hline 
 & Gap & & Conc & & Length & & Conc & & Precision & & Conc &    \\ 
 &  & & $ 2\%$ & & & & $ 2\%$ & & & & $ 2\%$ &    \\ 
& (1) & & (2) & & (3) & & (4) & & (5) & & (6) &  \\ 
\midrule
$BlockScaling$      &       -0.01&         ***&            &            &       -0.11&         ***&            &            &        0.12&         ***&            &            \\
                    &     (-5.01)&            &            &            &    (-13.11)&            &            &            &      (8.84)&            &            &            \\
$Gap$               &            &            &       -3.05&         ***&            &            &            &            &            &            &            &            \\
                    &            &            &     (-7.05)&            &            &            &            &            &            &            &            &            \\
$Length$            &            &            &            &            &            &            &       -0.16&         ***&            &            &            &            \\
                    &            &            &            &            &            &            &    (-17.55)&            &            &            &            &            \\
$Precision$         &            &            &            &            &            &            &            &            &            &            &        0.15&         ***\\
                    &            &            &            &            &            &            &            &            &            &            &     (15.52)&            \\
$Volume$            &       -0.01&         ***&        0.04&         ***&       -0.04&          **&        0.06&         ***&        0.15&         ***&        0.04&         ***\\
                    &     (-3.31)&            &      (5.73)&            &     (-2.10)&            &     (23.71)&            &      (5.42)&            &     (11.82)&            \\
$Volatility$        &        0.27&         ***&       -0.74&         ***&        1.36&         ***&       -1.34&         ***&       -5.42&         ***&       -0.72&         ***\\
                    &      (2.84)&            &     (-3.09)&            &      (3.13)&            &    (-17.31)&            &     (-6.55)&            &     (-6.48)&            \\
$|Return|$          &       -0.25&          **&       -0.80&         ***&       -3.53&         ***&       -0.62&         ***&        7.60&         ***&       -1.21&         ***\\
                    &     (-2.58)&            &     (-2.74)&            &     (-6.67)&            &     (-5.88)&            &      (7.43)&            &     (-7.36)&            \\
\midrule
$ Observations$     &      15,173&            &      15,174&            &      15,173&            &      15,174&            &      15,173&            &      15,174&            \\
$ HourFE$ & Yes & & Yes & & Yes & & Yes & & Yes & & Yes &  \\ 
$ DateFE$ & Yes & & Yes & & Yes & & Yes & & Yes & & Yes &  \\ 
\bottomrule
\end{tabularx}
}{\footnotesize\par}

\end{singlespace}

\pagebreak{}

\begin{table}[H]
\caption{\textbf{\footnotesize{}Slippage: summary statistics and multivariate analysis.}{\footnotesize{} Panel A of this table
presents summary statistics for slippage, $Slippage$ (in bp), on Uniswap v3 across Ethereum, Arbitrum and Polygon for two liquidity pools: ETH/USDC 0.05\% and ETH/USDC 0.3\%. We compute $Slippage$ for hypothetical trades of sizes [\$100, \$500, \$1K, \$5K, \$10K, \$50K, \$100K] for every minute in our sample
period (January 1, 2022 - June 30, 2023). Panels B and C present the results of OLS regressions with slippage, $Slippage$
(in bp), as the dependent variable. Model (1) reports the results for the total sample, with $BlockScaling$
equal to 1 for hypothetical trades on Arbitrum and Polygon, and zero otherwise.
Trades on Ethereum represent the benchmark sample. Models (2) and (3)
report results, conditioning on the trade size. Specifically, we define
$Large$ equal to 1 if a trade exceeds \$1K (Model 2) or \$5K (Model
3), and zero otherwise. Models (4) and (5) report results for the benchmark Model (2), separately for Arbitrum and Polygon. The vector of
control variables consists of $Size$, $Buy$, $Volume$,
$Volatility$, $|Return|$. See Appendix E for a detailed description of variable definitions.
All regressions include hour- and date-fixed effects, with standard
errors clustered at the day level. T-statistics of the two-tailed
t-test with the null-hypothesis of a coefficient equaling zero are
reported in parentheses. {*}{*}{*}, {*}{*}, and {*} indicate statistical
significance at the 1\%, 5\%, and 10\% level, respectively.}{\small{}\vspace{1cm}
}}

\label{Flo:slippage}

{\small{}\bigskip{}
}{\small\par}
\begin{centering}
{\footnotesize{}\begin{tabular}{cccccc}
\toprule
\multicolumn{6}{l}{ \textbf{Panel A: Summary statistics}} \\  

\hline
                    & \textbf{Mean} & \textbf{SD} & \textbf{25\%} & \textbf{50\%} & \textbf{75\%} \\ \hline
Slippage (bp)       &               &             &               &               &               \\
ETH/USDC 0.05\%     &               &             &               &               &               \\
Ethereum            & 0.30          & 0.61        & 0.01          & 0.05          & 0.37          \\
Arbitrum            & 8.80          & 27.81       & 0.07          & 0.60          & 4.25          \\
Polygon             & 4.25          & 8.05        & 0.09          & 0.73          & 5.92          \\
ETH/USDC 0.3\%      &               &             &               &               &               \\
Ethereum            & 0.55          & 0.90        & 0.01          & 0.11          & 0.85          \\
Arbitrum            & 23.25         & 51.95       & 0.35          & 2.62          & 20.63         \\
Polygon             & 34.24         & 62.18       & 0.72          & 5.40          & 38.23         \\ \hline

\end{tabular}}{\footnotesize\par}
\par\end{centering}
{\small{}\bigskip{}
}{\small\par}
\end{table}

\pagebreak{}

\begin{singlespace}
{\footnotesize{}\begin{tabularx}{\textwidth}{l *{5}{R@{ }l}} \\
\toprule
 \multicolumn{11}{l}{ \textbf{Panel B: ETH/USDC 0.05\%}} \\  \hline 
 & \multicolumn{6}{c}{BlockScaling} & \multicolumn{2}{c}{Arbitrum} & \multicolumn{2}{c}{Polygon}  \\ 
 & Total & & Large $>\$1K$ & & Large $>\$5K$ & & Large $>\$1K$ & & Large $>\$1K$ &  \\ 
& (1) & & (2) & & (3) & & (4) & & (5) &  \\ 
\midrule
$BlockScaling$      &        1.41&         ***&       -4.60&         ***&       -4.32&         ***&            &            &            &            \\
                    &      (3.18)&            &     (-9.10)&            &     (-8.65)&            &            &            &            &            \\
$Arbitrum$          &            &            &            &            &            &            &       -6.41&         ***&            &            \\
                    &            &            &            &            &            &            &     (-8.09)&            &            &            \\
$Polygon$           &            &            &            &            &            &            &            &            &       -1.62&         ***\\
                    &            &            &            &            &            &            &            &            &     (-8.73)&            \\
$BlockSc\cdot Large$&            &            &       10.53&         ***&       13.37&         ***&       14.37     & ***   &       6.67 &  		***\\
                    &            &            &     (19.79)&            &     (19.79)&            &      (15.12)&            &      (42.59)    &            \\
$Large$             &            &            &        0.51&         ***&        0.65&         ***&        0.51&         ***&        0.51&         ***\\
                    &            &            &     (37.00)&            &     (37.00)&            &     (37.00)&            &     (37.00)&            \\
$Size$              &        0.19&         ***&            &            &            &            &            &            &            &            \\
                    &     (20.63)&            &            &            &            &            &            &            &            &            \\
$Buy$               &       -0.00&            &       -0.00&            &       -0.00&            &        0.01&            &       -0.01&         ***\\
                    &     (-0.22)&            &     (-0.22)&            &     (-0.22)&            &      (0.27)&            &     (-3.42)&            \\
$Volume$            &       -0.01&         ***&       -0.01&         ***&       -0.01&         ***&       -0.01&         ***&       -0.00&         ***\\
                    &     (-9.06)&            &     (-9.06)&            &     (-9.06)&            &     (-8.02)&            &     (-9.61)&            \\
$Volatility$        &        0.86&         ***&        0.86&         ***&        0.86&         ***&        1.38&         ***&        0.41&         ***\\
                    &      (8.96)&            &      (8.96)&            &      (8.96)&            &      (7.68)&            &      (9.33)&            \\
$|Return|$          &        0.48&         ***&        0.48&         ***&        0.48&         ***&        0.39&         ***&        0.33&         ***\\
                    &      (9.26)&            &      (9.26)&            &      (9.26)&            &      (5.83)&            &     (10.60)&            \\
\midrule
$ R^2$              &        0.28&            &        0.20&            &        0.24&            &        0.24&            &        0.36&            \\
$ Observations$     &   6,603,702&            &   6,603,702&            &   6,603,702&            &   4,402,468&            &   4,402,468&            \\
$ HourFE$ & Yes & & Yes & & Yes & & Yes & & Yes &  \\ 
$ DateFE$ & Yes & & Yes & & Yes & & Yes & & Yes &  \\ 
\bottomrule
\end{tabularx}
}{\footnotesize\par}

{\footnotesize{}\begin{tabularx}{\textwidth}{l *{5}{R@{ }l}} \\
\toprule
 \multicolumn{11}{l}{ \textbf{Panel C: ETH/USDC 0.3\%}} \\  \hline 
 & \multicolumn{6}{c}{BlockScaling} & \multicolumn{2}{c}{Arbitrum} & \multicolumn{2}{c}{Polygon}  \\ 
 & Total & & Large $>\$1K$ & & Large $>\$5K$ & & Large $>\$1K$ & & Large $>\$1K$ &  \\ 
& (1) & & (2) & & (3) & & (4) & & (5) &  \\ 
\midrule
$BlockScaling$      &       21.61&         ***&       -5.78&         ***&       -4.48&         ***&            &            &            &            \\
                    &     (31.41)&            &     (-8.00)&            &     (-6.29)&            &            &            &            &            \\
$Arbitrum$          &            &            &            &            &            &            &       -9.74&         ***&            &            \\
                    &            &            &            &            &            &            &    (-12.45)&            &            &            \\
$Polygon$           &            &            &            &            &            &            &            &            &       -1.07&            \\
                    &            &            &            &            &            &            &            &            &     (-1.20)&            \\
$BlockSc\cdot Large$&            &            &       47.94&         ***&       60.89&         ***&      38.50 &    ***     &        57.38 & ***           \\
                    &            &            &     (50.21)&            &     (50.17)&            &      (29.78) &            &       (43.93)&            \\
$Large$             &            &            &        0.94&         ***&        1.19&         ***&        0.94&         ***&        0.94&         ***\\
                    &            &            &     (68.44)&            &     (68.45)&            &     (68.44)&            &     (68.44)&            \\
$Size$              &        0.81&         ***&            &            &            &            &            &            &            &            \\
                    &     (51.08)&            &            &            &            &            &            &            &            &            \\
$Buy$               &        0.18&           *&        0.18&           *&        0.18&           *&        0.29&         ***&       -0.03&            \\
                    &      (1.68)&            &      (1.68)&            &      (1.68)&            &      (3.22)&            &     (-0.25)&            \\
$Volume$            &       -0.12&         ***&       -0.12&         ***&       -0.12&         ***&       -0.19&         ***&       -0.03&          **\\
                    &     (-9.32)&            &     (-9.32)&            &     (-9.32)&            &    (-12.40)&            &     (-2.09)&            \\
$Volatility$        &        2.11&         ***&        2.11&         ***&        2.11&         ***&        3.20&         ***&        1.27&         ***\\
                    &     (10.90)&            &     (10.90)&            &     (10.90)&            &     (11.57)&            &      (4.50)&            \\
$|Return|$          &        1.00&         ***&        1.00&         ***&        1.00&         ***&        0.51&         ***&        0.98&         ***\\
                    &      (6.63)&            &      (6.63)&            &      (6.63)&            &      (2.89)&            &      (8.19)&            \\
\midrule
$ R^2$              &        0.49&            &        0.30&            &        0.41&            &        0.34&            &        0.36&            \\
$ Observations$     &   6,603,702&            &   6,603,702&            &   6,603,702&            &   4,402,468&            &   4,402,468&            \\
$ HourFE$ & Yes & & Yes & & Yes & & Yes & & Yes &  \\ 
$ DateFE$ & Yes & & Yes & & Yes & & Yes & & Yes &  \\ 
\bottomrule
\end{tabularx}
}{\footnotesize\par}

\end{singlespace}

\pagebreak{}

\begin{table}[H]
\caption{\textbf{\footnotesize{}Repositioning
and liquidity concentration: Arbitrum airdrop.}{\footnotesize{} This table presents
results of instrumental variable regressions that test causal effect
of repositioning intensity and precision, separately before and after the airdrop of ARB token on March 23, 2023. Panel A reports results for the low-fee pool (ETH/USDC 0.05\%), and Panel B for the high-fee pool (ETH/USDC 0.3\%).  Models (1)-(3) report results for the pre-airdrop sample, and Models (4)-(6) for the post-airdrop sample. Models (1) and (4) report the results for the first-stage regression,
with repositioning intensity of LPs, $Intensity$, as the dependent
variable and $Arbitrum$ as the main explanatory variable, which is used as an instrument for the endogenous
$Intensity$. $Arbitrum$ takes value of 1 for Arbitrum pools, and zero for Ethereum pools. Models (2) and (5)
report results for the second-stage regression, with liquidity concentration within 2\%  of the market price, $Conc$, as the dependent variable.  We estimate liquidity concentration at the end
of every 5-minute interval and the repositioning intensity over the
previous 5-minute interval. The set of instruments consists
of all explanatory variables, except that $Intensity$ is replaced
with $Arbitrum$. Models (3) and (6) report results for the second-stage regressions, with $Arbitrum$ used as an instrument for the endogenous
$Precision$. The vector of control variables consists of $Volume$, 
$Volatility$ and $|Return|$. See Appendix E for a detailed description of variable definitions.
All regressions include hour- and day-fixed effects, with standard
errors clustered at the day level. T-statistics of the two-tailed
t-test with the null-hypothesis of a coefficient equaling zero are
reported in parentheses. {*}{*}{*}, {*}{*}, and {*} indicate statistical
significance at the 1\%, 5\%, and 10\% level, respectively.}{\small{}\bigskip{}
}}

\label{Flo:intensity_airdrop}
\begin{centering}
{\footnotesize{}\begin{tabularx}{\textwidth}{l *{6}{R@{ }l}} \\
\toprule
 \multicolumn{13}{l}{ \textbf{Panel A: ETH/USDC 0.05\%}} \\  \hline 
 & \multicolumn{6}{c}{Before Airdrop} & \multicolumn{6}{c}{After Airdrop} \\  
 & Int & & Conc & & Conc & & Int & & Conc & & Conc &    \\ 
 &  & & $ 2\%$ & & $ 2\%$ & & &  & $ 2\%$ & & $ 2\%$ &    \\ 
& (1) & & (2) & & (3) & & (4) & & (5) & & (6) &  \\ 
\midrule
$Arbitrum$          &        0.22&         ***&            &            &            &            &        0.35&         ***&            &            &            &            \\
                    &     (31.23)&            &            &            &            &            &     (17.93)&            &            &            &            &            \\
$Intensity$         &            &            &        0.04&         ***&            &            &            &            &        0.29&         ***&            &            \\
                    &            &            &     (29.71)&            &            &            &            &            &     (35.37)&            &            &            \\
$Precision$         &            &            &            &            &        0.05&         ***&            &            &            &            &        0.68&         ***\\
                    &            &            &            &            &     (13.90)&            &            &            &            &            &      (7.11)&            \\
$Volume$            &        0.00&            &        0.01&         ***&        0.01&         ***&       -0.03&          **&        0.01&         ***&       -0.00&            \\
                    &      (1.17)&            &     (53.87)&            &     (28.77)&            &     (-2.59)&            &      (6.59)&            &     (-0.53)&            \\
$Volatility$        &       -0.90&            &       -1.30&         ***&       -1.22&         ***&        5.74&         ***&       -3.34&         ***&        3.07&          **\\
                    &     (-1.56)&            &    (-47.41)&            &    (-27.48)&            &      (3.28)&            &    (-12.26)&            &      (2.05)&            \\
$|Return|$          &       24.24&         ***&       -1.57&         ***&       -0.64&         ***&       20.11&         ***&       -6.55&         ***&       -6.67&         ***\\
                    &     (15.58)&            &    (-21.40)&            &     (-9.10)&            &      (6.88)&            &    (-10.33)&            &     (-4.74)&            \\
\midrule
$ Observations$     &      87,698&            &      87,698&            &      33,849&            &      27,119&            &      27,119&            &      14,138&            \\
$ HourFE$ & Yes & & Yes & & Yes & & Yes & & Yes & & Yes &  \\ 
$ DateFE$ & Yes & & Yes & & Yes & & Yes & & Yes & & Yes &  \\ 
\bottomrule
\end{tabularx}
}{\footnotesize\par}
\par\end{centering}
{\small{}\bigskip{}
}{\small\par}
\end{table}

\pagebreak{}

\begin{singlespace}
\centering{}{\footnotesize{}\begin{tabularx}{\textwidth}{l *{6}{R@{ }l}} \\
\toprule
 \multicolumn{13}{l}{ \textbf{Panel B: ETH/USDC 0.3\%}} \\  \hline 
 & \multicolumn{6}{c}{Before Airdrop} & \multicolumn{6}{c}{After Airdrop} \\  
 & Int & & Conc & & Conc & & Int & & Conc & & Conc &    \\ 
 &  & & $ 2\%$ & & $ 2\%$ & & &  & $ 2\%$ & & $ 2\%$ &    \\ 
& (1) & & (2) & & (3) & & (4) & & (5) & & (6) &  \\ 
\midrule
$Arbitrum$          &        0.07&         ***&            &            &            &            &        0.12&         ***&            &            &            &            \\
                    &      (2.74)&            &            &            &            &            &      (12.25)&            &            &            &            &            \\
$Intensity$         &            &            &        0.07&         ***&            &            &            &            &        0.65&         ***&            &            \\
                    &            &            &     (27.16)&            &            &            &            &            &      (3.69)&            &            &            \\
$Precision$         &            &            &            &            &        0.06&         ***&            &            &            &            &        0.31&          **\\
                    &            &            &            &            &      (7.04)&            &            &            &            &            &      (2.37)&            \\
$Volume$            &       -0.15&         ***&        0.06&         ***&        0.04&         ***&        0.20&            &       -0.08&            &       -0.04&            \\
                    &     (-4.08)&            &     (44.17)&            &     (11.10)&            &      (1.57)&            &     (-0.92)&            &     (-0.54)&            \\
$Volatility$        &        3.09&         ***&       -1.15&         ***&       -0.63&         ***&       -1.03&            &       -0.59&            &        0.00&            \\
                    &      (4.93)&            &    (-41.39)&            &     (-7.54)&            &     (-0.28)&            &     (-0.31)&            &         (.)&            \\
$|Return|$          &        8.90&         ***&       -0.81&         ***&       -0.34&         ***&       15.63&         ***&      -10.37&          **&        0.79&            \\
                    &      (8.10)&            &    (-11.25)&            &     (-3.71)&            &      (3.81)&            &     (-2.57)&            &      (0.50)&            \\
\midrule
$ Observations$     &      49,943&            &      49,943&            &       6,604&            &       3,202&            &       3,202&            &         549&            \\
$ HourFE$ & Yes & & Yes & & Yes & & Yes & & Yes & & Yes &  \\ 
$ DateFE$ & Yes & & Yes & & Yes & & Yes & & Yes & & Yes &  \\ 
\bottomrule
\end{tabularx}
}{\footnotesize\par}


\end{singlespace}

\pagebreak{}

\section*{Appendix A}

\section*{Estimating gas fees}

Any transaction on a blockchain requires the payment of gas fees, which are used to compensate validators for their work in verifying transactions and securing the network.

Gas fees are computed with the quantity of gas units required, which depends
on the transaction. For example, a trade on Uniswap v3 (Ethereum) requires on
average 120K gas units.

The price of one gas unit, the gas price, varies over our sample period,
as shown in Panel A of Figure A1. Panel B shows the gas
fees to pay for a hypothetical trade on Uniswap v3 (Ethereum) that requires
120K gas units. These fees vary from as little as \$1.30 in October 2022 to \$160
during the period of Terra collapse in early May 2022. On average,
a trade that requires 120K gas units on Uniswap v3 (Ethereum) costs \$14 in gas fees in 2022. 

\pagebreak{}

\setcounter{figure}{0} 
\renewcommand{\thefigure}{A\arabic{figure}}

\begin{figure}[H]
\caption{\textbf{\footnotesize{}Gas price and gas fees on Ethereum.}{\footnotesize{} Panel A of this figure shows historical gas price on Ethereum over 2022, downloaded from Etherscan. Panel B shows gas fees for a trade on Uniswap v3 (Ethereum), assuming that it requires on average 120K gas units.}}

{\small{}\vspace{0.5cm}
}{\small\par}

\label{Flo:gas}
\begin{centering}
\textbf{\small{}Panel A}{\small\par}
\par\end{centering}
{\small{}\vspace{0.5cm}
}{\small\par}
\begin{singlespace}
\begin{centering}
\includegraphics[scale=0.13]{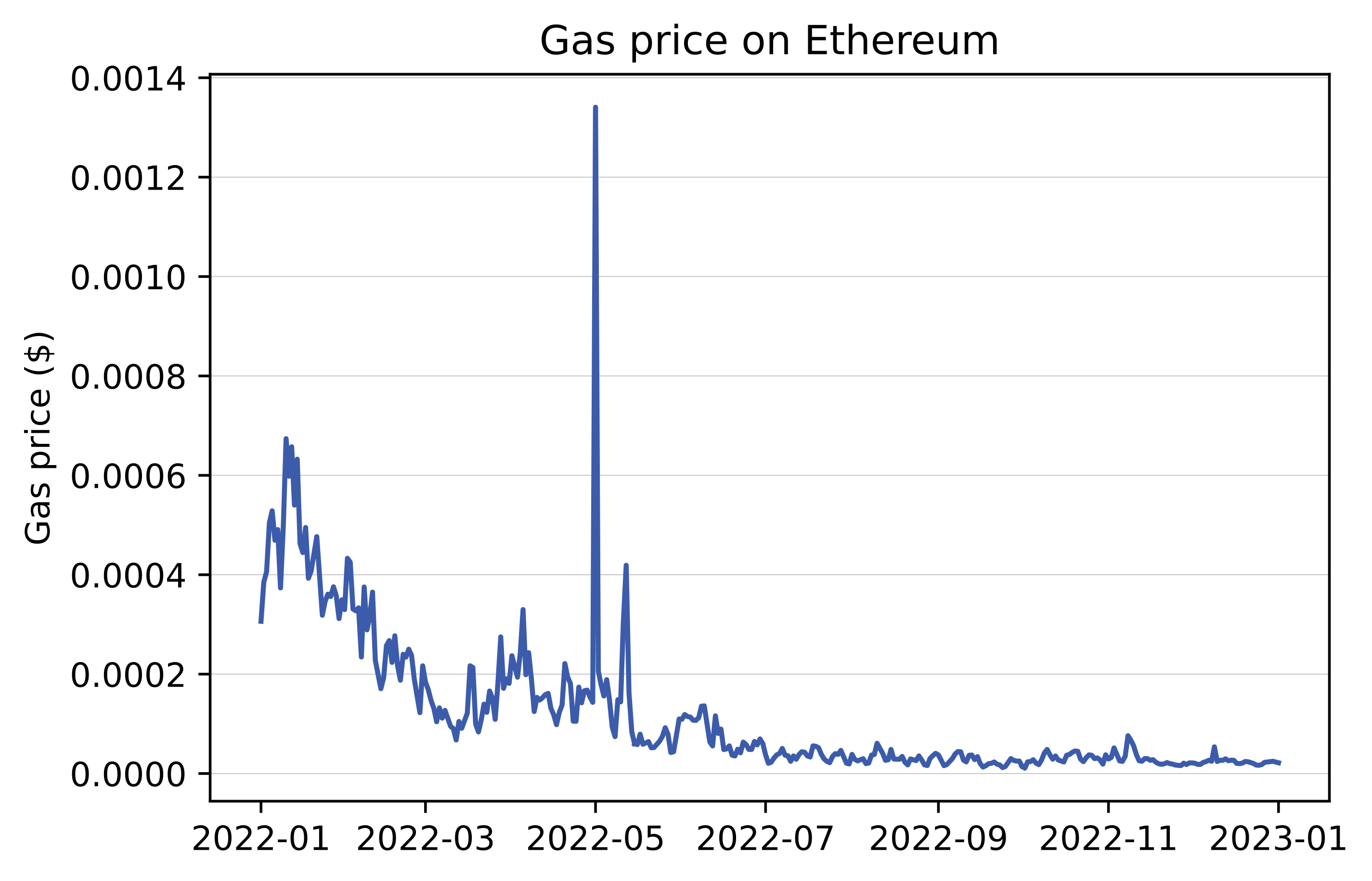}
\par\end{centering}
\end{singlespace}
{\small{}\vspace{0.5cm}
}{\small\par}
\begin{centering}
\textbf{\small{}Panel B}{\small{}\vspace{0.5cm}
}{\small\par}
\par\end{centering}
\centering{}\includegraphics[scale=0.13]{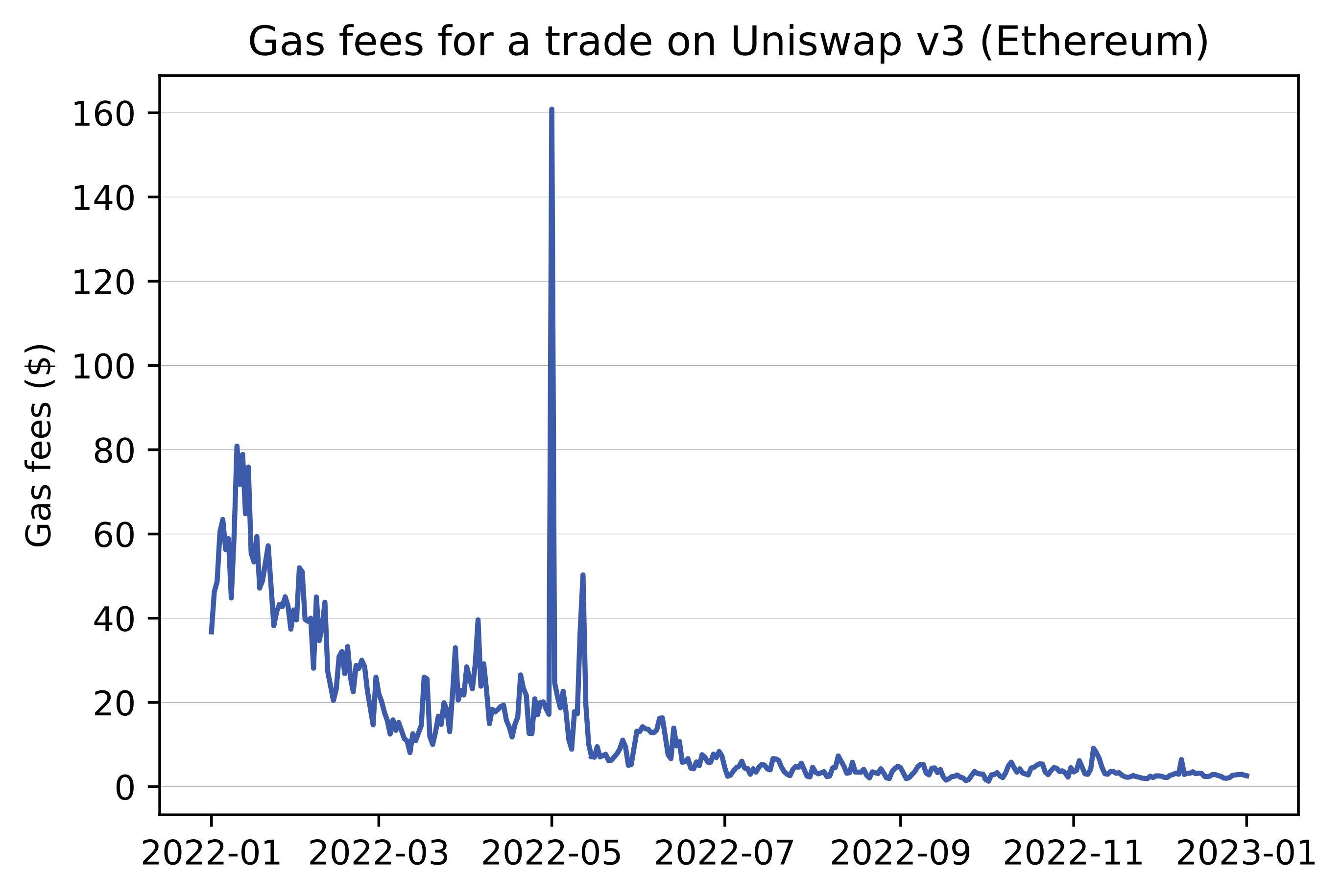}
\end{figure}

\pagebreak{}

\begin{onehalfspace}

\section*{Appendix B \label{sec:Appendix-Uniswapv2}}
\end{onehalfspace}

\section*{Trading mechanics on Uniswap v2 and Uniswap v3}

\textbf{Uniswap v2.} Uniswap v2 is a constant product market maker (CPMM), which is a widely adopted subclass of automated market makers (AMMs). The price curve, or so-called
bonding curve, is defined as:

\begin{equation}
x \cdot y=k
\end{equation}

where $x$ is the current quantity of token X, the base token (e.g., BTC for BTC/USDT pair), $y$ is the current quantity of
token Y, the quote token (e.g., USDT for BTC/USDT pair) and $k$ is
the pool's constant, showing the total liquidity available in the
pool. In absence of mints and burns, pool's constant $k$ has to preserve
its value. For example, a trade (or swap) that buys BTC with USDT will remove $\triangle x$ of BTC from the pool and add $\triangle y$
of USDT to the pool. Denote the new amount of BTC in the pool after the trade as $x'$,
i.e. $x'=x+\triangle x$, and the new amount of USDT as $y'$, i.e.
$y'=y+\triangle y,$ with $\triangle x<0$ and $\triangle y>0$. CPMM requires $k$ to be the same before and after the trade:

\begin{equation}
x\cdot y=(x+\triangle x)(y+\triangle y)=k
\end{equation}

\begin{figure}[H]
\caption{\textbf{\footnotesize{}Price curve for Uniswap v2: constant product
market maker (CPMM).}{\footnotesize{} This figure illustrates a swap
on the BTC/USDT pool on Uniswap v2. Pool reserves of token X, i.e.
BTC, are on the x-axis. Pool reserves of token Y, i.e. USDT, are on
y-axis. The price curve, $x\cdot y=k$, ensures the constant product
between the reserves of the pool's two assets (in absence of liquidity
deposits/withdrawals). A buy of BTC with USDT
reduces the reserves of BTC from $x$ to $x'$ and increases the reserves
of USDT from $y$ to $y'.$ The pool price after the buy moves
upwards from $p_{c}$ to $p_{b}.$ }}

{\small{}\vspace{1cm}
}{\small\par}

\label{Flo:uniswap_v2}
\begin{singlespace}
\centering{}\includegraphics[scale=0.75]{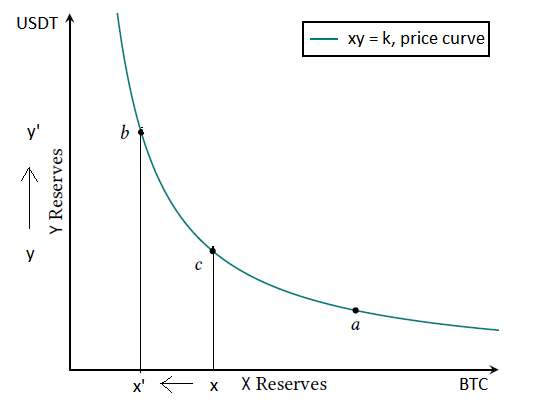}
\end{singlespace}
\end{figure}

Therefore, the product between the reserves of the pool's two tokens
is constant, provided there is no changes to total liquidity in the
pool. Figure \ref{Flo:uniswap_v2} illustrates this trade as a transition
along the price curve from state $c$ to state $b$. Note that the
convexity of the curve causes the trader to receive less BTC than
the actual BTC value of the USDT added to the pool, i.e. slippage
of the trade is determined by convexity of the price curve. 

\begin{center}
{[}Insert Figure \ref{Flo:uniswap_v2} approximately here{]}
\par\end{center}

The market price in Uniswap v2 pools is determined by the ratio of
the current quantities of two tokens in the pool:

\begin{equation}
p_{mkt}=\frac{y}{x}
\end{equation}

which shows the market price of token X (BTC) in units of
token Y (USDT). After the trade, as in example above, price of BTC
increases to $p'_{mkt}=\frac{y'}{x'}$, because removing some of BTC from
the pool makes it relatively scarce. In contrast, adding USDT to the
pool makes it a relatively abundant asset, such that it becomes relatively
cheaper. 

In Uniswap v2, LPs are required to deposit their liquidity in a 50-50
ratio, i.e. they deposit both tokens in equivalent quantities. If
liquidity providers add liquidity to the pool, the pool size
increases, i.e. the pool's constant $k$ increases. The larger the
pool size the lower the slippage of a given trade. However, trading
fees are then shared between a larger number of LPs. Recent studies
by \textcite{capponi2021adoption}, \textcite{lehar2021decentralized}, and \textcite{oneill2023} discuss the economics of Uniswap v2 in more detail. Specifically,
they show that LPs face a trade-off between the amount of fees earned
from liquidity providing and adverse selection cost, imposed on them
by informed arbitrageurs. \textcite{lehar2021decentralized} and \textcite{oneill2023}
further show that the pool size acts as an equilibrating mechanism
on DEXs, equivalent to the bid-ask spread on CEXs. 

\textbf{Uniswap v3.} In Uniswap v2, liquidity deposited in the pool is uniformly distributed
along the price curve, i.e. it is available for trading on the entire
price range, $[0,\infty[$. The advantage of this traditional implementation
of AMM is that LPs earn fees irrespective of the market price. However, it requires a great amount of capital to be ``locked
up'' in the pool without being ``active'', i.e. used in swaps. For
example, prices close to infinity are definitely unrealistic, but
the same amount of liquidity is reserved at these unrealistic prices
than at the market price.

Uniswap v3 seeks to improve ``capital efficiency'' of liquidity
providers, by allowing them to ``concentrate'' their liquidity on
smaller price ranges. When they open a new liquidity position, i.e.
deposit their tokens to a Uniswap v3 pool, they have to specify a
price range, $[p_{a},p_{b}]$, where $p_{a}$ is the minimum price
and $p_{b}$ the maximum price of token X in units of token Y at which their
position is active. According to Uniswap v3 whitepaper \parencite{adams2021uniswap}, ``a position
only needs to maintain enough reserves to support trading within its
range, and therefore can act like a constant product pool with larger
(virtual) reserves within that range''.

\begin{figure}[H]
\caption{\textbf{\footnotesize{}Price curve for Uniswap v3: constant product
market maker (CPMM) with virtual reserves.}{\footnotesize{} This figure
illustrates the virtual reserves curve on Uniswap v3. A position on
a range of $[p_{a},p_{b}]$ and a market price $p_{c} \in [p_{a},p_{b}]$
only needs to hold a smaller amount of real X reserves, $x_{real}$,
which are gradually depleted (i.e. swapped against Y reserves) until
the market price, $p_{c}$, reaches the upper price bound,
$p_{b}$. If the market price reaches $p_{b}$, then all X
reserves are depleted and the position is no longer active, i.e. stops
earning fees. The position then consists only of Y reserves. Similarly,
this position only needs to hold $y_{real}$ reserves, which are gradually
depleted until the market price reaches the lower price bound,
$p_{a}$, when Y reserves are depleted and the position becomes inactive.}}

{\small{}\vspace{1cm}
}{\small\par}

\label{Flo:uniswap_v3}
\begin{singlespace}
\centering{}\includegraphics[scale=0.7]{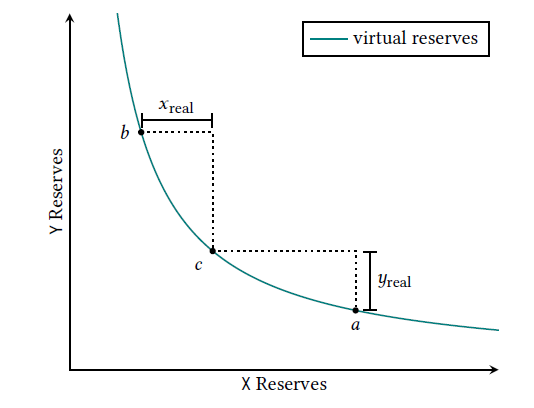}
\end{singlespace}
{\small{}\vspace{0.5cm}
}{\small\par}

{\small{}Source: Uniswap v3 whitepaper}{\small\par}

\end{figure}

Figure \ref{Flo:uniswap_v3} illustrates the notion of ``virtual
reserves''. A position
on a range of $[p_{a},p_{b}]$ and a market price $p_{c} \in [p_{a},p_{b}]$
only needs to hold a smaller amount of real X reserves, $x_{real}$,
which are gradually depleted (i.e. swapped against Y reserves) until
the market price, $p_{c}$, reaches the upper price bound,
$p_{b}$. If the market price reaches $p_{b}$, then all X
reserves are depleted and the position is no longer active, i.e. stops
earning fees. The position then consists only of Y reserves. Similarly,
this position only needs to hold $y_{real}$ reserves, which are gradually
depleted until the market price reaches the lower price bound,
$p_{a}$, when Y reserves are depleted and the position becomes inactive.
Should the pool price fall back to the specified price range, the
liquidity position becomes active again. Thus, virtual reserves ($x$ and $y$) magnify real reserves ($x_{real}$
and $y_{real}$) in a way that the liquidity in the pool matches that
of the market price range\footnote{\textcite{heimbach2022risks} provide mathematical relations between real and virtual reserves and discuss them in more detail.}.

\begin{center}
{[}Insert Figure \ref{Flo:uniswap_v3} approximately here{]}
\par\end{center}

The price curve for Uniswap v3 is a modification of $x \cdot y=k$, such that the position is solvent exactly within its price range, $[p_{lower},p_{upper}]$ \parencite{adams2021uniswap}:

\begin{center}
\begin{equation}
\left(x+\frac{L}{\sqrt{p_{upper}}}\right)\left(y+L\sqrt{p_{lower}}\right)=L^{2},
\label{eq:price_curve_v3}
\end{equation}
 
\par\end{center}

where $x$ and $y$ represent virtual reserves, and $L$ is a ``liquidity''
constant that shows the amount of liquidity deposited for each position. 

In contrast to Uniswap v2, the amount of tokens deposited in a liquidity
position is no longer in 50-50 ratio. In fact, it depends on the selected price range relative to the market price, with larger
reserves of Y required if the price range is skewed to prices lower
than the market price, $p_{mkt}$. If the selected price range is
strictly lower than the market price (and excludes it), then LP only
has to deposit token Y. This case corresponds to point $b$ being
below point $c$ on the price curve on Figure \ref{Flo:uniswap_v3},
i.e. $p_{mkt}>[p_{lower},p_{upper}]$. Similarly, if the preferred price range
is strictly higher than the market price, then LP only has to deposit
token X. This case corresponds to point $a$ being above point $c$
on Figure \ref{Flo:uniswap_v3}, i.e. $p_{mkt}<[p_{lower},p_{upper}]$. 



\textbf{Adding liquidity to Uniswap v3 pools.} On Uniswap v3, LPs
are able to specify the price range on which their liquidity position
will be active. To make this possible, the space of prices is divided into
discrete ticks, [$i$, $i \in \mathbb{Z}$]. According to Uniswap v3 whitepaper \parencite{adams2021uniswap},
there can be a tick at every price that is an integer power of
1.0001, such that the following relation holds for the price $p_i$
corresponding to tick $i$:
\begin{center}
\begin{equation}
p_{i}=1.0001^{i}
\end{equation}
\end{center}

Specifically, this relation implies that each tick is 1 bp (basis
point) away from its neighbouring ticks. However, not all ticks can
be initialized, but only those that are divisible by a pre-specified
pool parameter, the tick spacing $s$. For example, USDC/ETH 0.3\% pool has a $s$ of 60. Therefore, only ticks that are divisible by
60 can be initialized, i.e. (-120, -60, 0, 60, 120...)\footnote{In general, lower values for $s$ allow for more precise
price ranges. However, gas fees for swaps might be more expensive in
pools with low $s$, because the trader is required to
pay a constant gas fee for every initialized tick that the swap crosses.}. A tick range can then be defined as $[i,i+s]$. 

\begin{figure}[H]
\caption{\textbf{\footnotesize{}Examples of liquidity distributions on the
price space. }{\footnotesize{}Panel (I) shows that liquidity distribution
in Uniswap v2 is constant on the entire price range. Panel (II) shows
an example of a single position of a liquidity provider on Uniswap
v3 with the specified price range $[p_{lower},p_{upper}].$ Again, liquidity
provided is constant within the price range of this position. This
price range can cover one or more tick ranges $[i,i+s].$ Panel (III)
shows fragmentation of liquidity across multiple tick ranges on Uniswap
v3. Liquidity within each tick range is constant and represents an
aggregation of all LP positions active on this tick range.}}

{\small{}\vspace{1cm}
}{\small\par}

\label{Flo:uniswap_v3-ex}
\begin{centering}
\includegraphics[scale=0.5]{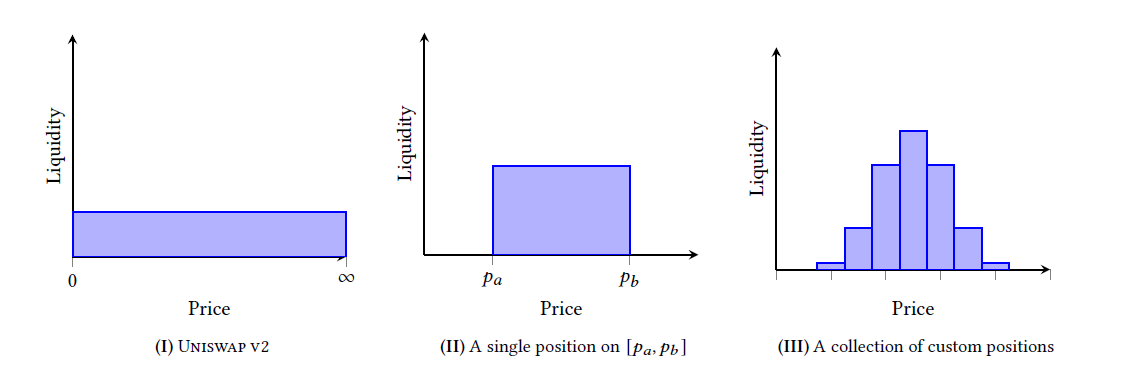}
\par\end{centering}
{\small{}\vspace{0.5cm}
}{\small\par}

{\small{}Source: \textcite{adams2021uniswap}.}{\small\par}
\end{figure}

One liquidity position of an LP can cover one or more tick ranges,
$[i,i+s]$. The liquidity on each tick range $[i,i+s]$, $L_{i}$,
is then an aggregation of all LP positions that are currently active
on it. Therefore, aggregated liquidity in a Uniswap v3 pool is no longer
constant (as in Uniswap v2), but fragmented across multiple tick ranges.
Figure \ref{Flo:uniswap_v3-ex} illustrates a stylized distribution
of liquidity on the price space for Uniswap v2 and Uniswap v3 \parencite{adams2021uniswap}.

\begin{center}
{[}Insert Figure \ref{Flo:uniswap_v3-ex} approximately here{]}
\par\end{center}

Let $p_{i}$ denote the price of token X (in units of token Y) that corresponds
to tick $i$. From Uniswap v3 whitepaper \parencite{adams2021uniswap}, we obtain the following
relations for $x_{i}$, quantity of tokens X locked in the tick range
$[i,i+s]$, and $y_{i},$ quantity of tokens Y locked in the same
tick range:

\begin{equation}
x_{i}=\frac{L_{i}}{\sqrt{z_{i}}}-\frac{L_{i}}{\sqrt{p_{i+s}}}\label{eq:xi}
\end{equation}

\begin{equation}
y_{i}=L_{i}\cdot(\sqrt{z_{i}}-\sqrt{p_{i}})\label{eq:yi}
\end{equation}

where
\begin{empheq}[left={z_{i}= \empheqlbrace}]{align*}
& p_{i}\text{ if }p_{mkt}\leq p_{i}\\
& p_{mkt}\text{ if }p_{i}<p_{mkt}<p_{i+s}\\
& p_{i+s}\text{ if }p_{i+s}\leq p_{mkt}
\end{empheq}
\par\leavevmode\par

Appendix D provides a numerical example of adding liquidity to USDC/ETH
0.3\% pool.

\pagebreak{}

\pagebreak{}

\begin{onehalfspace}

\section*{Appendix C}

\section*{Trading on Uniswap v3}
\end{onehalfspace}

\textbf{Trading on Uniswap v3 within a given tick range.} Suppose we would like to submit a
trade to USDC/ETH 0.3\% pool, i.e. swap tokens X and tokens Y. Let
$\triangle x$ and $\triangle y$ denote the total quantities of tokens
X and Y to be swapped in a trade (for example, if we would like to
buy tokens X with tokens Y, we have $\triangle x<0$ and $\triangle y>0$).
Since liquidity is fragmented along tick ranges, we have to consider
separately every tick range that is necessary for trade execution.
Let $\triangle x_{i}$ and $\triangle y_{i}$ denote the quantities of tokens
X and Y to be swapped within a given tick range $[i,i+s]$. Price
curve in Equation (\ref{eq:price_curve_v3}) imposes:

\begin{equation}
\left(x_{i}+\triangle x_{i}+\frac{L_{i}}{\sqrt{p_{i+s}}}\right)\cdot\left(y_{i}+\triangle y_{i}+L_{i}\cdot\sqrt{p_{i}}\right)=\left(x_{i}+\frac{L_{i}}{\sqrt{p_{i+s}}}\right)\cdot\left(y_{i}+L_{i}\cdot\sqrt{p_{i}}\right)=L_{i}^{2}
\end{equation}

from where we obtain the following for $\triangle x_{i}$ and $\triangle y_{i}$:

\begin{equation}
\Rightarrow
\begin{cases}
     \triangle x_i=-\frac{\triangle y_i \cdot (x_i+\frac{L_i}{\sqrt{p_{i+s}}})}{y_i+\triangle y_i+L_i \cdot \sqrt{p_i}}\\
     \triangle y_i=-\frac{\triangle x_i \cdot (y_i+L_i \cdot \sqrt{p_i})}{x_i+\triangle x_i+\frac{L_i}{\sqrt{p_{i+s}}}}
\end{cases} \label{eq:delta_xy}
\end{equation}

\par\leavevmode\par
\par\leavevmode\par
\par\leavevmode\par
\textbf{Tick-crossing trades on Uniswap v3.} Below we provide general formulas for trades that
are too large to be executed within a single tick range and involve
crossing several ticks. Suppose we would like to sell tokens X (USDC) for tokens Y (ETH). We are adding $\triangle x>0$ to the pool and withdrawing $\triangle y<0$. 

For a given tick range $[i,i+s]$, denote by $\triangle x_{i}^{max}$
the maximum quantity of tokens X that you can add to this range (by
taking reserves of token Y). Let $\triangle y_{i}^{min}$ denote
the minimum quantity of tokens Y that you can ``add'' to this range
- or, in other words, the maximum quantity of tokens Y than you can take
from this range by adding tokens X. From liquidity distribution
of the pool, we know reserves $y_{i}$ for each tick range, such that
$\triangle y_{i}^{min}=-y_{i}$. We can then compute $\triangle x_{i}^{max}$
using Equation (\ref{eq:delta_xy}):

\begin{empheq}[left=\empheqlbrace]{align*}
& \triangle y_{i}^{min}=-y_{i}\\
& \triangle x_{i}^{max}=-\frac{\triangle y_{i}^{min}\cdot(x_{i}+\frac{L_{i}}{\sqrt{p_{i+s}}})}{y_{i}+\triangle y_{i}^{min}+L_{i}\cdot\sqrt{p_{i}}}=\frac{x_{i}y_{i}}{L_{i}\cdot\sqrt{p_{i}}}+\frac{y_{i}}{\sqrt{p_{i}p_{i+s}}}
\end{empheq}
\par\leavevmode\par

Let $\{i_{start},i_{start}-s,...,i_{end}\}$ be the lower ticks of the consecutive
tick ranges that we use during the trade. We consume all available
liquidity in the tick ranges of lower ticks $i\in\{i_{start},i_{start}-s,...,i_{end}+s\}$,
and spend our remaining tokens X in the tick range of lower tick $i_{end}$,
such that $\triangle x$ and $\triangle y$ can be expressed as follows:

\begin{empheq}[left=\empheqlbrace]{align*}
& \triangle x=\sum_{i\in\{i_{start},i_{start}-s,...,i_{end}+s\}}\triangle x_{i}^{max}+\triangle x_{i_{end}}\\
& \triangle y=\sum_{i\in\{i_{start},i_{start}-s,...,i_{end}+s\}}\triangle y_{i}^{min}+\triangle y_{i_{end}}
\end{empheq}

\begin{equation}
\Rightarrow\boxed{\begin{cases}
\triangle x_{i_{end}}=\triangle x-\sum_{i\in\{i_{start},i_{start}-s,...,i_{end}+s\}}\triangle x_{i}^{max}\\
\triangle y\overset{(\ref{eq:delta_xy})}{=}-\sum_{i\in\{i_{start},i_{start}-s,...,i_{end}+s\}}y_{i}-\frac{\triangle x_{i_{end}}\cdot(y_{i_{end}}+L_{i_{end}}\cdot\sqrt{p_{i_{end}}})}{x_{i_{end}}+\triangle x_{i_{end}}+\frac{L_{i_{end}}}{\sqrt{p_{i_{end}+s}}}}
\end{cases}} \label{eq:delta_xie,delta_y}
\end{equation}

\par\leavevmode\par
\par\leavevmode\par

Appendix D provides numerical examples of calculating slippage for
trades within the current market range and for tick-crossing trades. Below we also provide formulas for the opposite trade direction, i.e. buying X with Y.

\textbf{Opposite trade direction: buy X with Y.} Suppose we would like to buy tokens X with tokens Y. We are adding $\triangle y>0$ to the pool and removing $\triangle x<0$.

For a given tick range $[i,i+s]$, denote by $\triangle y_{i}^{max}$
the maximum quantity of tokens Y that you can add to this range (by
removing reserves of token X). Let $\triangle x_{i}^{min}$ denote
the minimum quantity of tokens X that you can ``add'' to this range, i.e. the maximum quantity of tokens X than you can take
from this range by adding tokens Y. From liquidity distribution of the pool, we
know reserves $x_{i}$ for each tick range, such that $\triangle x_{i}^{min}=-x_{i}$. We can then compute $\triangle y_{i}^{max}$ using Equation (\ref{eq:delta_xy}):

\begin{empheq}[left=\empheqlbrace]{align*}
& \triangle x_{i}^{min}=-x_{i}\\
& \triangle y_{i}^{max}=-\frac{\triangle x_{i}^{min}\cdot(y_{i}+L_{i}\cdot\sqrt{p_{i}})}{x_{i}+\triangle x_{i}^{min}+\frac{L_{i}}{\sqrt{p_{i+s}}}}=\frac{x_{i}y_{i}\sqrt{p_{i+s}}}{L_{i}}+x_{i}\cdot\sqrt{p_{i}p_{i+s}}
\end{empheq}

Let $\{i_{start},i_{start}+s,...,i_{end}\}$ be the lower ticks of the consecutive tick ranges that we use during the trade. We consume all available
liquidity in the tick ranges of lower ticks $i\in\{i_{start},i_{start}+s,...,i_{end}-s\}$,
and spend our remaining tokens Y in the tick range of lower tick $i_{end}$,
such that $\triangle x$ and $\triangle y$ can be expressed as follows:

\begin{empheq}[left=\empheqlbrace]{align*}
\triangle y=\sum_{i\in\{i_{start},i_{start}+s,...,i_{end}-s\}}\triangle y_{i}^{max}+\triangle y_{i_{end}}\\
\triangle x=\sum_{i\in\{i_{start},i_{start}+s,...,i_{end}-s\}}\triangle x_{i}^{min}+\triangle x_{i_{end}}
\end{empheq}

\begin{equation}
\Rightarrow\boxed{\begin{cases}
\triangle y_{i_{end}}=\triangle y-\sum_{i\in\{i_{start},i_{start}+s,...,i_{end}-s\}}\triangle y_{i}^{max}\\
\triangle x\overset{(\ref{eq:delta_xy})}{=}-\sum_{i\in\{i_{start},i_{start}+s,...,i_{end}-s\}}x_{i}-\frac{\triangle y_{i_{end}}\cdot(x_{i_{end}}+\frac{L_{i_{end}}}{\sqrt{p_{i_{end}+s}}})}{y_{i_{end}}+\triangle y_{i_{end}}+L_{i_{end}}\cdot\sqrt{p_{i_{end}}}}.
\end{cases}}
\end{equation}

\pagebreak{}

\begin{onehalfspace}

\section*{Appendix D}

\section*{Numerical example}
\end{onehalfspace}

\textbf{Adding liquidity to USDC/ETH 0.3\% pool.} To illustrate, suppose that an LP would
like to add liquidity to USDC/ETH 0.3\% pool (with a $s$
of 60). Assume the market tick is 200618, such that the market
price is:
\begin{center}
$p_{mkt}=1.0001^{200618}$
\par\end{center}

To convert prices to a human-readable format, we have to scale $p_{mkt}$
by $10^{(d_{y}-d_{x})}$, where $d_{y}$ is the number of decimals
for token Y and $d_{x}$ is the number of decimals for token X. In
our example, $d_{y}=18$ for ETH and $d_{x}=6$ for USDC:
\begin{center}
$p_{mkt_{adj}}=\frac{1.0001^{200618}}{10^{(18-6)}}=0.00051558$ ETH per USDC
\end{center}

\begin{center}
which corresponds to $1/0.00051558=1939.56$ USDC per ETH
\end{center}

\begin{figure}[H]
\caption{\textbf{\footnotesize{}Numerical example: Adding liquidity to USDC/ETH
0.3\% pool.}}

{\small{}\vspace{0.5cm}
}{\small\par}

\label{Flo:uniswap_v3-ex-1}
\begin{centering}
\textbf{\small{}Panel A: Liquidity mints}{\small\par}
\par\end{centering}
{\small{}\vspace{0.5cm}
}{\small\par}
\begin{singlespace}
\begin{centering}
\includegraphics[scale=0.6]{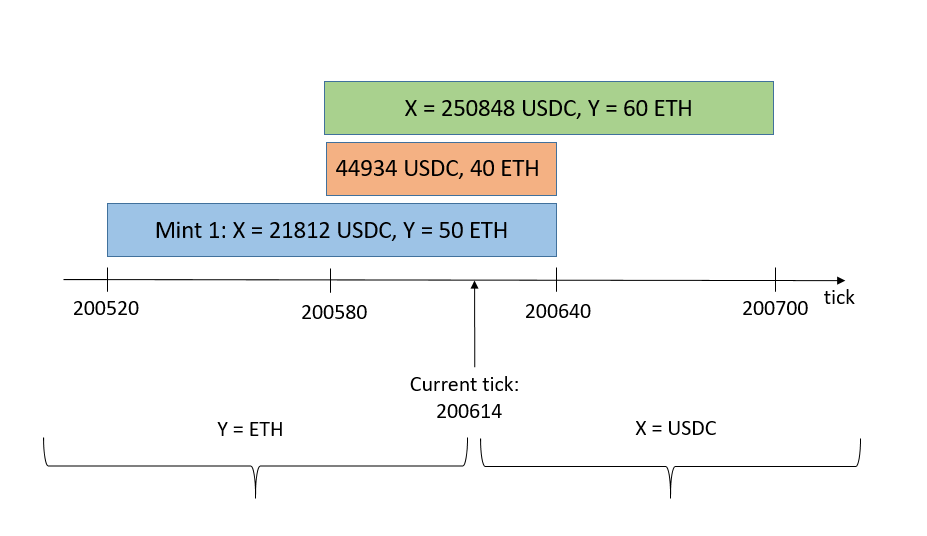}
\par\end{centering}
\end{singlespace}
{\small{}\vspace{0.5cm}
}{\small\par}
\begin{centering}
\textbf{\small{}Panel B: Liquidity by tick range}{\small\par}
\par\end{centering}
{\small{}\vspace{0.5cm}
}{\small\par}
\centering{}\includegraphics[scale=0.6]{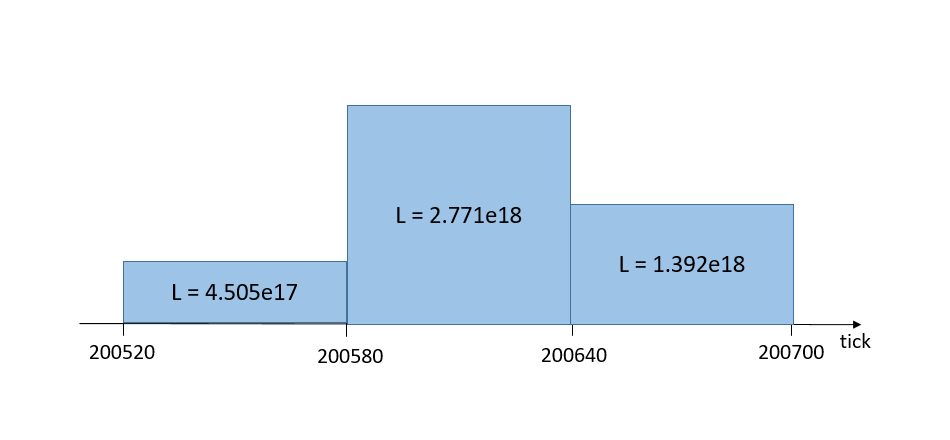}
\end{figure}

Suppose an LP would like to add 50 ETH (token Y) to the price range
$[p_{lower},p_{upper}]$ that corresponds to tick range $[i_{lower},i_{upper}]=[200520,200640]$, illustrated
as Mint 1 on Panel A of Figure \ref{Flo:uniswap_v3-ex-1}. Note that
this range includes two corresponding elementary tick ranges: $[200520,200580]$
and $[200580,200640].$ How many USDC (token X) does the LP
have to add to complete their liquidity position?

\begin{center}
{[}Insert Figure \ref{Flo:uniswap_v3-ex-1} approximately here{]}
\par\end{center}

To compute the required quantity of tokens X, we first infer the liquidity $L_{pos}$ of this position from Equation (\ref{eq:yi}):
\begin{center}
$L_{pos}=\frac{y_{pos}}{\sqrt{p_{mkt}}-\sqrt{p_{lower}}}=\frac{50\cdot10^{18}}{\sqrt{1.0001^{200618}}-\sqrt{1.0001^{200520}}}=4.505\cdot10^{17}$
\par\end{center}

with $z_{p}=p_{mkt},$ because the position includes the market tick,
$p_{lower}<p_{mkt}<p_{upper}.$ Note that we have to rescale 50 ETH to ``Uniswap v3''
format as follows: $y_{pos}=50\cdot10^{d_{y}}=50\cdot10^{18}$. Given
$L_{pos}$, we can compute $x_{pos}$ from Equation (\ref{eq:xi}):
\begin{center}
$x_{pos}=\frac{L_{pos}}{\sqrt{p_{mkt}}}-\frac{L_{pos}}{\sqrt{p_{upper}}}=L_{pos}\frac{\sqrt{p_{upper}}-\sqrt{p_{mkt}}}{\sqrt{p_{mkt}}\cdot\sqrt{p_{upper}}}$
\par\end{center}

Again to convert to ``human readable'' format, we have to scale
$x_{pos}$ by $10^{d_{x}}$, i.e. $x_{{pos}_{adj}}=x_{pos}/10^{6}=21812$ USDC.

Suppose there are two additional mints to the pool, as illustrated
on Panel A of Figure \ref{Flo:uniswap_v3-ex-1}. Table 2 below summarizes all three liquidity positions in the pool:

\vspace{0.5cm}

\begin{center}
{\small{}Table 2:}\textbf{\small{} Adding liquidity to USDC/ETH 0.3\%
pool}{\small\par}
\end{center}

\begin{center}
{\small{}}%
\begin{tabular}{|c|c|c|c|c|c|}
\hline 
 & $i_{lower}$ & $i_{upper}$ & {\small{}$x_{pos}$} & $y_{pos}$ & $L_{pos}$\tabularnewline
\hline 
{\small{}Mint 1} & {\small{}200520} & {\small{}200640} & {\small{}21812} & {\small{}50} & {\small{}$4.505e17$}\tabularnewline
\hline 
{\small{}Mint 2} & {\small{}200580} & {\small{}200640} & {\small{}44934} & {\small{}40} & {\small{}$9.281e17$}\tabularnewline
\hline 
{\small{}Mint 3} & {\small{}200580} & {\small{}200700} & {\small{}250848} & {\small{}60} & {\small{}$1.392e18$}\tabularnewline
\hline 
\end{tabular}{\small\par}
\par\end{center}

\vspace{0.5cm}

To compute aggregated liquidity on each tick range, $L_{i}$, we
have to add up liquidity of all active liquidity positions, $L_{pos}$,
on a given tick range. Panel B of Figure \ref{Flo:uniswap_v3-ex-1} shows
distribution of aggregated liquidity by tick range. For example, there
is only one active liquidity position on tick range $[200520,200580]$.
Therefore, $L_{i}=L_{pos}=4.505e17$. On the next tick range, $[200580,200640]$, all three liquidity positions are active, such that $L_{i}=\sum_{j=1}^{3} L_{pos_{j}} =2.771e18$.
From Equations (\ref{eq:xi}) and (\ref{eq:yi}), we can then compute aggregated $x_{i}$ and $y_{i}$
liquidity reserves for each tick range $[i,i+s]$. Table 3 below
summarizes distribution of liquidity and reserves by tick range:

\par\leavevmode\par
\par\leavevmode\par

\begin{center}
{\small{}Table 3:}\textbf{\small{} Liquidity distribution by tick
range}{\small\par}
\par\end{center}
\begin{center}
\begin{tabular}{|c|c|c|c|}
\hline 
 & {\small{}{[}200520,200580{]}} & {\small{}{[}200580,200640{]}} & {\small{}{[}200640,200700{]}}\tabularnewline
\hline 
{\small{}$L_{i}$} & {\small{}4.505e17} & {\small{}2.771e18} & {\small{}1.392e18}\tabularnewline
\hline 
{\small{}$x_{i}$ (USDC)} & {\small{}0} & {\small{}134159} & {\small{}183427}\tabularnewline
\hline 
{\small{}$y_{i}$ (ETH)} & {\small{}30.58} & {\small{}119.42} & {\small{}0}\tabularnewline
\hline 
\end{tabular}{\small\par}
\par\end{center}

\vspace{0.5cm}

Note that tick ranges that do not include the market tick, $i_{mkt}=200618$,
have only reserves in one of the tokens, either X (if $i>i_{mkt}$),
or Y (if $i+s<i_{mkt}$). 

\par\leavevmode\par

\textbf{Trading on Uniswap v3 within a given tick range.} Assume the current distribution
of liquidity in USDC/ETH 0.3\% pool as on Panel B of Figure \ref{Flo:uniswap_v3-ex-1}.
Suppose we would like to buy ETH with 200K USDC. How
many ETH can we get? 

We know that $\triangle x_{i}=200$K. Further, from Table 3 we know
$L_{i}$, $x_{i}$ and $y_{i}$ for the market tick range, {[}200580,
200640{]}. From Equation (\ref{eq:delta_xy}), we obtain for $\triangle y_{i}$:
\begin{center}
$\triangle y_{i}=-\frac{200000\cdot10^{6}\left(119.42\cdot10^{18}+2.771\cdot10^{18}\cdot\sqrt{1.0001^{200580}}\right)}{\left(134159\cdot10^{6}+200000\cdot10^{6}+\frac{2.771\cdot10^{18}}{\sqrt{1.0001^{200640}}}\right)\cdot10^{18}}=-102.94$ ETH
\par\end{center}

Overall, we obtain 102.94 ETH with 200K USDC. Therefore,
the average execution price is $p_{avg}=\frac{\triangle y}{\triangle x}=\frac{102.94}{200000}=0.0005147$ ETH per USDC. The slippage of this trade is:
\begin{center}
$Slippage=\left|\frac{p_{avg}}{p_{mkt}}-1\right|=\left|\frac{0.00051470}{0.00051558}-1\right|=0.0017=17$ bp
\par\end{center}

\par\leavevmode\par

\textbf{Tick-crossing trades.} In the numerical example
above, we could execute the whole trade within the market tick range,
because the reserves of Y in this tick range were sufficient, i.e.
$119.42>102.94$. To illustrate a tick-crossing
trade, suppose we would now like to buy ETH with a larger quantity of 250K
USDC. How many ETH can we get? 

We first start computing $\triangle x_{i}^{max}$ for the market
tick range, [200580,200640], given the current reserves
$y_{i}=119.42$ ETH and $x_{i}=134159$ USDC:
\begin{center}
$\triangle x_{i}^{max}=\frac{x_{i}y_{i}}{L_{i}\cdot\sqrt{p_{i}}}+\frac{y_{i}}{\sqrt{p_{i}p_{i+s}}}=\frac{134159\cdot10^{6}\cdot119.42\cdot10^{18}}{2.771\cdot10^{18}\cdot\sqrt{1.0001^{200580}}}+\frac{119.42\cdot10^{18}}{\sqrt{1.0001^{200580}\cdot1.0001^{200640}}}$\\
$\triangle x_{i,adj}^{max}=\frac{\triangle x_{i}^{max}}{10^{6}}=232064$ USDC
\par\end{center}

This means that we can spend 232064 USDC in the market tick range,
swapping them for the full amount of 119.42 ETH reserves. Therefore,
we are left with $\triangle x_{i_{end}}=250K-232064=17936$ USDC to spend
in the next (lower) tick range, [200520,200580]\footnote{In this simple example, our trade of 250K only crosses one tick. Therefore,
there is only one term in the sum $\sum_{i\in\{i_{start},i_{start}-s,...,i_{end}+s\}}\triangle x_{i}^{max}.$}.

From Equation (\ref{eq:delta_xie,delta_y}), we can now compute $\triangle y:$
\begin{center}
$\triangle y\overset{}{=}-119.42\cdot10^{18}-\frac{17936\cdot10^{6}(30.58\cdot10^{18}+4.505\cdot10^{17}\cdot\sqrt{1.0001^{200520}})}{0+17936\cdot10^{6}+\frac{4.505\cdot10^{17}}{\sqrt{1.0001^{200580}}}}$\\
$\triangle y_{adj}=\frac{\triangle y}{10^{18}}=-128.63$ ETH
\par\end{center}

Overall, we obtain 128.63 ETH with 250K USDC. Therefore,
the average execution price is $p_{avg}=\frac{\triangle y}{\triangle x}=\frac{128.63}{250000}=0.00051452$
ETH per USDC. The slippage of this trade is:
\begin{center}
$Slippage=|\frac{0.00051452}{0.00051558}-1|=0.0021=21$ bp
\par\end{center}

\pagebreak{}

\pagebreak{}

\section*{Appendix E}
\fontsize{11pt}{10pt}\selectfont
\begin{spacing}{2.5}
\vspace{-1cm}
\label{tab:variables}
\end{spacing}
\setlength\extrarowheight{4pt}
\begin{longtable}{lXl}\\ \toprule
Variable & Description & Source \\\hline

$Arbitrum$&A variable that takes the value 1 if the transaction takes place on Arbitrum, and zero otherwise. & The Graph\\

$BlockScaling$&A variable that takes the value 1 if the transaction takes place either on Arbitrum or on Polygon, and zero otherwise. & The Graph\\

$Buy$& A variable that takes the value 1 if the trade is a buy of tokens X with tokens Y, and zero otherwise. & The Graph \\

$Intensity$& Repositioning intensity of liquidity providers, measured as the value of repositioning mints (in \$) over a 5-minute interval divided by the total value minted (in \$) over the same interval. We define a repositioning mint as a mint preceded by a burn in the same pool by the same liquidity provider in the previous 5 minutes. & The Graph\\

$IntensFreq$& Alternative measure of repositioning intensity, calculated as the number of repositioning mints, divided by the total number of mints over a 5-minute interval.& The Graph\\ 

$Large$&A variable that takes the value 1 if the size of the trade exceeds a specified cutoff (e.g., \$1K or \$5K), and zero otherwise. &\\

$Conc$& Liquidity concentration, measured as market depth within 1\%/2\%/10\% of $p_{mkt}$, divided by $TVL$. Market depth within x\% of $p_{mkt}$ is computed as the \$ value of the liquidity between $\frac{p_{mkt}}{1+x\%}$ and $p_{mkt} \cdot (1+x\%)$ & The Graph\\

$p_{mid}$ & Mid price of a position, computed as $\sqrt{p_{lower} \cdot p_{upper}}$. & The Graph\\

$p_{mkt}$&Current price of the pool & The Graph\\

$Polygon$&A variable that takes the value 1 if the transaction takes place on Polygon, and zero otherwise. & The Graph\\

$Precision$& Precision of repositioning mints around $p_{mkt}$, computed and scaled as $1-1.0001^{\frac{-1}{Gap \cdot Length}}$. &The Graph\\

$Gap$ & Position gap of repositioning mints, measured as a relative difference between $p_{mid}$ and $p_{mkt}$, computed as $\left|\frac{p_{mid}}{p_{mkt}}-1 \right|$. & The Graph\\

$Length$ & Price range length of repositioning mints, computed as $\frac{\lvert p_{upper}-p_{lower} \rvert}{p_{mid}}$. & The Graph\\

$Return$ & 1-minute return of ETH/USD. & Binance\\

$Size$ & Hypothetical trade size in 100, 500, 1K, 5K, 10K, 50K, 100K (in \$). &\\

$Slippage$ & Slippage, computed as $\left|\frac{p_{avg}}{p_{mkt}}-1\right|$. Please refer to Appendix X for details on $p_{avg}$ computation on Uniswap v3. & The Graph\\

$TVL$ & Total value locked, i.e. \$ value of all the liquidity in the pool. & The Graph\\

$Volatility$ & Realized volatility of ETH/USD over the previous 24 hours, computed as the square root of the sum of squared 1-minute returns over previous 24 hours. & Binance\\

$Volume$ & Traded volume over previous 24 hours (in \$). & The Graph\\
\bottomrule
\end{longtable}
\setlength\extrarowheight{0pt}


\pagebreak{}

\section*{Internet Appendix}

\setcounter{table}{0} 
\renewcommand{\thetable}{IA\arabic{table}}
\pagestyle{empty}

\begin{table}[H]
\caption{\textbf{\footnotesize{}Repositioning intensity and liquidity concentration: Large LPs.}{\footnotesize{} 
This table replicates our analysis from Table \ref{Flo:intensity} for a subset of large LPs. For each pool and chain, we sum up total minted and burned liquidity (in \$) for each individual LP over our entire sample period. We then define large LPs as those in the top quartile of our sample distribution. Panel A reports results for the low-fee pool (ETH/USDC 0.05\%), and Panel B for the high-fee pool (ETH/USDC 0.3\%). $Arbitrum$ ($Polygon$) takes value of 1 for Arbitrum (Polygon) pools, and zero for Ethereum pools. Model (1) reports the results for the first-stage regression,
with repositioning intensity of LPs, $Intensity$, as the dependent
variable and $Arbitrum$ as the main explanatory variable, which is used as an instrument for the endogenous
$Intensity$. Model (2)
reports results for the second-stage regression, with liquidity concentration within 2\% of the market price, $Conc$, as the dependent variable. The set of instruments consists
of all explanatory variables, except that $Intensity$ is replaced
with $Arbitrum$. The table shows corresponding results for Polygon pools in Models (3) and (4), with $Polygon$ used as an instrument for the endogenous
$Intensity$. Models (5) and (6) use $BlockScaling$ that takes value of 1 for both Arbitrum and Polygon pools, and zero for Ethereum pools, as an instrument. 
The vector of control variables consists of $Volume$, 
$Volatility$ and $|Return|$. See Appendix E for a detailed description of variable definitions.
All regressions include hour- and day-fixed effects, with standard
errors clustered at the day level. T-statistics of the two-tailed
t-test with the null-hypothesis of a coefficient equaling zero are
reported in parentheses. {*}{*}{*}, {*}{*}, and {*} indicate statistical
significance at the 1\%, 5\%, and 10\% level, respectively.
}}

\label{Flo:intensity_act}
\begin{centering}
{\footnotesize{}\begin{tabularx}{\textwidth}{l *{6}{R@{ }l}} \\
\toprule
 \multicolumn{13}{l}{ \textbf{Panel A: ETH/USDC 0.05\%}} \\  \hline 
 & \multicolumn{4}{c}{Arbitrum} & \multicolumn{4}{c}{Polygon} & \multicolumn{4}{c}{BlockScaling}  \\ 
 & Int & & Conc & & Int & & Conc & & Int & & Conc &    \\ 
 &  & & 2\% & &  & & 2\% & &  & & 2\% &    \\ 
& (1) & & (2) & & (3) & & (4) & & (5) & & (6) &  \\ 
\midrule
$Arbitrum$          &        0.25&         ***&            &            &            &            &            &            &            &            &            &            \\
                    &     (32.87)&            &            &            &            &            &            &            &            &            &            &            \\
$Polygon$           &            &            &            &            &        0.19&         ***&            &            &            &            &            &            \\
                    &            &            &            &            &     (27.33)&            &            &            &            &            &            &            \\
$BlockScaling$      &            &            &            &            &            &            &            &            &        0.21&         ***&            &            \\
                    &            &            &            &            &            &            &            &            &     (35.16)&            &            &            \\
$Intensity$         &            &            &        0.08&         ***&            &            &        0.10&         ***&            &            &        0.10&         ***\\
                    &            &            &     (44.84)&            &            &            &     (49.92)&            &            &            &     (52.41)&            \\
$Volume$            &       -0.01&            &        0.01&         ***&       -0.01&         ***&        0.01&         ***&        0.01&          **&        0.01&         ***\\
                    &     (-1.58)&            &     (46.49)&            &     (-3.70)&            &     (50.86)&            &      (2.15)&            &     (53.91)&            \\
$Volatility$        &       -0.53&            &       -1.56&         ***&        0.18&            &       -1.48&         ***&       -1.53&         ***&       -1.37&         ***\\
                    &     (-0.91)&            &    (-39.64)&            &      (0.38)&            &    (-41.87)&            &     (-3.17)&            &    (-47.51)&            \\
$|Return|$          &       22.56&         ***&       -2.51&         ***&       17.12&         ***&       -2.26&         ***&       21.25&         ***&       -2.80&         ***\\
                    &     (16.61)&            &    (-24.52)&            &     (16.72)&            &    (-26.56)&            &     (19.10)&            &    (-34.40)&            \\
\midrule
$ Observations$     &      90,401&            &      90,401&            &     119,075&            &     119,075&            &     181,095&            &     181,095&            \\
$ HourFE$ & Yes & & Yes & & Yes & & Yes & & Yes & & Yes &  \\ 
$ DateFE$ & Yes & & Yes & & Yes & & Yes & & Yes & & Yes &  \\ 
\bottomrule
\end{tabularx}
}{\footnotesize\par}
\par\end{centering}
{\small{}\bigskip{}
}{\small\par}
\end{table}

\pagebreak{}

\begin{singlespace}
\centering{}{\footnotesize{}\begin{tabularx}{\textwidth}{l *{6}{R@{ }l}} \\
\toprule
 \multicolumn{13}{l}{ \textbf{Panel B: ETH/USDC 0.3\%}} \\  \hline 
 & \multicolumn{4}{c}{Arbitrum} & \multicolumn{4}{c}{Polygon} & \multicolumn{4}{c}{BlockScaling}  \\ 
 & Int & & Conc & & Int & & Conc & & Int & & Conc &    \\ 
 &  & &  2\% & & & & 2\% & & & & 2\% &    \\ 
& (1) & & (2) & & (3) & & (4) & & (5) & & (6) &  \\ 
\midrule
$Arbitrum$          &        0.17&         ***&            &            &            &            &            &            &            &            &            &            \\
                    &     (14.44)&            &            &            &            &            &            &            &            &            &            &            \\
$Polygon$           &            &            &            &            &        0.21&         ***&            &            &            &            &            &            \\
                    &            &            &            &            &     (28.36)&            &            &            &            &            &            &            \\
$BlockScaling$      &            &            &            &            &            &            &            &            &        0.18&         ***&            &            \\
                    &            &            &            &            &            &            &            &            &     (25.46)&            &            &            \\
$Intensity$         &            &            &        0.06&         ***&            &            &        0.17&         ***&            &            &        0.11&         ***\\
                    &            &            &     (27.21)&            &            &            &     (40.20)&            &            &            &     (41.73)&            \\
$Volume$            &       -0.19&         ***&        0.06&         ***&       -0.02&            &        0.04&         ***&       -0.02&            &        0.07&         ***\\
                    &     (-4.74)&            &     (37.89)&            &     (-1.10)&            &     (18.73)&            &     (-1.07)&            &     (48.48)&            \\
$Volatility$        &        3.27&         ***&       -1.13&         ***&       -0.05&            &       -0.95&         ***&        0.22&            &       -1.47&         ***\\
                    &      (4.64)&            &    (-33.73)&            &     (-0.10)&            &    (-14.01)&            &      (0.41)&            &    (-33.85)&            \\
$|Return|$          &        9.76&         ***&       -0.76&         ***&       18.27&         ***&       -3.23&         ***&       18.37&         ***&       -2.15&         ***\\
                    &      (7.63)&            &     (-9.61)&            &     (11.36)&            &    (-16.39)&            &     (12.84)&            &    (-17.85)&            \\
\midrule
$ Observations$     &      34,264&            &      34,264&            &      38,424&            &      38,424&            &      54,151&            &      54,151&            \\
$ HourFE$ & Yes & & Yes & & Yes & & Yes & & Yes & & Yes &  \\ 
$ DateFE$ & Yes & & Yes & & Yes & & Yes & & Yes & & Yes &  \\ 
\bottomrule
\end{tabularx}
}{\footnotesize\par}


\end{singlespace}

\pagebreak{}

\begin{table}[H]
\caption{\textbf{\footnotesize{}Repositioning intensity and liquidity concentration: Other pairs.}{\footnotesize{} 
This table replicates our analysis from Table \ref{Flo:intensity} for four other pairs that are actively traded across all three chains (i.e. 12 additional pools): BTC/ETH 0.05\%, BTC/ETH 0.3\%, UNI/ETH 0.3\% and LINK/ETH 0.3\%. $Arbitrum$ ($Polygon$) takes value of 1 for Arbitrum (Polygon) pools, and zero for Ethereum pools. Model (1) reports the results for the first-stage regression,
with repositioning intensity of LPs, $Intensity$, as the dependent
variable and $Arbitrum$ as the main explanatory variable, which is used as an instrument for the endogenous
$Intensity$. Model (2)
reports results for the second-stage regression, with liquidity concentration within 2\% of the market price, $Conc$, as the dependent variable. The set of instruments consists
of all explanatory variables, except that $Intensity$ is replaced
with $Arbitrum$. The table shows corresponding results for Polygon pools in Models (3) and (4), with $Polygon$ used as an instrument for the endogenous
$Intensity$. Models (5) and (6) use $BlockScaling$ that takes value of 1 for both Arbitrum and Polygon pools, and zero for Ethereum pools, as an instrument. 
The vector of control variables consists of $Volume$, 
$Volatility$ and $|Return|$. See Appendix E for a detailed description of variable definitions.
All regressions include hour- and day-fixed effects, with standard
errors clustered at the day level. T-statistics of the two-tailed
t-test with the null-hypothesis of a coefficient equaling zero are
reported in parentheses. {*}{*}{*}, {*}{*}, and {*} indicate statistical
significance at the 1\%, 5\%, and 10\% level, respectively.
}}

\label{Flo:intensity_other}
\begin{centering}
{\footnotesize{}\begin{tabularx}{\textwidth}{l *{6}{R@{ }l}} \\
\toprule
 & \multicolumn{4}{c}{Arbitrum} & \multicolumn{4}{c}{Polygon} & \multicolumn{4}{c}{BlockScaling}  \\ 
 & Int & & Conc & & Int & & Conc & & Int & & Conc &    \\ 
 &  & & $ 2\%$ & &  & & $ 2\%$ & &  & & $ 2\%$ &    \\ 
& (1) & & (2) & & (3) & & (4) & & (5) & & (6) &  \\ 
\midrule
$Arbitrum$          &        0.18&         ***&            &            &            &            &            &            &            &            &            &            \\
                    &     (26.16)&            &            &            &            &            &            &            &            &            &            &            \\
$Polygon$           &            &            &            &            &        0.12&         ***&            &            &            &            &            &            \\
                    &            &            &            &            &     (22.47)&            &            &            &            &            &            &            \\
$BlockScaling$      &            &            &            &            &            &            &            &            &        0.15&         ***&            &            \\
                    &            &            &            &            &            &            &            &            &     (28.36)&            &            &            \\
$Intensity$         &            &            &        0.11&         ***&            &            &        0.44&         ***&            &            &        0.29&         ***\\
                    &            &            &     (34.47)&            &            &            &     (40.52)&            &            &            &     (48.86)&            \\
$Volume$            &        0.01&          **&        0.03&         ***&        0.03&         ***&        0.03&         ***&        0.02&         ***&        0.03&         ***\\
                    &      (2.55)&            &     (40.08)&            &      (8.06)&            &     (23.79)&            &      (5.27)&            &     (46.91)&            \\
$Volatility$        &       -0.81&          **&       -1.76&         ***&        0.08&            &       -2.52&         ***&       -0.26&            &       -2.49&         ***\\
                    &     (-2.58)&            &    (-66.80)&            &      (0.34)&            &    (-38.92)&            &     (-0.96)&            &    (-58.44)&            \\
$|Return|$          &       20.35&         ***&       -3.15&         ***&       17.51&         ***&       -9.14&         ***&       20.68&         ***&       -7.50&         ***\\
                    &      (7.08)&            &     (-8.07)&            &     (12.06)&            &    (-15.23)&            &     (11.57)&            &    (-13.11)&            \\
\midrule
$ Observations$     &      57,904&            &      57,904&            &      98,608&            &      98,608&            &     131,461&            &     131,461&            \\
$ HourFE$ & Yes & & Yes & & Yes & & Yes & & Yes & & Yes &  \\ 
$ DateFE$ & Yes & & Yes & & Yes & & Yes & & Yes & & Yes &  \\ 
\bottomrule
\end{tabularx}
}{\footnotesize\par}
\par\end{centering}
{\small{}\bigskip{}
}{\small\par}
\end{table}

\pagebreak{}

\begin{table}[H]
\caption{\textbf{\footnotesize{}Repositioning
precision and liquidity concentration: Arbitrum and Polygon.}{\footnotesize{} This table presents
results of instrumental variable regressions that test causal effect
of repositioning precision of LPs, separately for Arbitrum and Polygon. We use three measures of repositioning precision: $Gap$, $Length$ and $Precision$. See Appendix E for
a detailed description of variable definitions. Panel A reports results for the low-fee pool (ETH/USDC 0.05\%), and Panel B for the high-fee pool (ETH/USDC 0.3\%). Model (1)-(3) report the results of the second-stage regressions with liquidity concentration within 2\% as the
dependent variable. $Arbitrum$ is used as an instrument for the endogenous repositioning
precision. Models (4)-(6) present corresponding results, with $Polygon$ used as an instrument for repositioning
precision.
The vector of control variables
consists of $Volume$, $Volatility$ and $|Return|$. All regressions include
hour- and day-fixed effects, with standard errors clustered at the
day level. T-statistics of the two-tailed t-test with the null-hypothesis
of a coefficient equaling zero are reported in parentheses. {*}{*}{*},
{*}{*}, and {*} indicate statistical significance at the 1\%, 5\%,
and 10\% level, respectively.}{\small{}\bigskip{}
}}

\label{Flo:precision_ia}
\begin{centering}
{\footnotesize{}\begin{tabularx}{\textwidth}{l *{6}{R@{ }l}} \\
\toprule
 \multicolumn{13}{l}{ \textbf{Panel A: ETH/USDC 0.05\%}} \\  \hline 
 & \multicolumn{6}{c}{Arbitrum} & \multicolumn{6}{c}{Polygon} \\  
 & \multicolumn{6}{c}{Conc, $ 2\%$} & \multicolumn{6}{c}{Conc, $ 2\%$}    \\ 
& (1) & & (2) & & (3) & & (4) & & (5) & & (6) &  \\ 
\midrule
$Gap$               &       -2.73&         ***&            &            &            &            &       -4.76&         ***&            &            &            &            \\
                    &    (-17.52)&            &            &            &            &            &    (-12.54)&            &            &            &            &            \\
$Length$            &            &            &       -0.35&         ***&            &            &            &            &       -0.50&         ***&            &            \\
                    &            &            &    (-28.48)&            &            &            &            &            &    (-24.57)&            &            &            \\
$Precision$         &            &            &            &            &        0.12&         ***&            &            &            &            &        0.22&         ***\\
                    &            &            &            &            &     (24.31)&            &            &            &            &            &     (18.00)&            \\
$Volume$            &        0.01&         ***&        0.01&         ***&        0.01&         ***&        0.01&         ***&        0.01&         ***&        0.01&         ***\\
                    &     (15.97)&            &     (35.36)&            &     (25.24)&            &      (8.27)&            &     (18.51)&            &     (16.40)&            \\
$Volatility$        &       -1.15&         ***&       -1.41&         ***&       -1.28&         ***&       -0.65&         ***&       -1.07&         ***&       -0.89&         ***\\
                    &    (-11.91)&            &    (-22.52)&            &    (-19.44)&            &     (-4.53)&            &    (-14.87)&            &    (-10.39)&            \\
$|Return|$          &       -1.64&         ***&       -2.37&         ***&       -1.15&         ***&       -2.12&         ***&       -2.75&         ***&       -1.76&         ***\\
                    &    (-11.24)&            &    (-16.20)&            &    (-11.20)&            &     (-8.86)&            &    (-16.42)&            &    (-11.02)&            \\
\midrule
$ Observations$     &      47,987&            &     47,987&            &      47,987&            &      65,146&            &     65,146&            &      65,146&            \\
$ HourFE$ & Yes & & Yes & & Yes & & Yes & & Yes & & Yes &  \\ 
$ DateFE$ & Yes & & Yes & & Yes & & Yes & & Yes & & Yes &  \\ 
\bottomrule
\end{tabularx}
}{\footnotesize\par}
\par\end{centering}
{\small{}\bigskip{}
}{\small\par}
\end{table}

\pagebreak{}

\begin{singlespace}
\centering{}
{\footnotesize{}\begin{tabularx}{\textwidth}{l *{6}{R@{ }l}} \\
\toprule
 \multicolumn{13}{l}{ \textbf{Panel B: ETH/USDC 0.3\%}} \\  \hline 
 & \multicolumn{6}{c}{Arbitrum} & \multicolumn{6}{c}{Polygon} \\  
 & \multicolumn{6}{c}{Conc, $ 2\%$} & \multicolumn{6}{c}{Conc, $ 2\%$}    \\ 
& (1) & & (2) & & (3) & & (4) & & (5) & & (6) &  \\ 
\midrule
$Gap$               &       -0.77&         ***&            &            &            &            &       -8.76&         ***&            &            &            &            \\
                    &     (-7.29)&            &            &            &            &            &     (-3.71)&            &            &            &            &            \\
$Length$            &            &            &       -0.06&         ***&            &            &            &            &       -0.30&         ***&            &            \\
                    &            &            &     (-9.89)&            &            &            &            &            &    (-16.98)&            &            &            \\
$Precision$         &            &            &            &            &        0.07&         ***&            &            &            &            &        0.24&         ***\\
                    &            &            &            &            &      (8.06)&            &            &            &            &            &     (15.80)&            \\
$Volume$            &        0.04&         ***&        0.05&         ***&        0.04&         ***&       -0.07&            &        0.02&         ***&       -0.00&            \\
                    &      (9.37)&            &     (17.81)&            &     (10.15)&            &     (-1.60)&            &      (4.26)&            &     (-0.13)&            \\
$Volatility$        &       -0.81&         ***&       -0.97&         ***&       -0.59&         ***&        1.79&            &       -0.61&         ***&        0.24&            \\
                    &     (-8.89)&            &    (-15.25)&            &     (-6.23)&            &      (1.52)&            &     (-4.41)&            &      (1.29)&            \\
$|Return|$          &       -0.45&         ***&       -0.27&         ***&       -0.31&         ***&       -2.19&          **&       -1.35&         ***&       -2.26&         ***\\
                    &     (-3.91)&            &     (-3.65)&            &     (-3.01)&            &     (-2.00)&            &     (-6.66)&            &     (-7.58)&            \\
\midrule
$ Observations$     &       7,153&            &       7,153&            &       7,153&            &      10,746&            &      10,746&            &      10,746&            \\
$ HourFE$ & Yes & & Yes & & Yes & & Yes & & Yes & & Yes &  \\ 
$ DateFE$ & Yes & & Yes & & Yes & & Yes & & Yes & & Yes &  \\ 
\bottomrule
\end{tabularx}
}{\footnotesize\par}

\end{singlespace}

\pagebreak{}

\begin{table}[H]
\caption{\textbf{\footnotesize{}Repositioning
precision and liquidity concentration: Other pairs.}{\footnotesize{} This table replicates our analysis from Table \ref{Flo:precision} for four other pairs that are actively traded across all three chains (i.e. 12 additional pools): BTC/ETH 0.05\%, BTC/ETH 0.3\%, UNI/ETH 0.3\% and LINK/ETH 0.3\%. We use three measures of repositioning precision: $Gap$, $Length$ and $Precision$. See Appendix E for
a detailed description of variable definitions. Model (1) reports the results of the first-stage IV regression with $Gap$ as the dependent variable. Model (2) reports the results of the second-stage regression with liquidity concentration within 2\% as the
dependent variable. $Blockscaling$ is used as an instrument for the endogenous repositioning
precision (i.e. $Gap$). Models (3) and (4) present corresponding results for $Length$ as a measure of repositioning precision. Models (5) and (6) present results for $Precision$. 
The vector of control variables
consists of $Volume$, $Volatility$ and $|Return|$.  All regressions include
hour- and day-fixed effects, with standard errors clustered at the
day level. T-statistics of the two-tailed t-test with the null-hypothesis
of a coefficient equaling zero are reported in parentheses. {*}{*}{*},
{*}{*}, and {*} indicate statistical significance at the 1\%, 5\%,
and 10\% level, respectively.}{\small{}\bigskip{}
}}

\label{Flo:precision_ia_other}
\begin{centering}
{\footnotesize{}\begin{tabularx}{\textwidth}{l *{6}{R@{ }l}} \\
\toprule
 & Gap & & Conc & & Length & & Conc & & Precision & & Conc &    \\ 
 &  & & $ 2\%$ & & & & $ 2\%$ & & & & $ 2\%$ &    \\ 
& (1) & & (2) & & (3) & & (4) & & (5) & & (6) &  \\ 
\midrule
$BlockScaling$      &       -0.003&          **&            &            &       -0.14&         ***&            &            &        0.13&         ***&            &            \\
                    &     (-2.25)&            &            &            &    (-15.45)&            &            &            &     (10.64)&            &            &            \\
$Gap$               &            &            &      -13.64&         ***&            &            &            &            &            &            &            &            \\
                    &            &            &     (-4.69)&            &            &            &            &            &            &            &            &            \\
$Length$            &            &            &            &            &            &            &       -0.30&         ***&            &            &            &            \\
                    &            &            &            &            &            &            &    (-43.83)&            &            &            &            &            \\
$Precision$         &            &            &            &            &            &            &            &            &            &            &        0.28&         ***\\
                    &            &            &            &            &            &            &            &            &            &            &     (19.27)&            \\
$Volume$            &       -0.00&         ***&       -0.02&           *&       -0.07&         ***&        0.02&         ***&        0.08&         ***&        0.02&         ***\\
                    &     (-6.80)&            &     (-1.68)&            &    (-10.74)&            &     (23.07)&            &      (8.95)&            &      (9.20)&            \\
$Volatility$        &        0.59&         ***&        5.23&         ***&        5.19&         ***&       -1.03&         ***&      -10.12&         ***&        0.04&            \\
                    &     (11.53)&            &      (2.98)&            &     (12.68)&            &    (-19.20)&            &    (-14.26)&            &      (0.26)&            \\
$|Return|$          &       -0.02&            &       -1.27&            &       -6.14&         ***&       -3.29&         ***&        5.06&         ***&       -2.40&         ***\\
                    &     (-0.19)&            &     (-0.95)&            &     (-5.63)&            &    (-10.40)&            &      (3.97)&            &     (-6.80)&            \\
\midrule
$ Observations$     &      36,726&            &      36,726&            &     36,726&            &     36,726&            &      36,726&            &      36,726&            \\
$ HourFE$ & Yes & & Yes & & Yes & & Yes & & Yes & & Yes &  \\ 
$ DateFE$ & Yes & & Yes & & Yes & & Yes & & Yes & & Yes &  \\ 
\bottomrule
\end{tabularx}
}{\footnotesize\par}
\par\end{centering}
{\small{}\bigskip{}
}{\small\par}
\end{table}

\end{document}